\documentclass[apj]{emulateapj}
\pdfoutput=1
\usepackage{graphicx}

\usepackage{tabularx}    
\usepackage{subfigure}  
\usepackage{amsmath, amssymb}
\usepackage{natbib}		
\usepackage[figuresright]{rotating}
\usepackage{hyperref}
\usepackage{multirow}
\usepackage{booktabs}

\makeatletter

\newcommand{\Rmnum}[1]{\expandafter\@slowromancap\romannumeral #1@}
\makeatother
\newcommand\T{\rule{0pt}{2.6ex}}
\newcommand\B{\rule[-1.2ex]{0pt}{0pt}}
\newcommand{\oii}{[\ion{O}{2}]~}
\newcommand{\smass}{log($M_*$/$M_{\sun}$)}

\newcommand{\sfr}{log($\psi$)}
\newcommand{\ssfr}{log($\psi$/$M_{*}$)}
\newcommand{\zrange}{0.74$<$$z$$<$1.4~}
\newcommand{\corf}{$\xi(r)$}
\newcommand{\corfp}{$\xi(r_p, \pi)$}
\newcommand{\wprp}{$\omega_{p}(r_{p})$}

\newcommand{\z}{$z$}

\begin{document}

 \title{The DEEP2 Galaxy Redshift Survey: Clustering Dependence on Galaxy Stellar Mass 
 	and Star Formation Rate at $z\sim1$}

\author{Nick Mostek\altaffilmark{1}} 
\altaffiltext{1}{Space Sciences Laboratory, University of California, Berkeley, CA, 94720}
\email{njmostek@lbl.gov}
 
\author{Alison L. Coil\altaffilmark{2}}
\altaffiltext{2}{Center for Astrophysics and Space Sciences, University of California, San Diego, La Jolla, CA, 92093}
    
\author{Michael Cooper\altaffilmark{3}}
\altaffiltext{3}{Center for Galaxy Evolution, Department of Physics and Astronomy, University of California, Irvine, CA, 92697}

\author{Marc Davis\altaffilmark{4}}
\altaffiltext{4}{Department of Astronomy, University of California at Berkeley, Berkeley, CA, 94720}

\author{Jeffrey A. Newman\altaffilmark{5}}
\altaffiltext{5}{Department of Physics and Astronomy and PITT-PACC, University of Pittsburgh, Pittsburgh, PA, 15620}    
    
\author{Benjamin J. Weiner\altaffilmark{6}}
\altaffiltext{6}{Steward Observatory, Department of Astronomy, University of Arizona, Tucson, AZ, 85721}

\begin{abstract}
We present DEEP2 galaxy clustering measurements at 
$z\sim1$ as a function of stellar mass, star formation rate (SFR), and 
specific SFR (sSFR).   
We find a strong positive correlation between stellar mass and 
clustering amplitude on 1-10 $h^{-1}$ Mpc scales for blue, star-forming galaxies with 
9.5$<$\smass$<$11 and no dependence for red, quiescent galaxies 
with 10.5$<$\smass$<$11.5.  Using
recently re-calibrated DEEP2 SFRs from restframe $B$-band magnitude and 
optical colors, we find that 
within the blue galaxy population at $z\sim1$ the clustering
amplitude increases strongly with increasing SFR and decreasing sSFR. 
For red galaxies there is no significant correlation between clustering amplitude and 
either SFR or sSFR.  Blue galaxies with high SFR or
low sSFR 
%resemble red galaxies in stellar mass, similar to previously observed trends,
 are as clustered on large scales as red galaxies. 
We find that the clustering trend observed with SFR can be explained mostly, but not entirely, 
by the correlation between stellar mass and clustering amplitude for blue galaxies.
We also show that galaxies above the star-forming ``main sequence" 
are less clustered than galaxies below the main sequence, at a given
stellar mass.  These results are not consistent with the high-sSFR 
population being dominated by major mergers. 
We also measure the clustering amplitude on small scales 
($\leq$0.3~$h^{-1}$ Mpc) and find an enhanced clustering signal relative to the best-fit large-scale 
power law for red galaxies with high stellar mass, blue galaxies with high SFR,
and both red and blue galaxies with high sSFR. The increased small-scale 
clustering for galaxies with high sSFRs is likely linked to triggered 
star formation in interacting galaxies. These measurements provide strong constraints on 
galaxy evolution and halo occupation distribution models at $z\sim1$.
\end{abstract}
\keywords{galaxies: evolution --- galaxies: active --- galaxies: high-redshift}

\section{Introduction}
\label{sec:Intro}

The non-uniform spatial distribution of galaxies in the Universe, known as
large-scale structure, likely reflects the non-uniform spatial distribution of
dark matter.  \cite{Kaiser84} showed that galaxies are biased tracers of
the total underlying mass distribution.  Measurements of galaxy
clustering therefore reflect, in part, the clustering of dark matter
at a given cosmic epoch.  Further, measurements of galaxy clustering
over cosmic time constrain the evolving relationship between dark
matter, which is governed by gravity and reflects cosmological
parameters, and galaxy properties, which are governed by processes
associated with baryonic physics.  As such, galaxy clustering
measurements have the ability to constrain both cosmology \citep{Peacock01,
Seljak04, Eisenstein05} and galaxy evolution physics \citep{Zheng07,
Conroy09, Zehavi12}
and are thus a powerful tool in the study of large-scale structure formation.

The standard method used to measure galaxy clustering is the two-point
correlation function (2PCF), which quantifies the excess probability
of finding a galaxy in a given volume relative to a random
distribution \citep{Davis83, Landy93}. Combined
with redshift information, the 2PCF produces a statistical
representation of the three-dimensional galaxy density distribution as
a function of scale.  
In recent years, the clustering of galaxies with respect to
dark matter has been interpreted using the ``halo model''
framework.  The clustering amplitude of both dark matter particles and
dark matter halos can be analytically fit, for a given cosmology,
using $N$-body simulations \citep{Mo96,Sheth01}. 
Simple proposed analytic models describe how galaxies
populate dark matter halos as a function of scale and mass \citep{Jing98,
Peacock00, Seljak00, Benson00,  Cooray02}.  In these halo occupation
distribution (HOD) models, galaxies populate dark matter halos using a
probability distribution generated to match measured galaxy clustering
statistics, therefore statistically connecting the baryonic matter of
galaxies to the dark matter halos \citep{Berlind02}. Such
HOD models typically include descriptions of the halo structure in
terms of the ``central" galaxy within a given dark matter halo and
the sub-halo ``satellite" galaxies that reside within the
parent dark matter halo \citep{Kravtsov04}.

Modern day redshift surveys are uniquely suited
to characterize the large-scale galaxy distribution through the 2PCF
over a wide range of cosmic time, including surveys of local
\citep[\z$<$0.15,][]{York00, Colless01, DR7}, intermediate \citep[$z\sim0.5$,][]{Blake09,
Eisenstein11, Coil11}, and higher redshift galaxies 
\citep[0.7$<$\z$<$2,][]{VVDS05, Lilly07, NMBS, Newman12}.
Such surveys have shown that the clustering of galaxies is correlated with 
various galaxy properties such as luminosity and color, 
where the clustering amplitude increases for galaxies with redder color 
and/or higher luminosity at least to \z$<$1 (\citealt{Coil08}, hereafter C08; 
\citealt{Zehavi11}).

While there is a strong, demonstrable relationship between galaxy
luminosity and color with clustering amplitude, interpreting this relationship in terms of the underlying 
physics can be complicated, as there are competing effects due to varying 
star formation histories, mass-to-light ratios, metallicity, dust, 
and other complex galaxy physics.  A cleaner physical
relationship in the mapping between galaxies and dark matter halos 
may be derived from measuring clustering as a function of 
properties that are easier to interpret, such as stellar mass and star formation rate (SFR). 
Using these galaxy properties, one may obtain direct constraints on 
the baryonic processes involved in galaxy evolution rather than with 
intermediate dependencies on luminosity and color.
For example, \cite{Zheng07} constrained the HOD of the luminosity-dependent
clustering results from the SDSS and DEEP2 surveys, but the interpretation
is complicated by uncertainties in connecting luminosity and stellar mass for 
different galaxy populations.  \citeauthor{Zheng07} showed that while one can 
use the HOD to constrain the evolution of the stellar-halo mass relationship, further
model refinement requires clustering measurements 
for galaxy samples selected by stellar mass. 

Measuring clustering with respect to galaxy properties has been made possible by the advent of large redshift 
surveys in the last decade.  Using the SDSS, \cite{Li06} found that 
on large scales ($>1~h^{-1}$ Mpc), the dependence of galaxy clustering on 
stellar mass closely mirrors that of luminosity; the clustering bias is 
relatively flat below the characteristic stellar mass scale $M^*$ and increases 
exponentially above $M^*$. They further investigate how the clustering 
amplitude depends on other galaxy color, morphology, 
and stellar mass, and find significant scale-dependent
differences at low stellar mass (\smass$\lesssim$10) between red and blue galaxy samples.
At higher redshift ($z\sim0.85$), \cite{Meneux08} measured the 
clustering of $\approx$3200 galaxies in the VVDS \citep{VVDS05} 
selected by stellar mass.  Similar 
to the local behavior, \citeauthor{Meneux08} found that the clustering amplitude 
increases with stellar mass and that the 
clustering amplitude at 9.5$<$\smass$<$10.5 is systematically lower 
than in local samples. The clustering 
amplitude at higher stellar masses (\smass$>$10.5) has not evolved in the 
same span of time within the measurement errors. 

While it is preferable to measure the 2PCF using spectroscopic redshifts, several studies have used photometric redshifts
to measure the two-dimensional angular correlation function with respect to stellar mass.  Because photometric redshifts
depend only on accurate panchromatic photometry, larger sample sizes are more easily obtained, particularly at
higher redshifts where spectroscopy requires extensive telescope resources. \cite{Foucaud10} used $K_{s}$-band imaging
from the Palomar Observatory Wide-field Infrared Survey \citep{Bundy06} and CFHT optical photometry
to measure the angular clustering of stellar mass-limited galaxy samples and found that a strong positive correlation exists between
stellar mass and halo mass to \z$<2$. This result was later confirmed with additional angular
clustering measurements with more precise photometric redshifts from the NEWFIRM Medium Band Survey \citep[NMBS,][]{NMBS} 
between 1$<$\z$<$2 \citep{Wake11}.
\cite{Alexi12} also used the angular correlation function along with
weak lensing maps from COSMOS \citep{Capak07, Ilbert09} 
to fit an HOD model between 0.2$<$\z$<$1. 
They found that the peak of the ratio of $M_*$/$M_{\rm{halo}}$ is a 
constant with cosmic time, indicating a fundamental connection
between the fraction of baryons within a given halo mass and the 
conversion rate of those baryons into stars. While angular clustering though photo-$z$s is less precise
than the 2PCF, some of the most advanced HOD models to date have been built off such data.

In contrast to stellar mass, there has been very little study of how 
galaxy clustering depends on SFR. For local galaxies, \cite{Li08} used SFRs of SDSS galaxies to
study the clustering dependence on the specific star-formation rate
(sSFR, the SFR divided by stellar mass), though not the dependence on 
SFR alone.  At higher redshifts
($z\sim2$), a recent study by \cite{Lin12} measured the angular
correlation function for $BzK$-selected star-forming galaxies in 160 arcmin$^2$ of the GOODS-North field \citep{GOODS}. 
Lin et al. found a significant large-scale clustering dependence with increasing SFR and decreasing sSFR. However, 
no clustering analyses using SFR or sSFR have been performed at 
$z\sim1$ where the global SFR is rapidly declining  \citep{Hopkins06, Zhu09}. 

The effects of galaxy clustering on SFR are interesting for HOD models as the 
parent dark matter halos of galaxies may influence the SFR through 
various mechanisms. For galaxy populations that are highly clustered 
and therefore reside in massive halos, the 
available gas in the galaxy may have been stripped, depleting the galaxy of the 
material necessary to sustain further star formation. Alternatively, the gas 
may be heated to the virial temperature of the host halo, such that the cooling
time becomes long enough to suppress star formation.  However, such
star formation ``quenching'' could arise from secular evolution, which
would not necessarily be dependent upon host halo properties.  
Similarly, the sSFR is a measure of the SFR relative 
to the existing stellar mass in a galaxy and 
 reflects how \emph{efficiently} a galaxy is processing gas into stars. 
Determining the clustering properties and galaxy bias with respect to SFR and 
sSFR as a function of redshift is therefore crucial to understanding how the 
host dark matter halo and surrounding environment of a galaxy play a role 
in the evolution of star formation.

While clustering analyses of the 2PCF as a function of stellar mass and SFR are 
relatively rare at intermediate redshifts, there are measurements of the 
relationship between these parameters and galaxy environment. 
Environment studies generally measure the galaxy 
overdensity for a given population relative to the mean density at the same
redshift, often using the N$^{th}$ nearest neighbor statistic (see \cite{Muldrew12} 
for a review of various environmental measures).  The advantage 
of using the local environment is that it provides better statistics in what is 
essentially a cross-correlation between a given galaxy sample and the mean 
density field, rather than the auto-correlation that is used in 2PCF analyses. 
However, environmental analyses can be difficult to model and compare to mock 
galaxy simulations, in part because the mean observed galaxy 
density is subject to selection effects that depend on the particular survey. 
More importantly, density measurements often average across a
range of physical scales (typically with a mean scale of 
$\gtrsim1 h^{-1}$ Mpc at \z$\sim1$) to infer correlations.
Because the ``density-defining" population is generally measured on 
scales that are roughly the size of a halo, such measurements preclude 
smaller sub-halo scales. For example, \cite{Li06} measured the 2PCF
of SDSS galaxies with respect to their structural parameters and found 
significant clustering dependence on scales below $<$1 $h^{-1}$ Mpc, a result
that was not previously seen in environment studies.  
The scale-dependence of clustering as a function of galaxy properties 
can yield important clues 
as to the physical processes involved in galaxy formation and evolution.

As environment and 2PCF analyses are complementary, it is important 
to compare the trends found in both kinds of studies to test for 
consistency, particularly on large scales.
\citeauthor{Cooper08} (2008, hereafter CP08) used data from SDSS and DEEP2 to quantify 
the evolution of the relationships between SFR and sSFR with 
environment. They found that the relation between sSFR and overdensity is essentially unchanged 
with redshift, suggesting that mergers did not drive the decline
in the global SFR at \z$<$1.  Confirming the results from \cite{Elbaz07}, CP08 also found that galaxies with high SFRs 
are found in high density environments at $z\sim1$,
which leads to an ``inversion" of the SFR-density relationship between 
$z\sim1$ and $z\sim0$, where high-SFR galaxies are found in relatively 
sparse environments. 

Other studies have measured the evolution of galaxy environment with
respect to SFR at a fixed stellar mass between 0.2$<$\z$<$1. Several studies report that
the average color-density relation does not
appear to have evolved since \z$<$1, indicating that the sSFR-density 
relationship is relatively constant with cosmic time since $z=1$ \citep{Cucciati06, Scodeggio09, 
Peng10, Sobral11, Grutz11}. These findings
generally agree with the studies within CP08, with the caveat that
finer sub-sampling of the most extreme density regions can
reveal a statistically significant difference in galaxy restframe
colors \citep{Cooper10}. Similarly, \cite{Scodeggio09} and \cite{Peng10} 
find that bulk trends in the SFR-density relation for field
galaxies (typically with \smass$<$11) can also be explained by the
change in stellar mass at a fixed SFR. \cite{Sobral11} 
suggests that most, but not all, of the SFR-density correlation is due
to changes in stellar mass in highly star-forming galaxies.  The general
consensus from these studies is that there is a fundamental relationship 
between stellar mass and large-scale environment and that correlations between
environment and other galaxy properties are due to second-order effects.

In this paper, we quantify the relationship between dark matter halos and 
galaxy properties at $z\sim1$ by measuring the clustering of DEEP2 galaxy 
samples selected by stellar mass, SFR, and sSFR, on scales of 
$\sim0.1-20~h^{-1}$ Mpc. Our results will enable future HOD modeling 
studies at $z\sim1$ as a function of these galaxy properties and will further 
be useful for projecting the statistical power expected from 
large-area baryon acoustic oscillation (BAO) surveys at these redshifts 
(e.g. BigBOSS, \citealt{Schlegel11,BigBOSS}).
The structure of the paper is as follows. 
In Section 2, we present the DEEP2 dataset and derived galaxy properties used 
in this study. In Section~\ref{sec:Sample}, we describe the galaxy samples defined by 
stellar mass, SFR, and sSFR. Section~\ref{sec:method} details the 
clustering analysis methods using the 2PCF, while Section~\ref{sec:results} 
presents the results of our clustering analysis 
with respect to the various galaxy properties.  We compare our findings to 
the recent literature and discuss the broader implications of our results in 
Section~\ref{sec:discuss}, and we 
present our conclusions in Section~\ref{sec:conclusions}. Throughout this work, 
we assume a $\Lambda$CDM cosmology with $\Omega_M$=0.3, $\Omega_\Lambda$=0.7, 
$\sigma_{8}$=0.8, and $h=0.7$. All quoted magnitudes are in the AB magnitude 
system.

\section{Data}
\label{sec:data}

The DEEP2 Galaxy Redshift Survey \citep{Newman12} is 
one of the largest, most complete spectroscopic redshift surveys of the 
Universe to $z\sim1$. The DEEP2 survey was conducted on the Keck II telescope 
using the DEIMOS spectrograph \citep{Faber03} and 
measured reliable redshifts for 31,656 galaxies from 0$<$\z$<$1.4 over four 
fields dispersed across the Northern Sky.  The DEEP2 survey targeted galaxies 
to a magnitude limit of $R_{\rm{AB}}<$24.1, and in three of the four fields 
used a $BRI$ color cut to efficiently target galaxies with \z$>$0.7.

For this study, we initially select galaxies with
0.74$<$\z$<$1.4 and a high confidence redshift ($>$95\%, $\z_{\rm{quality}}$$\geq$3). 
As we are interested in measuring clustering properties as a function of
several galaxy properties, we retain only those galaxies with well-measured values (i.e 
not flagged as an invalid measurement) for restframe $B$-band magnitude ($M_B$), color ($U-B$), stellar mass ($M_*$), 
and star-formation rate (SFR or $\psi$). Because the average SFRs are now estimated 
from the galaxy luminosity and restframe color \citep{Mostek12}, we do not exclude
red galaxies in our sample.  These combined selections result in 
a parent sample of 22,331 galaxies.

\subsection{Calibrations of Stellar Mass and SFR}
\label{sec:datacal}

\begin{figure}[h]
\centering
\includegraphics[width=1.0\columnwidth]{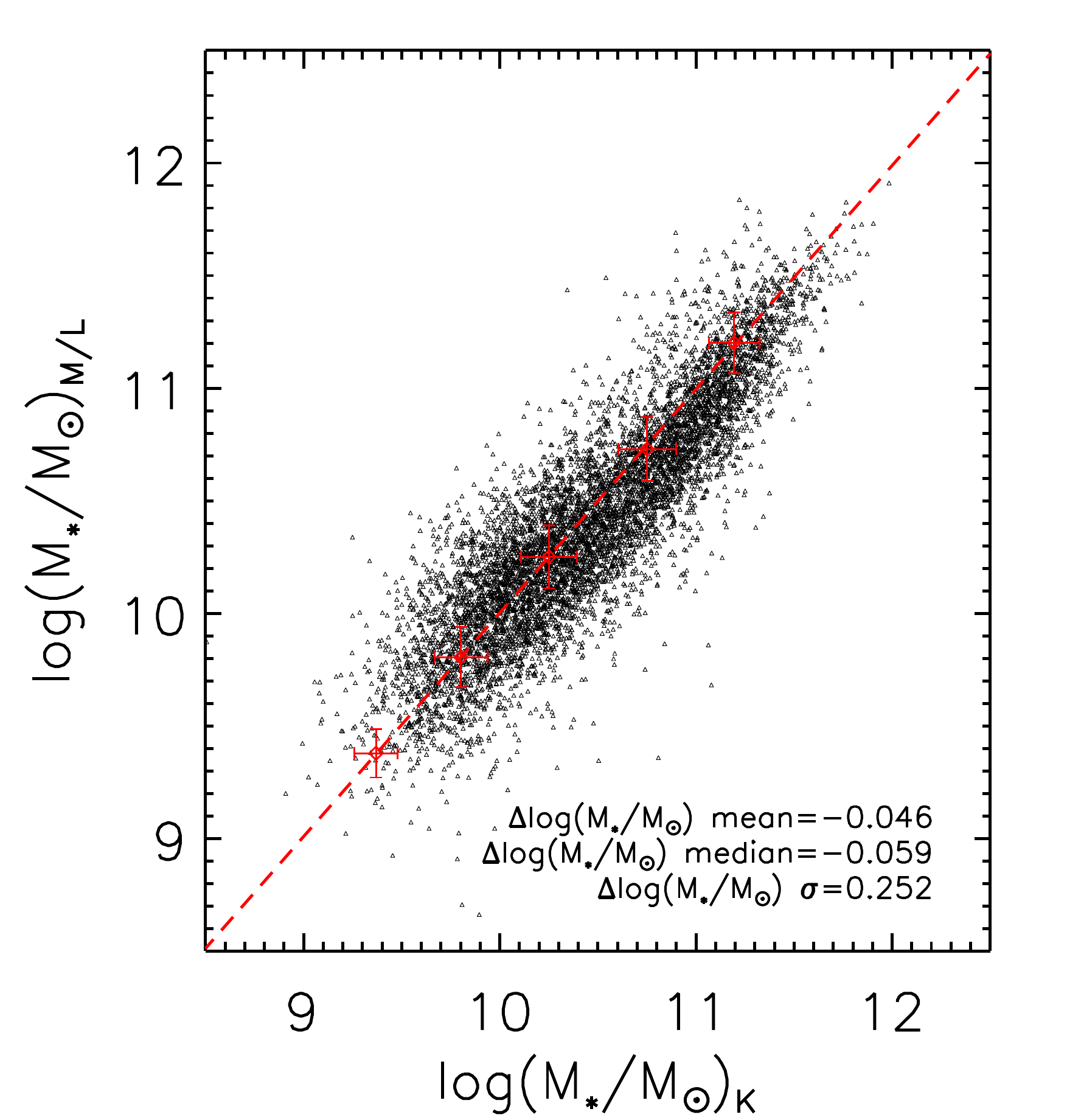}
\caption{Comparison of DEEP2 color-M/L stellar masses and 
$K$-band stellar masses.  The rms scatter between the 
stellar mass estimates is 0.25 dex and the sample mean offset is -0.05 dex 
for a Chabrier IMF.}
\label{fig:masscheck}
\end{figure}

The stellar masses used in this study are constructed from the prescription 
described in \cite{Lin07} and the Appendix of \cite{Weiner09}, where the local restframe 
color-M/L relation \citep{Bell03} is extended to higher redshift. 
The prescription depends on redshift, restframe $B$-band luminosity, and 
the restframe $U-B$ and $B-V$ colors, and is calibrated to a matched sample 
of stellar masses derived from $K$-band luminosity measurements \citep{Bundy06}.  
Figure~\ref{fig:masscheck} compares stellar masses from the 
DEEP2 calibration and matched to a new, extended sample of
7,267 $K$-band stellar masses (K. Bundy, priv. communication) for galaxies
with 0.74$<$\z$<$1.4. Independently, we confirm that the DEEP2 stellar masses used here are 
in agreement with this larger sample of $K$-band derived masses with a rms scatter of 0.25 dex 
and a mean offset of -0.05 dex. 
The DEEP2 stellar masses are based on a Chabrier IMF, in keeping 
with the \citeauthor{Weiner09} calibration. 

\begin{figure*}[ht]
\centering
\includegraphics[width=1.0\textwidth]{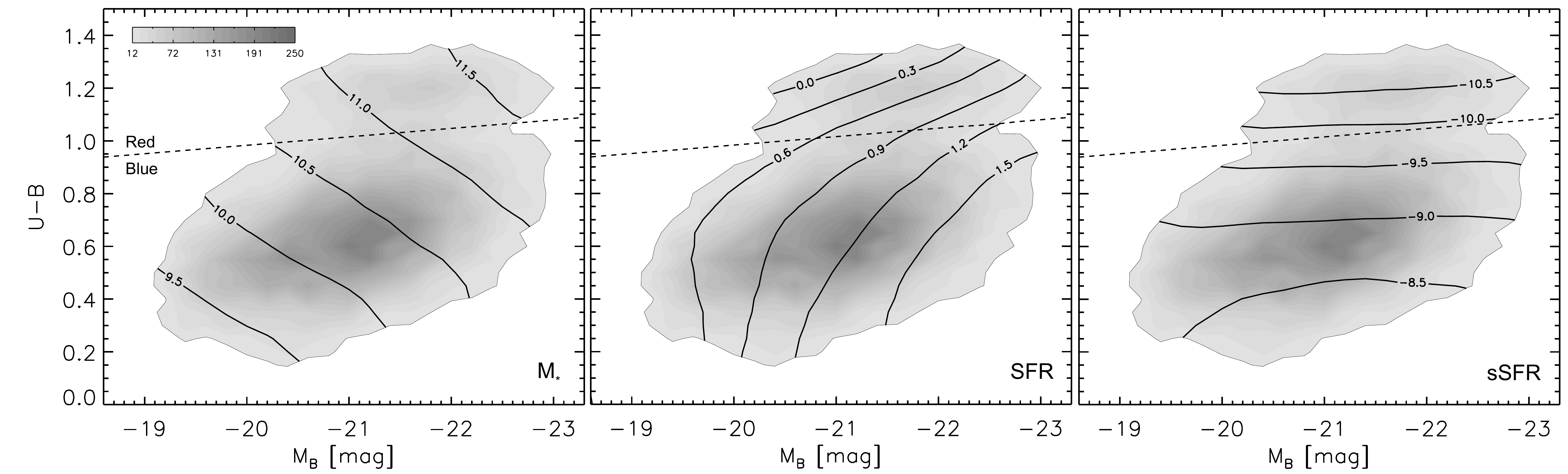}
\caption{The greyscale in all three panels shows the restframe $U-B$ color 
and $B$-band magnitude distribution for the \zrange DEEP2 galaxy sample. 
(left) Stellar mass contours span 9.0$<$\smass$<$11.5 with levels of 
$\Delta$\smass=0.5. More massive galaxies tend to be redder and brighter. The dashed line shows the 
DEEP2 restframe galaxy color selection from
 Equation~\ref{eq:color} splitting the red sequence and blue cloud. 
(middle) SFR contours calibrated from 
\cite{Mostek12} spanning 0$<$\sfr$<$1.5 and 
levels of $\Delta$\sfr=0.3 $M_{\sun}$ yr$^{-1}$. 
Galaxies with higher SFR are bluer and brighter.  SFRs are correlated in restframe $M_B$ and $U-B$ 
partially by construction, where optical photometry has been used in the calibration.
(right) Contours of sSFR 
spanning $-8.5$$>$\ssfr$>$$-11$ with levels of $\Delta$\ssfr=0.5 yr$^{-1}$. 
Galaxies with higher sSFR are bluer; $U-B$ color and sSFR correlate tightly,
as evidenced by the horizontal contours. 
}
\label{fig:colormasscontour}
\end{figure*}

In CP08, SFRs for DEEP2 galaxies were 
derived from measurements of the \oii emission line and an empirical calibration 
to galaxy $B$-band luminosity and SFR performed at low redshift 
\citep{Moustakas06}.  Applying this local calibration to redshifts 
$z\sim1$ poses some unique problems, including accounting for 
luminosity and dust evolution as well as line emission from AGN 
in red sequence galaxies (Yan et al., 2006). More recent work 
by \cite{Mostek12} established a new empirical SFR 
calibration for DEEP2 galaxies using broadband optical color and UV/optical 
SED-based SFRs \citep{Salim09}. \citeauthor{Mostek12} showed that the strong observed 
evolution in the restframe $M_B$ from local redshifts 
($\sim1.3$ mag to $z=1$) caused the SFRs in CP08 to be 
overestimated by an average of $\sim0.5$ dex. However, as this evolution
causes to first order a zeropoint offset in the SFR calibration, the 
relative trends between SFR and overdensity presented in CP08 are still valid. 
Here we use the \citeauthor{Mostek12} SFR calibration and a Salpeter IMF, in keeping 
with previous SFR calibrations commonly used in the literature \citep{Kennicutt98,
Moustakas06}.

The sSFRs used here are constructed from the DEEP2 stellar masses and SFRs 
described above, using \ssfr~$\equiv$~\smass~$-~$\sfr.  We convert the 
stellar masses from a Chabrier IMF to a Salpeter IMF by adding a constant 
0.2 dex offset and quote sSFRs for a Salpeter IMF. We note that both the \cite{Moustakas06} and \cite{Mostek12}
SFR calibrations rely on correlated mean trends in large galaxy samples, and therefore 
may not be highly accurate on a individual galaxy-by-galaxy basis 
(typically 0.2-0.3 dex rms scatter). However, the \emph{samples} selected from such empirical SFR 
calibrations are statistically representative of the mean SFR in the galaxy population. 

As clustering measurements are more commonly performed as a function of 
galaxy luminosity and color, we show in Figure~\ref{fig:colormasscontour} 
how our derived stellar masses, SFRs, and sSFRs depend on restframe color 
and magnitude, showing contours of each derived property in the $U-B$ - $M_B$ 
plane. The contours of constant stellar mass generally run perpendicular to 
the contours of constant SFR in this plane. When converted to sSFR, these 
opposing trends cancel out to produce a largely luminosity-independent 
trend with color, reflecting the fact that sSFR and restframe color 
generally reflect the light-averaged age of the stellar population.

\section{Galaxy Samples}
\label{sec:Sample}

In this section, we describe the construction of samples for 
\zrange DEEP2 galaxies selected by stellar mass, SFR, and sSFR. 
For galaxy clustering measurements, it is preferable to use complete 
samples for which galaxies of the same `type' (selected by magnitude, 
color, stellar mass, etc.) are included to the highest redshift of the sample.
  In this way, within a given sample there is no redshift dependence of the 
galaxy property of interest, which would necessitate weighting each galaxy 
by the volume over which it could be observed.  Complete galaxy samples also 
facilitate comparison with simulations and HOD modeling, where a galaxy population
can be well described by a few simple selection criteria and a redshift limit.  

We create both binned and threshold samples in this study. Constructing samples with independent bins 
facilitates identification of clustering trends with respect to each galaxy 
parameter, while constructing threshold samples facilitates future HOD 
modeling applications. The binned sample sizes are chosen such that the bin 
size is greater than or equal to the rms scatter in the galaxy property and so 
that there is a sufficient number of galaxies per bin to minimize Poisson 
errors. For each galaxy parameter, we treat the bin with the lowest number 
density, which generally corresponds to the more massive or higher SFR 
galaxies, as a threshold lower limit. 

Our galaxy samples cover two primary redshift ranges: 0.74$<$\z$<$1.05 for 
samples that are complete for both blue and red galaxies, 
and \zrange for samples that are complete only for blue galaxies. 
The former redshift range allows us to compare clustering trends between 
galaxies with different star-formation histories - including both quiescent 
and star-forming galaxies - while the latter allows us to obtain the best 
statistical measurements for blue galaxies.

\subsection{Stellar Mass}
\label{sec:smasssample}

As shown in the left panel of Figure~\ref{fig:colormasscontour}, galaxy samples
 selected by stellar mass naturally include both red and blue galaxies, but 
the ratio of red to blue galaxies strongly depends on stellar mass. 
In addition, the $R$-band selection of DEEP2 corresponds to a different 
restframe 
wavelength selection with redshift (bluer at higher redshift), such that 
DEEP2 is complete to different stellar masses for red and blue galaxies as 
a function of redshift. To probe the full range of stellar masses allowed by 
the data, we construct both color-independent (e.g. `all' galaxies, blue and 
red) and color-dependent stellar mass samples for \z$<$1.05, and we 
construct mass samples
for blue galaxies only for \z$<$1.4.  Following \cite{Willmer06}, we use 
the standard DEEP2 red and blue galaxy separation defined by
\begin{equation}
(U-B)=-0.032(M_B+21.62)+1.035.
\label{eq:color}
\end{equation}

\begin{figure*}[t]
\centering
\includegraphics[width=1.0\columnwidth]{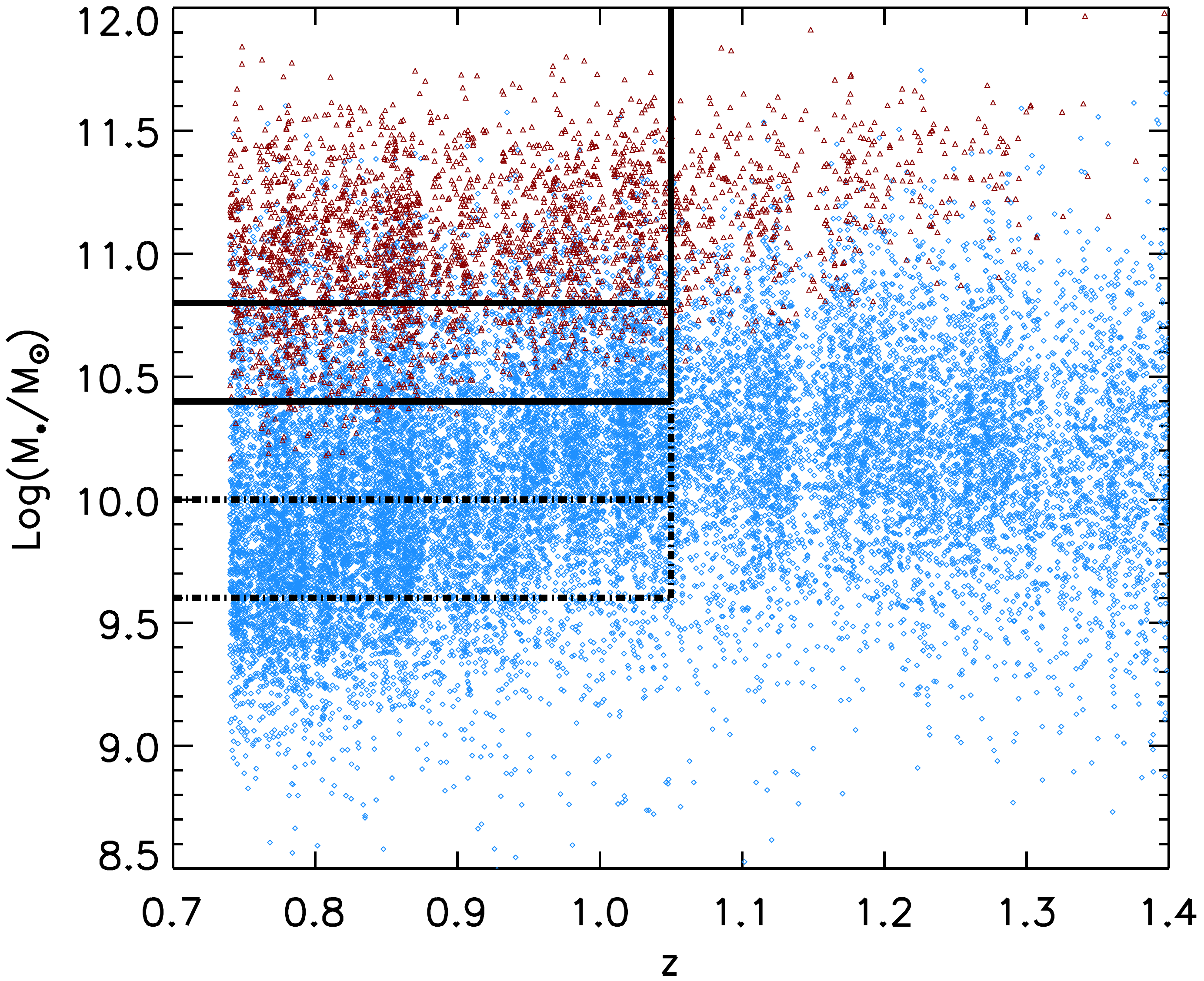}
\includegraphics[width=1.0\columnwidth]{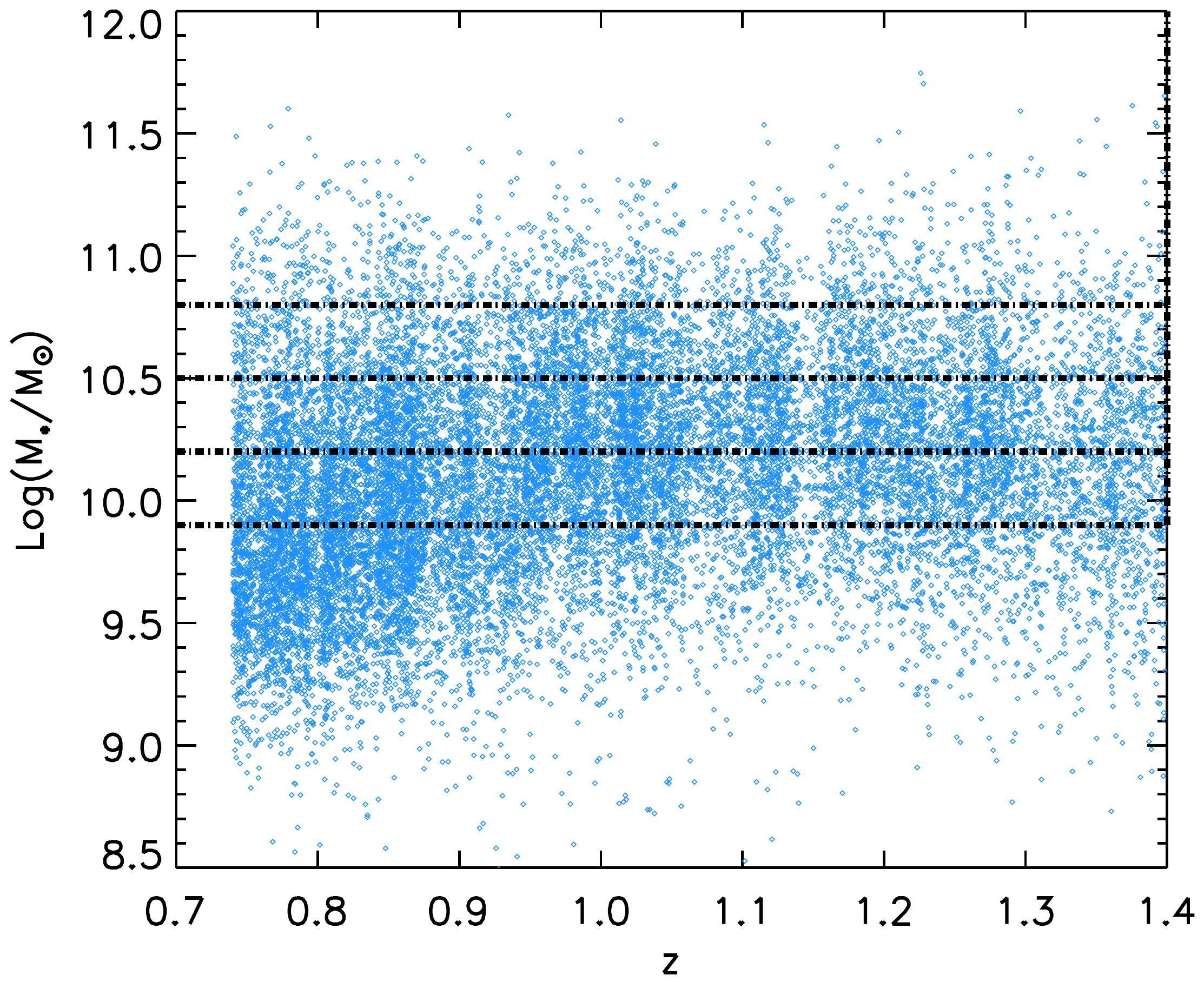}
\caption{Stellar mass selected galaxy samples. (left) Colored points show all 
galaxies in the parent sample, where red galaxies are defined to be redder
than the color cut given in Equation~\ref{eq:color}. Solid lines show the mass limits
for color-independent galaxy samples, where the highest mass sample is roughly volume-limited 
for both red and blue galaxies. Dot-dashed lines show the lower mass sample limits for blue galaxies only 
at a \z$<$1.05 redshift limit. Red galaxy mass sample limits are given in Tables~\ref{tab:smassresults}
and \ref{tab:smassbinresults}.
(right) Stellar mass bin samples for blue galaxies only, defined 
with finer $\Delta$\smass=0.3 bins and approximately complete in DEEP2 to \z$<$1.4.
The highest mass bin in both samples are implemented as a mass 
threshold in order to allow sufficient statistics in the clustering 
measurement.}
\label{fig:massbins}
\end{figure*}

Using this color separation, the left panel of Figure~\ref{fig:massbins} 
shows stellar masses for red and blue DEEP2 galaxies as a function of 
redshift, with boxes indicating our binned samples to \z=1.05. 
As a starting point, we define color-independent samples including both
red and blue galaxies in two bins in stellar mass.
For the highest stellar mass sample, we use a mass threshold of \smass$>$10.8
to ensure a large enough sample size such that clustering measurements can be 
performed with reasonable accuracy. This high-stellar-mass sample is roughly volume-limited 
for both red and blue galaxies in this stellar mass range.  The lower stellar mass 
color-independent mass bin spans 10.4$<$\smass$<$10.8 but is somewhat 
incomplete for red galaxies at $z\sim1$ due to the 
R-band magnitude limit in DEEP2.  In this mass bin, 30\% of the galaxies 
between 0.75$<$\z$<$0.85 are red while only 12\% of galaxies between
 0.95$<$\z$<$1.05 are red, indicating that 
red galaxies are increasingly excluded at the highest redshifts. 
Therefore, we can construct only one sample that is
 complete for both red and blue galaxies (with \smass$>$10.8), and
while we quote results from the lower mass color-independent bin, we 
refrain from calculating a number density in this sample.  We further 
caution that this sample underestimates the contribution from low mass, 
high redshift red galaxies missing in DEEP2, and we keep this in mind when
interpreting the results.

In order to compare the clustering of red and blue galaxies separately as a 
function of stellar mass, we also create red and blue mass-selected samples.
The mass and redshift limits of each sample are given in Table 2, and the 
blue samples are shown in Figure~\ref{fig:massbins}.  For red galaxies,  
we create two samples with 10.5$<$\smass$<$11.0 and \smass$>$11.0 at 
\z$<$1.05.  To facilitate comparison with red galaxies over the same volume, 
we create three blue mass-limited bin samples down to a lower
mass of \smass$>$9.6, where the blue population is complete at $z=1.05$.
We further create blue mass-limited bin samples over a wider redshift range, to 
$z=1.4$, down to a mass of \smass$=$9.9 (right panel 
of Figure~\ref{fig:massbins}). The blue galaxy mass bin samples to \z$=$1.4 
have a slightly smaller width (0.3 dex) than the lower redshift samples 
(0.4 dex), but the width is still greater than the rms scatter
in the stellar mass estimates (0.25 dex, see Figure~\ref{fig:masscheck}). 
An upper threshold limit of \smass$>$10.8 is used to 
define the highest mass bin.

We also define stellar 
mass-selected \emph{threshold} samples (not shown in Figure~\ref{fig:massbins})
 with which to measure clustering properties.  These threshold samples are useful for comparing
with simulations and semi-analytic models, where samples may be defined by a 
given mass limit, and for performing HOD modeling.  For the color-independent
stellar mass threshold samples, we only report the number density for stellar masses above
\smass$>$10.5 due to the limited number of red galaxies in the measured volume. 
The sample properties for each stellar mass-selected sample are listed in 
Tables~\ref{tab:smassresults} and \ref{tab:smassbinresults}.

\begin{figure}[h]
\centering
\includegraphics[width=1.0\columnwidth]{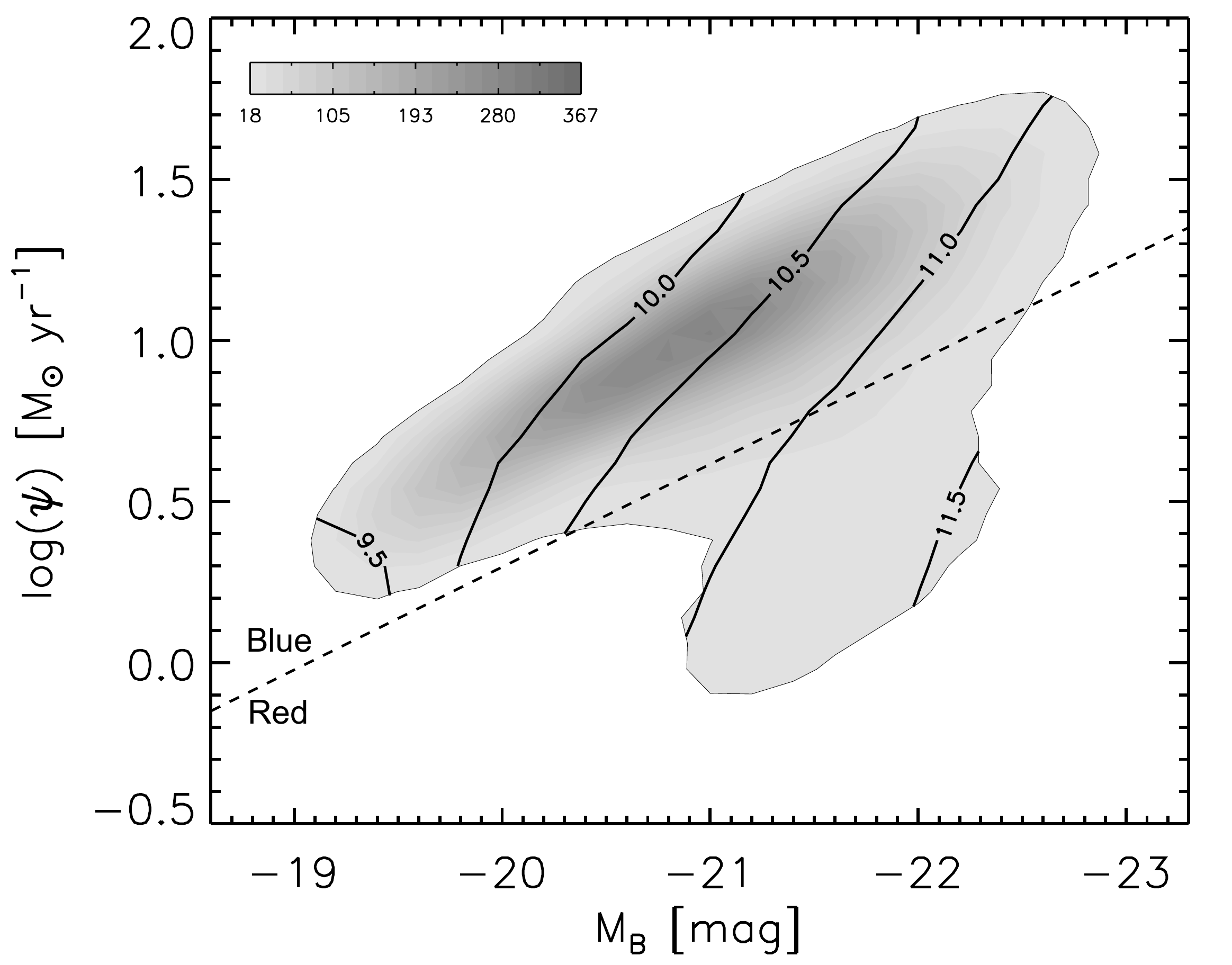}
\caption{SFRs using the \cite{Mostek12} calibration versus $M_B$. 
Greyscale shows the distribution of DEEP2 galaxies between 0.74$<$\z$<$1.4.  
Contours show lines of constant stellar mass, with the same mass levels as 
in Figure~\ref{fig:colormasscontour}. The dashed line shows the approximate 
demarcation between red and blue galaxies, as defined in the color-magnitude
diagram and Equation~\ref{eq:color}.}
\label{fig:sfrcontours}
\end{figure}

\begin{figure*}[ht]
\centering
\includegraphics[width=0.93\columnwidth]{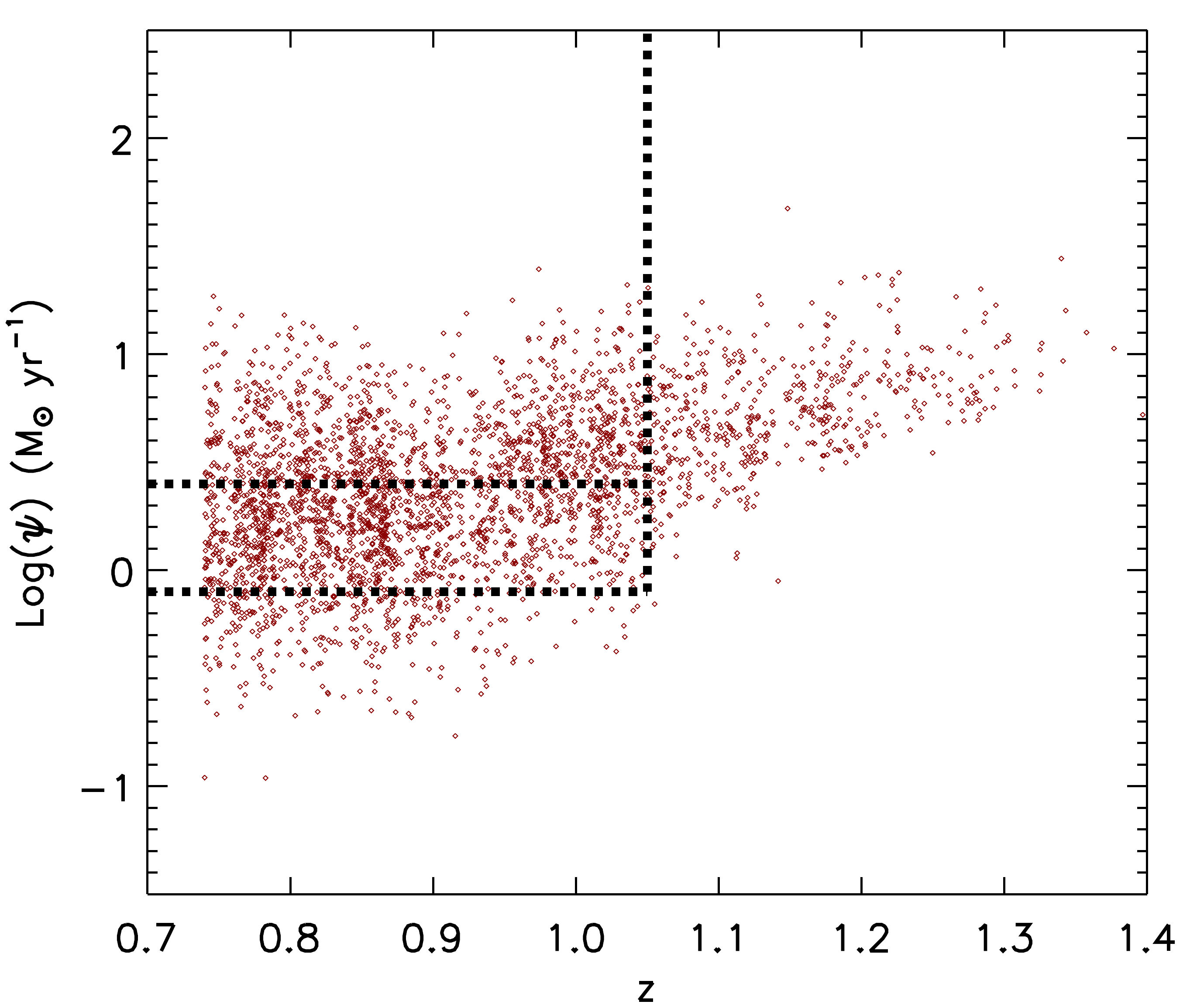}
\includegraphics[width=0.93\columnwidth]{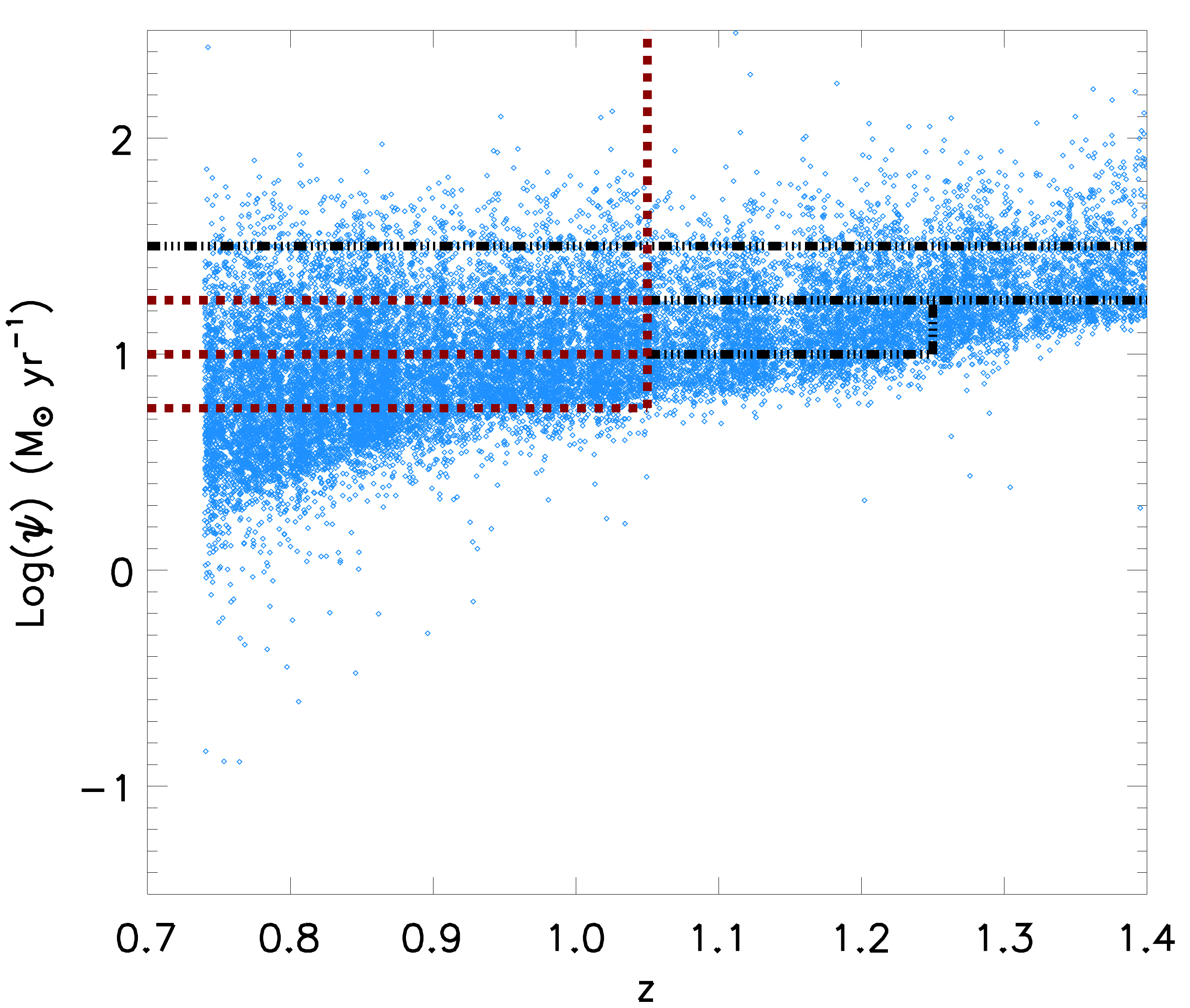}
\caption{Galaxy samples defined by star formation rate. (left) SFR versus 
redshift for all red galaxies in the parent sample.  SFR-selected bin samples 
for red galaxies are shown with dashed lines; these samples are 
complete at \z=1.05 for \sfr$>$0.3 and approximately complete for
\sfr$>$-0.1. (right) SFR versus redshift for all blue galaxies in the parent sample.  
SFR-selected bin samples for blue galaxies complete to \z=1.05
are shown with red, dashed lines, while samples complete to 
either \z=1.25 or \z=1.4 are shown with black, dot-dashed lines.  As with
stellar mass, the highest SFR samples are treated as thresholds.}
\label{fig:sfrbins}
\end{figure*}

\subsection{Star Formation Rate}
\label{sec:sfrsample}

As with stellar mass,  galaxies at a given SFR can have a range of restframe
color and magnitude. Figure~\ref{fig:sfrcontours} shows the SFRs for DEEP2 
galaxies as a function of $M_B$; the dashed line indicates the separation of 
red and blue galaxies following Equation~\ref{eq:color}. While most red 
galaxies have an estimated SFR below \sfr$<$0.5 $M_{\sun}$ yr$^{-1}$, only those blue 
galaxies with $M_B<$-20 have similar SFRs. Figure~\ref{fig:sfrcontours} 
also shows that SFR samples will select across a wide range 
of stellar mass (shown with contours), particularly for blue galaxies where 
$M_*$~can differ by 1.5 dex at a given SFR. 
Therefore, our analysis using SFR-selected samples not only probes 
the relationship between clustering and SFR but also has a weak dependence 
on stellar mass. We will explicitly remove the stellar mass dependence 
by constructing specific SFR samples in Section~\ref{sec:ssfrsample}. 

While one could in theory construct color-independent samples in SFR given
a large enough survey, 
DEEP2 does not have a sufficient number of both red and blue galaxies at a constant 
SFR and complete within the same volume to obtain a
reliable color-independent clustering measurement.  We therefore create color-dependent SFR samples that 
are approximately complete within their restframe color as shown 
in Figure~\ref{fig:sfrbins}. To mirror the selection with stellar mass, we 
choose red galaxies with \z$<$1.05 in two SFR samples limited to -0.1$<$\sfr$<$0.3 $M_\sun$ yr$^{-1}$
and \sfr$\ge$0.3 $M_\sun$ yr$^{-1}$. The SFR bin dividing line at 0.3 $M_\sun$ yr$^{-1}$ is chosen
such that the two samples each have $\sim1000$ galaxies, which is roughly the minimum
sample size needed to produce a reasonable clustering signal within the measured volume of this dataset.
As SFR is strongly dependent on $M_B$, particularly in blue galaxies, the redshift completeness limits are 
clear; blue galaxies are complete for \sfr$>$0.75$M_\sun$ yr$^{-1}$ below
 \z$<$1.05, \sfr$>$1.0$M_\sun$ yr$^{-1}$ for \z$<$1.25, and 
\sfr$>$1.25$M_\sun$ yr$^{-1}$ for \z$<$1.4.  The SFR bin sizes are selected to 
roughly reflect the SFR accuracy quoted in \citeauthor{Mostek12} (0.5 dex and 0.25 dex rms 
scatter for red and blue galaxies, respectively), while maintaining sufficient sample sizes 
in each bin. The sample properties for each SFR sample are listed in Tables~\ref{tab:sfrthreshresults} and \ref{tab:sfrbinresults}.

\subsection{Specific Star Formation Rate}
\label{sec:ssfrsample}

\begin{figure}[ht]
\centering
\includegraphics[width=0.97\columnwidth]{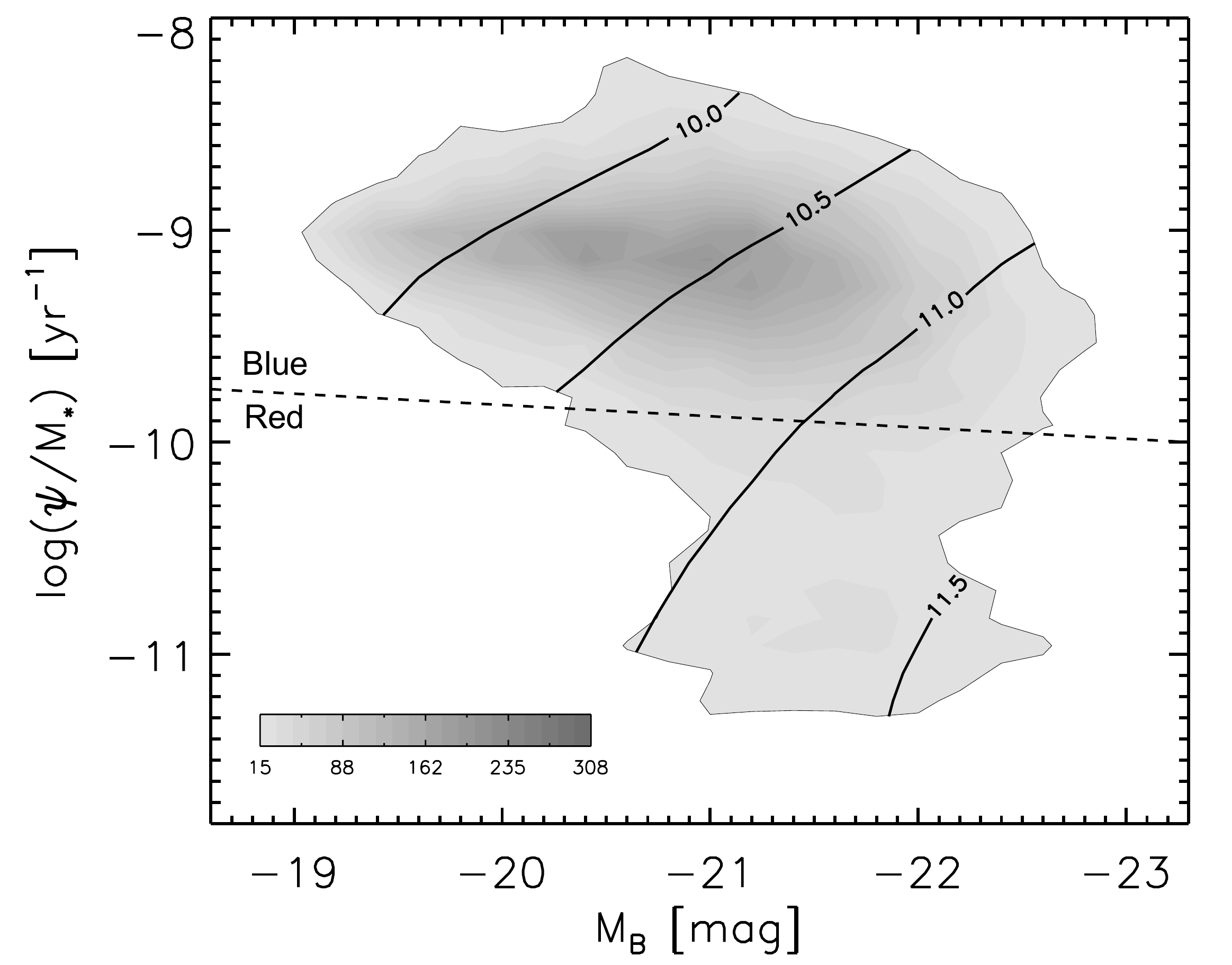}
\caption{Galaxy specific SFR versus the restframe $B$-band magnitude, $M_B$. Contours correspond to constant stellar mass, 
with the same levels shown as in Figure~\ref{fig:colormasscontour}. }
\label{fig:ssfrcontours}
\end{figure}

To construct highly-complete samples selected by sSFR, we first emphasize 
that sSFR, as shown in Figure~\ref{fig:colormasscontour}, 
is tightly correlated with restframe color.  Similarly, we find that plotting
the sSFR against the restframe $B$-band magnitude (Figure~\ref{fig:ssfrcontours}) shows that nearly all 
DEEP2 blue galaxies with \zrange have sSFRs between $-10$$<$\ssfr$<$$-8$ yr$^{-1}$,
while red galaxies have lower sSFRs of \ssfr$<$$-10$ yr$^{-1}$. 
Figure~\ref{fig:ssfrcontours} also shows that like SFR, galaxies at a fixed sSFR 
 have a wider range of stellar masses (represented by contour lines) 
in the blue galaxy population than red galaxy population. Further, a sSFR-selected
blue galaxy sample will have a residual correlation between the mean stellar mass and luminosity of the sample. 
This residual correlation indicates that galaxy luminosity is only a rough approximation
to the stellar mass within the blue population at $z\sim1$.

\begin{figure*}[ht]
\centering
\includegraphics[width=0.85\columnwidth]{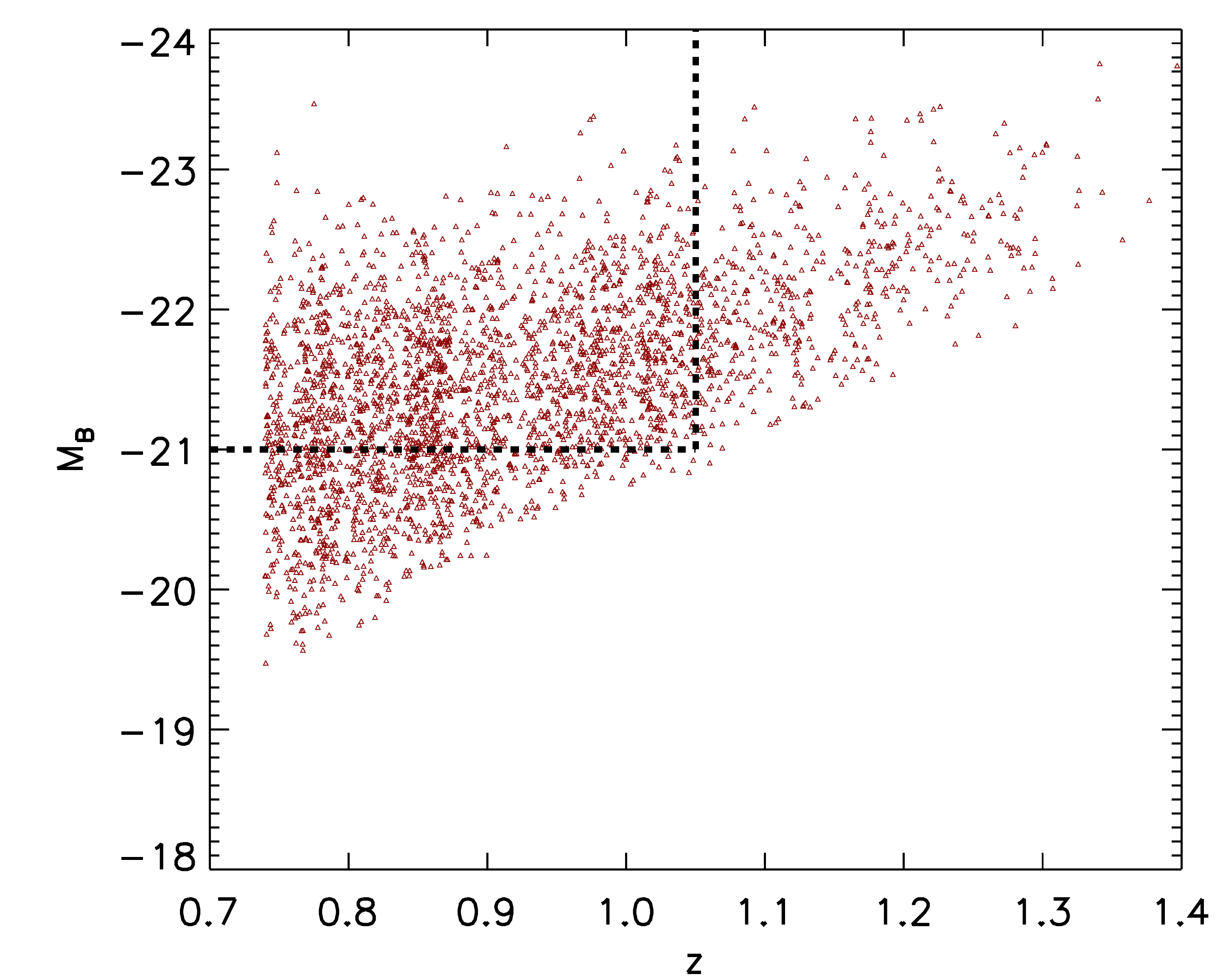}
\includegraphics[width=0.85\columnwidth]{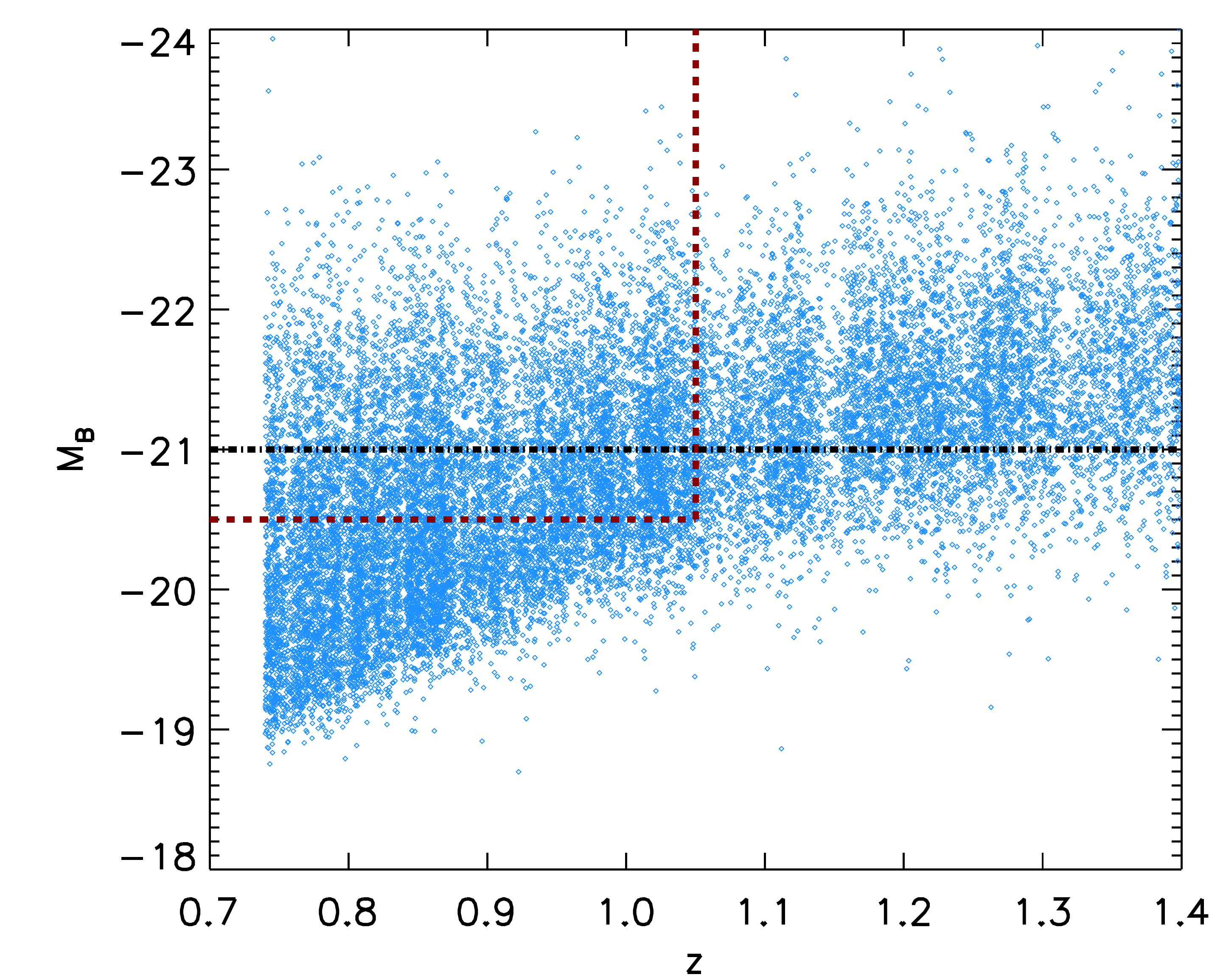}
\includegraphics[width=0.85\columnwidth]{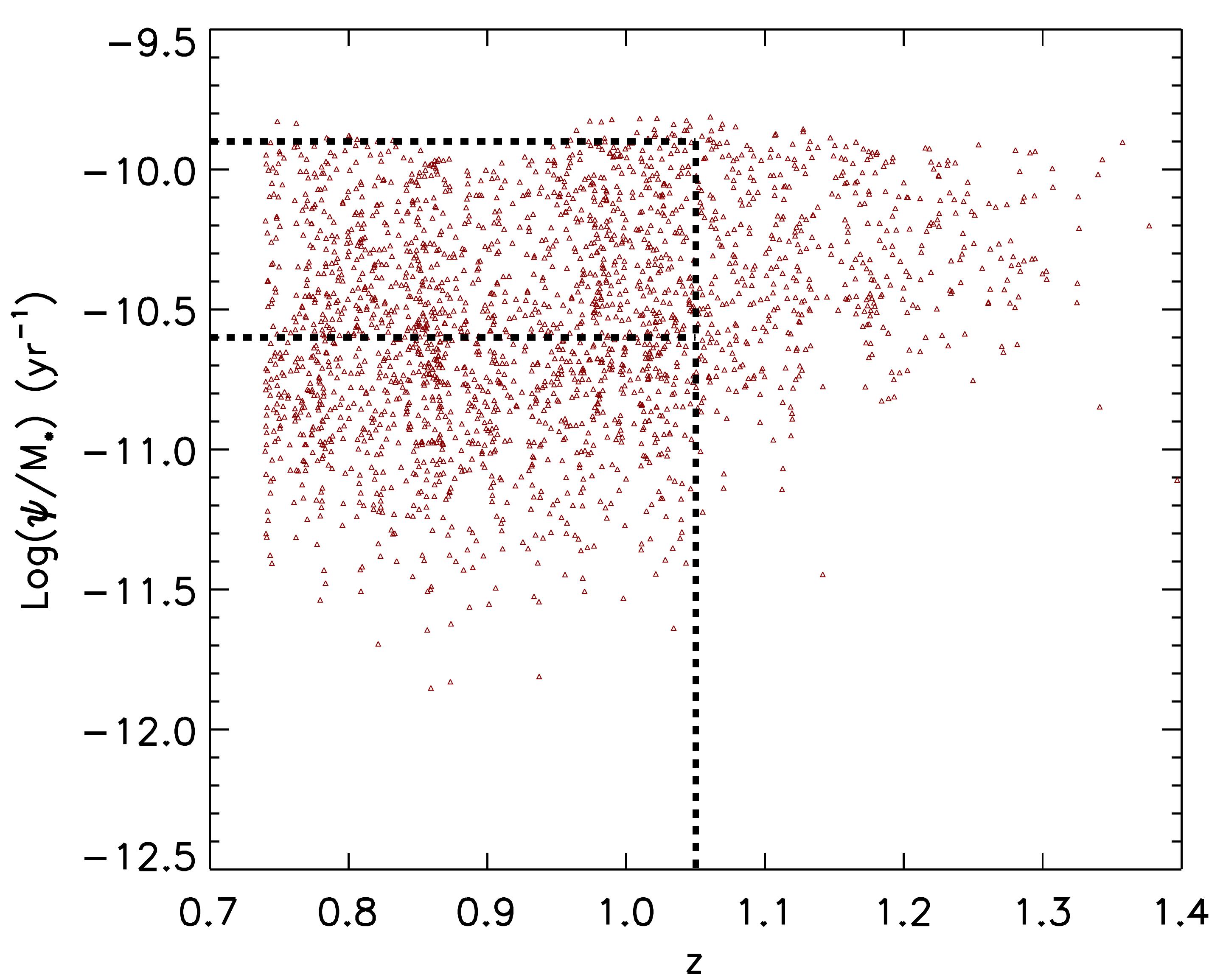}
\includegraphics[width=0.85\columnwidth]{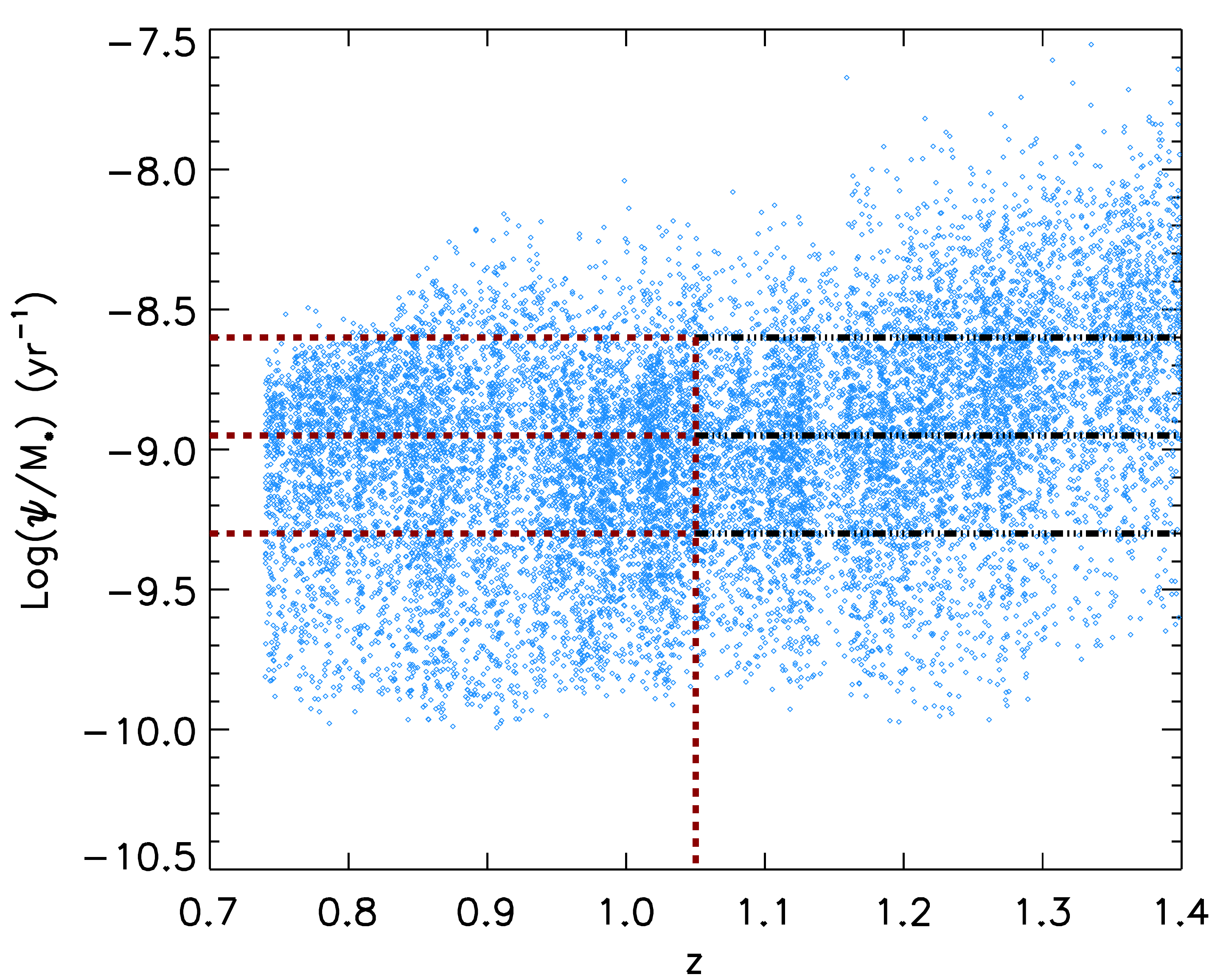}
\caption{Galaxy samples defined by specific star formation rate.
(Top row) Restframe $B$-band magnitude for red (left) and blue (right) galaxies in the parent sample are 
shown as a function of redshift.  To \z=1.05, the red galaxies are 
complete to $M_B$=$-21$ while blue galaxies are complete to 
$M_B$=$-20.5$.  To \z=1.4, blue galaxies are limited to $M_B$=$-21.0$.
(Bottom row) From the magnitude-limited samples defined in the top row, 
we create sSFR-selected samples with \ssfr$<$$-9.9$ for 
red galaxies and \ssfr$<$$-8.6$ for blue galaxies.}
\label{fig:ssfrbins}
\end{figure*}

Stellar mass and SFR each have different completeness limits in DEEP2
as a function of redshift, color, and magnitude, which could in principle be
 difficult to disentangle when combined as sSFR. However, we can avoid these 
complications by recognizing that sSFR closely traces color.  We can 
therefore first limit the parent sample by restframe $B$-band magnitude, $M_B$,
 for red and blue galaxies separately, and then define sSFR-selected
subsamples free from incompleteness expected at lower luminosities within each
restframe color.  The final color-dependent sSFR samples will then be complete \emph{to 
the magnitude limit of the sample}. 

The upper panels of Figure~\ref{fig:ssfrbins} show $M_B$ as a function of 
redshift for red and blue galaxies separately. At \z$<$1.05, the parent sample 
is complete for $M_B$$<$$-21$ and  $M_B$$<$$-20.5$ for red and blue galaxies, 
respectively.  At \z$<$1.4 the blue galaxy sample is limited to 
$M_B$$<$$-21$. We choose not to use a more complicated, color-dependent 
magnitude limit \citep[see][]{Gerke07} so that our samples can be easily 
interpreted in comparison to other sSFR clustering studies.

After making these initial cuts on $M_B$ and redshift, we can construct 
 sSFR samples that are complete to the given $M_B$ limit. The bottom row of 
Figure~\ref{fig:ssfrbins} shows our sSFR-selected samples.  In general, the 
lowest luminosity galaxies have the highest sSFRs. We define upper limits on
 sSFR of \ssfr$<$$-9.9$ yr$^{-1}$ for red galaxies and 
\ssfr$<$$-8.6$ yr$^{-1}$ for blue galaxies. We also limit the blue 
galaxy population at \z$<$1.05 and \z$<$1.4 in order to facilitate comparisons 
with the stellar mass and SFR-selected samples.  The individual sSFR sample 
properties are given in Tables~\ref{tab:ssfrthreshresults} and \ref{tab:ssfrbinresults}.

\section{Analysis Method}
\label{sec:method}

\subsection{The two-point correlation function}
\label{sec:cf}

The clustering analysis performed in this study mirrors that of C08, 
and we summarize the methodology here. We measure the clustering of DEEP2 
galaxies by calculating the two-point correlation function, \corf, 
in each 
of the galaxy samples. In a given volume, \corf~measures the excess 
probability of finding galaxy pairs as a function of separation  
over that of a random distribution. The excess probability can be 
calculated using the \cite{Landy93} estimator,
\begin{equation}
\xi = \frac{1}{RR}\left [DD\left(\frac{n_R}{n_D}\right)^2 - 2DR\left(\frac{n_R}{n_D}\right) + RR\right ].
\label{eq:LSz}
\end{equation}
Here RR, DD, and DR are the random-random, data-data, and data-random
pairs, while $n_R$ and $n_D$ are the mean number densities of random
points and galaxies in each sample. Each pair is separated by a
three-dimensional distance $r$. As such, \corf~reflects the
three-dimensional clustering, which must be inferred through
measurements of the angular separation between galaxies on the sky and
a distance measurement along the line of sight.

The redshift of a galaxy reflects both its distance (assuming a
given cosmology) and its peculiar velocity.  Due to the latter effect, the
observed clustering signal has additional power that appears in redshift space 
but is not attributable to clustering in real space.
On small scales, the peculiar velocities of individual galaxies within
collapsed overdensities results in an ``elongation" of the clustering
signal along the line of sight, known as the ``Fingers of God''.  On
larger scales, the coherent infall of galaxies streaming into
potential wells causes a ``squashing" of the signal along the line of
sight. To remove these redshift-space distortions from the real-space
clustering signal, we separate the correlation function using two
separate coordinate vectors. We define two vectors,
\textit{\textbf{s}}=(\textit{\textbf{v}}$_1-$\textit{\textbf{v}}$_2$)
for the redshift-space separation and
\textit{\textbf{l}}=$\frac{1}{2}$(\textit{\textbf{v}}$_1+$\textit{\textbf{v}}$_2$)
for the mean coordinate of the pair, and calculate
\begin{equation}
\pi=\frac{\textit{\textbf{s}} \cdot \textit{\textbf{l}}}{\left\vert \textit{\textbf{l}}\,\right\vert},
\end{equation}
\begin{equation}
r_p=\sqrt{\textit{\textbf{s}} \cdot \textit{\textbf{s}}-\pi^2}.
\end{equation}
We then can express the three-dimensional \corf~as a two-dimensional 
projection, \corfp, where $r_p$ is the distance across the line of 
sight and $\pi$ is the distance along the line of sight.  We apply the estimator in
 Equation~\ref{eq:LSz} to \corfp~by counting galaxy pairs.  The random samples are
 drawn from the measured selection function for each galaxy sample spatial distribution 
 (see Section~\ref{sec:masks}). 

In order to recover the correlation function signal without redshift 
space distortions, we project the two-dimensional correlation function 
\corfp~onto the $r_p$ coordinate across the line of sight. The projection 
requires integrating along the line of sight, following \cite{Davis83},
\begin{equation}
\omega_{p}(r_p)=2\int_{0}^{\infty}{\!\!d\pi\,\xi(r_p, \pi)} = 2\int_{0}^{\infty}{\!\!dy\,\xi\left(r_p^2+y^2\right )^{1/2}},
\label{eq:wprp}
\end{equation}
where $y$ is the real-space separation along the line of sight. By modeling 
\corf~as a power law, $\xi$=$(r/r_0)^{-\gamma}$, Equation~\ref{eq:wprp} has 
an analytical solution,
\begin{equation}
\omega_{p}(r_p)=r_p\left(\frac{r_0}{r_p}\right)^{\gamma}\frac{\Gamma(1/2)\Gamma[(\gamma-1)/2]}{\Gamma(\gamma/2)},
\label{eq:wprpinf}
\end{equation}
which can be evaluated for given values of the clustering length, $r_0$, and 
correlation function slope, $\gamma$. Following C08, we integrate 
the projected correlation function \wprp~to 
a limit of $\pi_{\rm{max}}=20$ $h^{-1}$ Mpc, beyond 
which the DEEP2 sample is not large enough to measure an accurate signal. 
Implementing the $\pi_{\rm{max}}$ limit requires that we also correct Equation~\ref{eq:wprpinf}
 to be truncated to the same integration limit. However, the resulting \wprp~is accurate for measured scales
 where $r_p$/$\pi_{\rm{max}}\lesssim0.25$  (see C08 for details).
 
We fit for a power law of the form $\xi$=$(r/r_0)^{-\gamma}$ on scales of 
1-10 $h^{-1}$ Mpc. As we will show, these scales 
mitigate the effects of non-linear behavior at smaller $r_p$ and avoid 
noisy data at larger scales. The final values of $r_0$ and $\gamma$ are 
obtained by performing a $\chi^2$ minimization between the measured \wprp~and 
Equation~\ref{eq:wprpinf}, after correcting for the truncation of 
$\pi_{\rm{max}}$. 

For each sampled scale, galaxy pair counts are averaged over 10 separate DEEP2 pointings
 and errors on \wprp~are computed using the standard error among these pointings. While formal errors for 
$r_0$ and $\gamma$ are computed from the $\chi^2$ fit, we report the rms 
errors for each parameter using 10 jackknife samples of the data. The 
jackknife error estimates on $r_0$ and $\gamma$ encapsulate the cosmic 
variance inherent within the DEEP2 data and take into account the covariance
among the $r_p$ bins. Because some DEEP2 pointings are correlated with
each other on the sky, the cosmic variance may be slightly underestimated. Based on calculations with QUICKCV \citep{Newman02}, 
the actual rms error due to cosmic variance could be up to $\sim6$\% larger than the estimates 
from the 10 jackknife samples. We do not include this estimated error contribution in our analysis, electing
instead to work with the error given directly from the data.

In order to facilitate interpretation of the results and to avoid relying 
solely on power-law fits which may not describe the data well over all 
scales, we further estimate the galaxy bias for each sample.  The bias is 
a measure of the clustering of galaxies relative to that of dark matter 
at the same redshift. We generate a dark matter \corfp~for the mean 
redshift of each galaxy sample using code from \cite{Smith03}, integrating 
to the same $\pi_{\rm{max}}$ as in the data. We then calculate the average bias over 
scales of 1-10 $h^{-1}$ Mpc using $b^2$=[\wprp]$_{\rm{gal}}$/[\wprp]$_{\rm{dark~matter}}$ and 
weighting each scale equally in the average bias. The average scale
of the mean bias is $\bar{r}_p$=4.1 $h^{-1}$ Mpc.
The dark matter correlation function assumes a $\Lambda$CDM cosmology with 
$\Omega_m$(0)=0.3 and $\sigma_8$=0.8. A different assumed value 
of $\sigma_8$ will linearly scale the bias value (e.g. a $\sigma_8$=0.9 will increase the absolute 
bias by $\sim12$\%). The bias error is estimated from the mean bias rms measured from 10 jackknife 
samples, similar to the errors on $r_0$ and $\gamma$ using power law fits. 

\begin{figure*}[ht]
\centering
\includegraphics[width=0.7\textwidth]{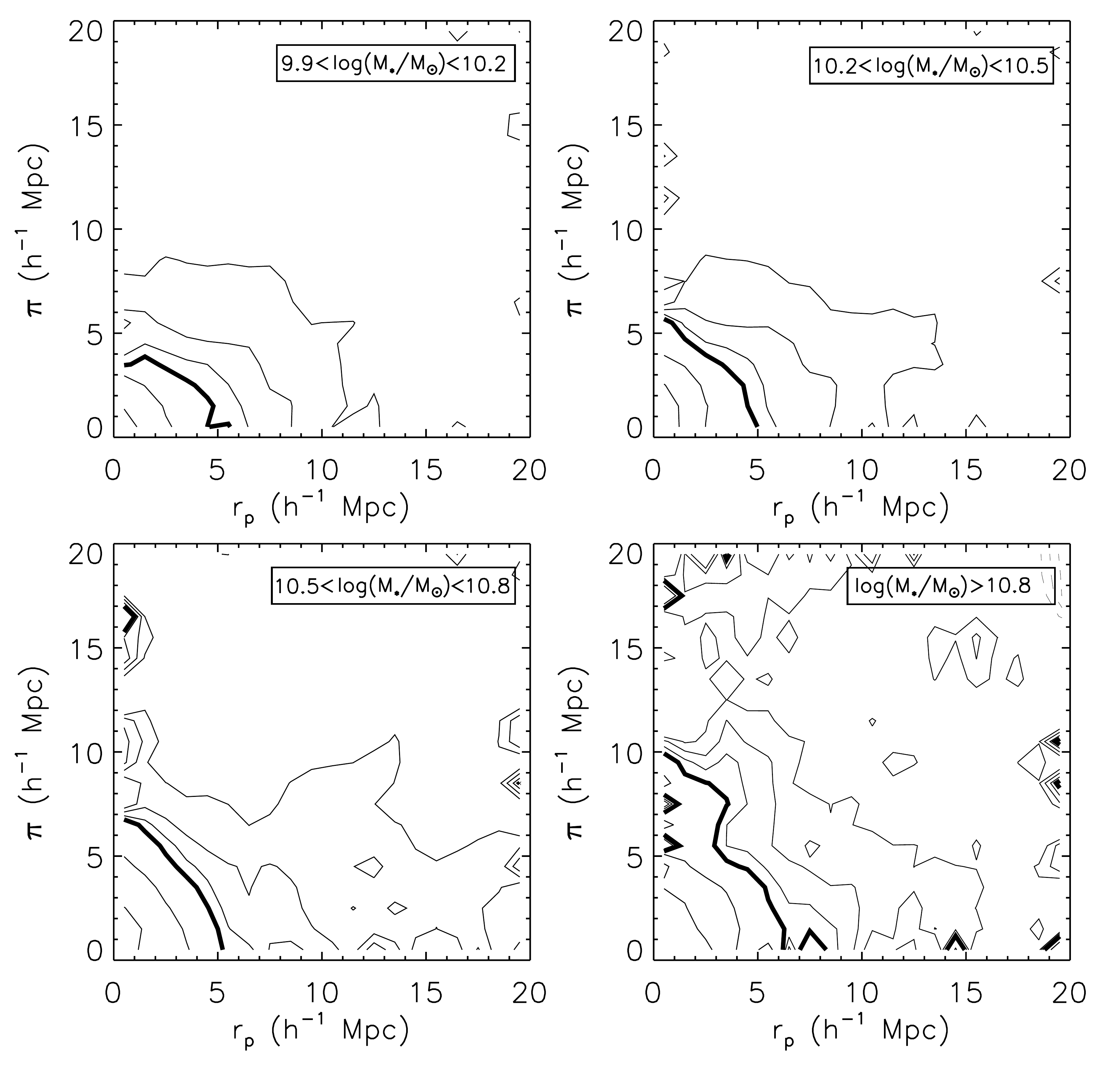}
\caption{Contours of constant correlation strength for the two-dimensional 
correlation function, \corfp, for blue galaxies with 0.74$<$\z$<$1.4 binned by 
stellar mass. The contour lines correspond to correlation amplitudes of 0.25, 
0.5, 0.75, 1.0, 2.0, and 5.0, with \corfp=1 shown as a thick line. The number of galaxies in each bin is as follows: 4774 (upper left), 
4279 (upper right), 2435 (lower left), and 1134 (lower right). The contours have been smoothed by 
a $1\times1~h^{-1}$ Mpc boxcar for clarity in the figure.}
\label{fig:massxisp}
\end{figure*}

\subsection{\rm{DEEP2}~\it{Masks and 1/V$_{\rm{max}}$}}
\label{sec:masks}

When calculating \corfp~for a given galaxy sample, we generate random catalogs 
of uniformly distributed points spanning the same volume as the data. There 
are several observational selection effects that must be taken into account 
in the random catalog in order to match the spatial distribution of the data. 
Examples of these effects include the overall survey geometry, bright star 
masks, incompleteness due to DEIMOS slit design and 
placement, and variations in the 
redshift success rate due to observing conditions for each slitmask. 
The effects of survey geometry, data gaps, and the varying redshift success 
rates are all accounted for in the DEEP2 window function. 

To remove the window function and any redshift selection effects, we generate a selection function 
from the galaxy redshift distribution and their estimated comoving distance in a $\Lambda$CDM cosmology. 
We smooth and spline fit the one-dimension spatial distribution such that local overdensities due to cosmic variance are removed while preserving
the overall shape of the distribution. The galaxy distributions are calculated in comoving bins of $\Delta r$=0.011 $h^{-1}$ Mpc, 
and smoothing of the spline fit occurs over scales of $\Delta r$=0.03-0.05 $h^{-1}$ Mpc. All selection functions are checked by eye such
that no power on smaller scales is retained in the smoothed selection function, which
could reduce inferred correlations on those scales. Random samples are then generated 
from the probability distribution of the smoothed selection function. The procedure 
ensures that the computed correlation function statistics are not affected by the survey design 
or completeness of our selected galaxy samples

Although the correlation function statistics are robust, a systematic error is still imprinted on small scales where the 
correlation function is more likely to be incomplete due to slit collisions. 
Using DEEP2 mock catalogs \citep{Yan04}, C08 found that the bias varies 
smoothly as a function of scale and varies in size from 25\% on the smallest 
scales to 2\% on the largest scales. We apply a correction for this systematic 
bias directly to the measured \corfp~and \wprp, and we include an additional 
rms error term for the correction which is measured from the mocks and added 
in quadrature to \wprp.

To aid in the application of our results to HOD models, we calculate the 
number density for each of our complete samples using the nonparametric 
1/$V_{\rm{max}}$ method \citep{Felten76} employed in previous DEEP2 studies 
\citep{Blanton06, Zhu09}.  The volume over which a given galaxy 
can be observed is given by 

\begin{equation}
V_{\rm{max}}=\frac{1}{3}\int d\Omega \int_{z_{\rm{min}}}^{z_{\rm{max}}} dz \frac{d[D_{c}(z)^3]}{dz} f(z),
\label{eq:vmax}
\end{equation}
where $D_{c}(z)$ is the comoving distance and a spatially flat universe 
\citep{Hogg99}, $d\Omega$ is the solid angle of sky covered in the survey, 
and $f(z)$ is the probability that a given galaxy is targeted and a 
successful redshift is produced in a given DEEP2 mask.  We calculate 
$d\Omega$ using the same DEEP2 masks as those used in our clustering 
analysis, which excludes regions of bright stars and accounts for mask 
edges, and we find that the integrated survey area covered in our slit masks is 
2.54 deg$^2$. 

The DEEP2 redshift success rate, $f(z)$, is a function of both  
$R$-band magnitude and restframe color; these are discussed in detail 
in \cite{Newman12} (see also \citealt{Blanton06}, \citealt{Willmer06}, and \citealt{Zhu09}). 
For each observed galaxy, we generate 1200 simulated galaxies with similar observational properties 
between $z_{\rm{min}}$ and $z_{\rm{max}}$ and calculate $k$-corrected 
magnitudes for each simulated redshift. We then apply the selection 
function $f(z)$ to these simulated galaxies and produce a fraction 
which is multiplied by the comoving volume to produce $V_{\rm{max}}$.

Once $V_{\rm{max}}$ is obtained, we calculate the effective number of 
galaxies in the comoving volume,

\begin{equation}
N_{\rm{eff}} = \left[ \sum_{i} \frac{1}{(V_{\rm{max}})_i} \right]^2 / \left[ \sum_{i} \frac{1}{(V_{\rm{max}})_i^2} \right],
\label{eq:neff}
\end{equation}
and divide by the comoving volume between $z_{\rm{min}}$ and $z_{\rm{max}}$ 
to derive the number density, $n$. Upper and lower limits for $n$ are 
computed by assuming a Poisson error distribution with approximations 
provided by \cite{Gehrels86}. However, because DEEP2 is a small area survey, 
cosmic variance ($\sigma_{\rm{cv}}$) 
will be a significant source of error in the number density. 
Following \cite{Zhu09}, we calculate the variance in number density 
from each of the four DEEP2 fields and report the estimated 
$\sigma_{\rm{cv}}$ along with each of our samples. In selected stellar mass cases where DEEP2 
is complete, we have cross-checked our number densities with the published galaxy mass function between
0.75$<$\z$<$1.0 given in \cite{Bundy06} and find good statistical agreement with their measurements.

\begin{figure}[ht]
\centering
\includegraphics[width=0.95\columnwidth]{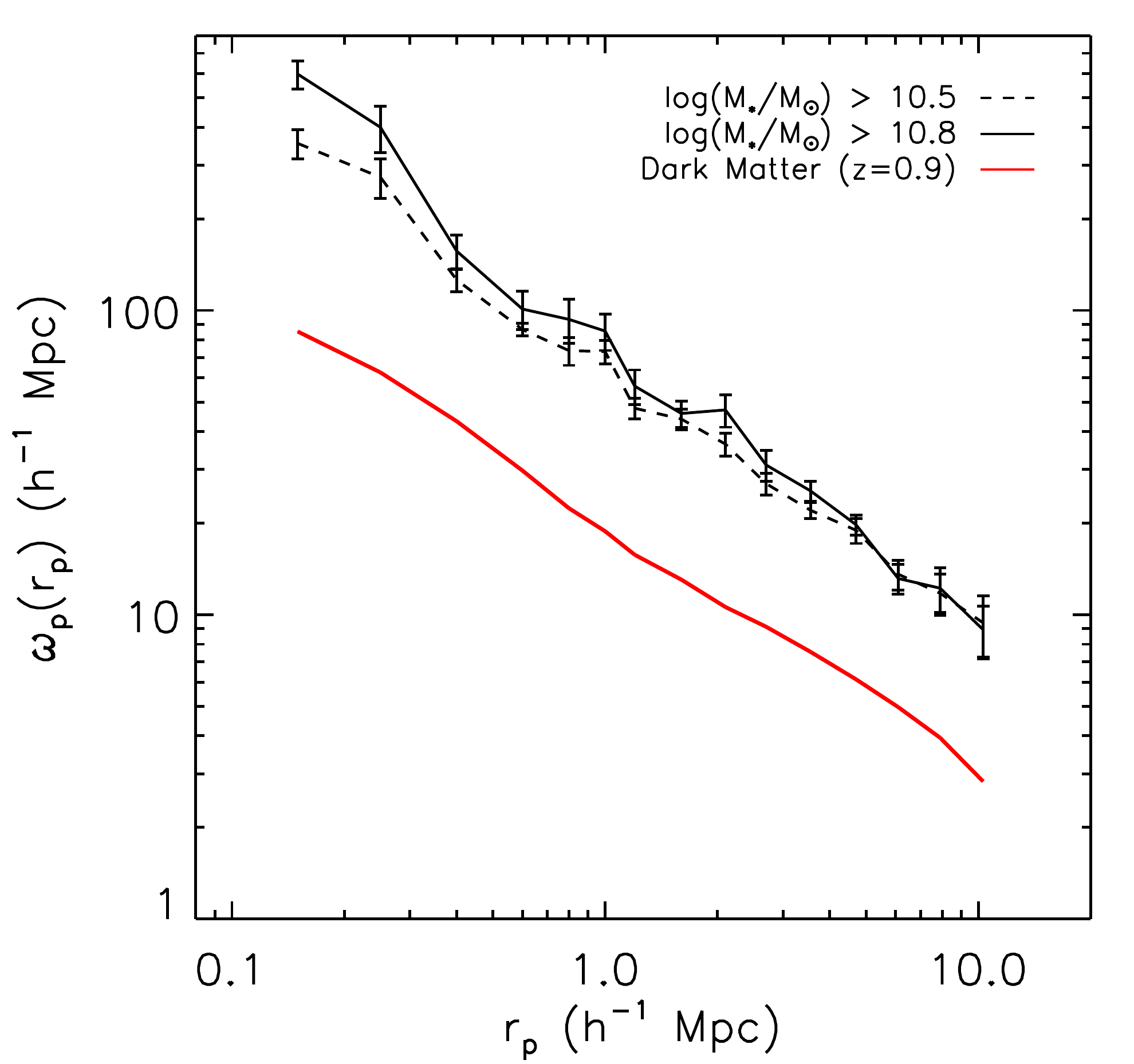}
\caption{The projected correlation function, \wprp, for color-independent 
galaxy samples selected by stellar mass thresholds and limited to \z$<$1.05. The plot
shows the two highest mass threshold samples with \smass$>$10.5; the highest mass sample
is complete for both red and blue galaxies and the lower mass sample is slightly incomplete
for red galaxies beyond \z$>$0.9.  
The projected clustering of dark matter (solid, red line) is generated from the prescription of
\citealt{Smith03} and shown for a redshift of \z=0.9, which is similar to the 
mean redshift of the galaxy samples. }
\label{fig:masscorrall}
\end{figure}

\begin{figure*}[ht]
\centering
\includegraphics[width=0.85\columnwidth]{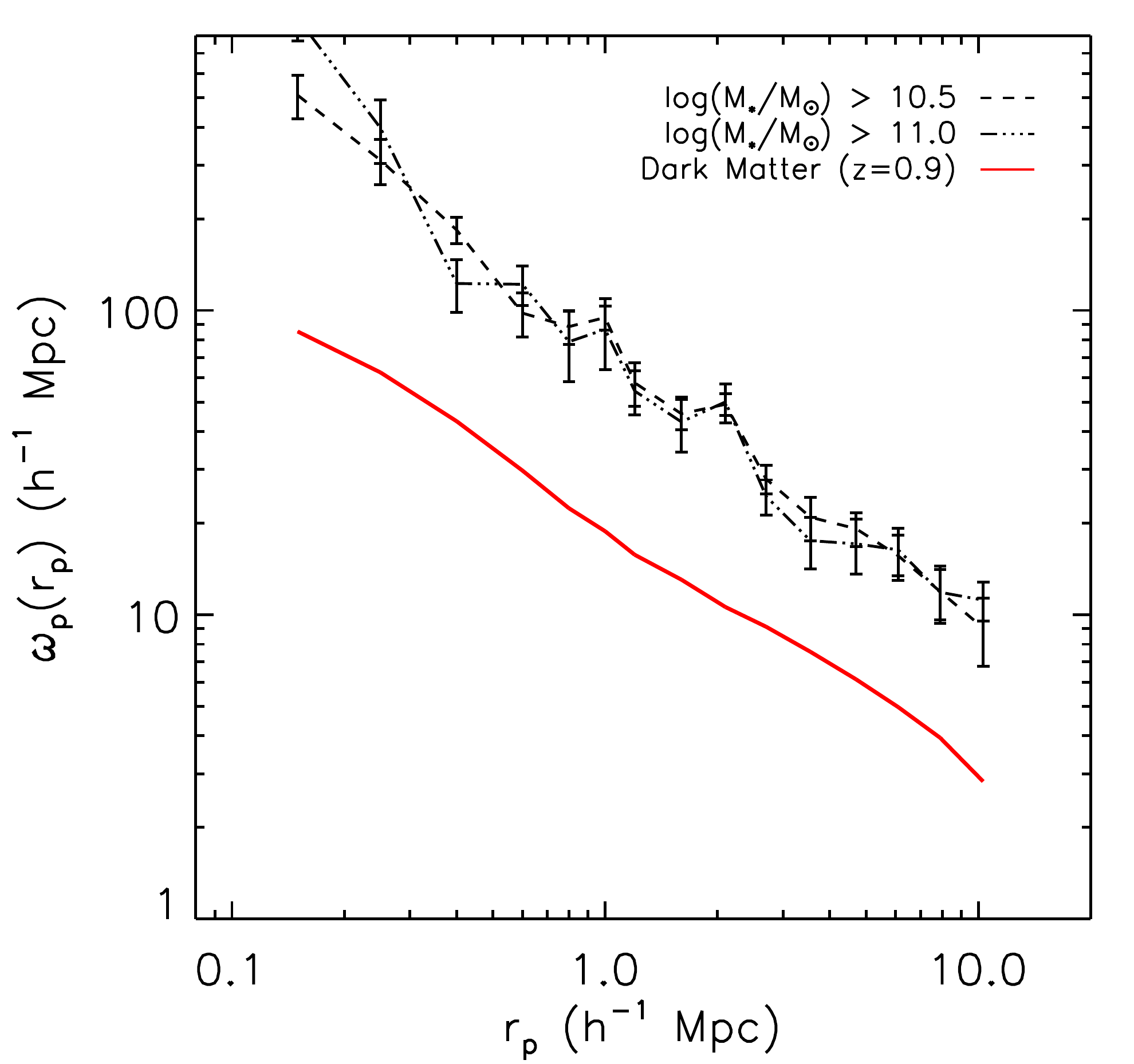}
\hspace{0.3in}
\includegraphics[width=0.85\columnwidth]{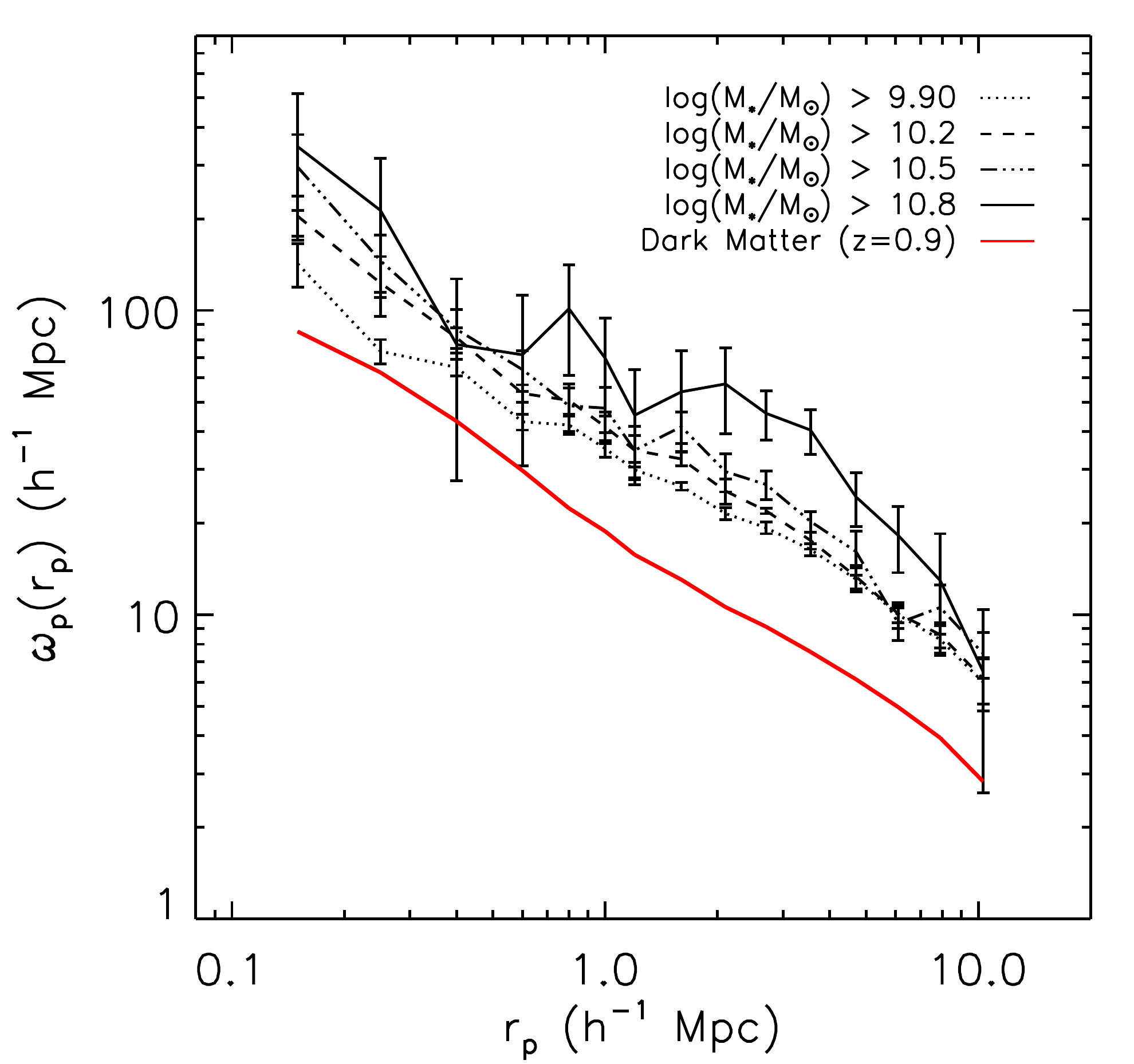}
\caption{The projected correlation function, \wprp, for color-dependent 
galaxy samples selected by stellar mass. 
Stellar mass threshold samples for red galaxies at \z$<$1.05 are shown on 
the left, while stellar mass threshold samples for blue galaxies at \z$<$1.4 
are shown on the right. }
\label{fig:masscorrcolor}
\end{figure*}

\section{Galaxy Clustering Results}
\label{sec:results}

\subsection{Clustering Dependence on Stellar Mass}
\label{sec:smass}

\begin{figure*}[ht]
\centering
\includegraphics[width=0.95\columnwidth]{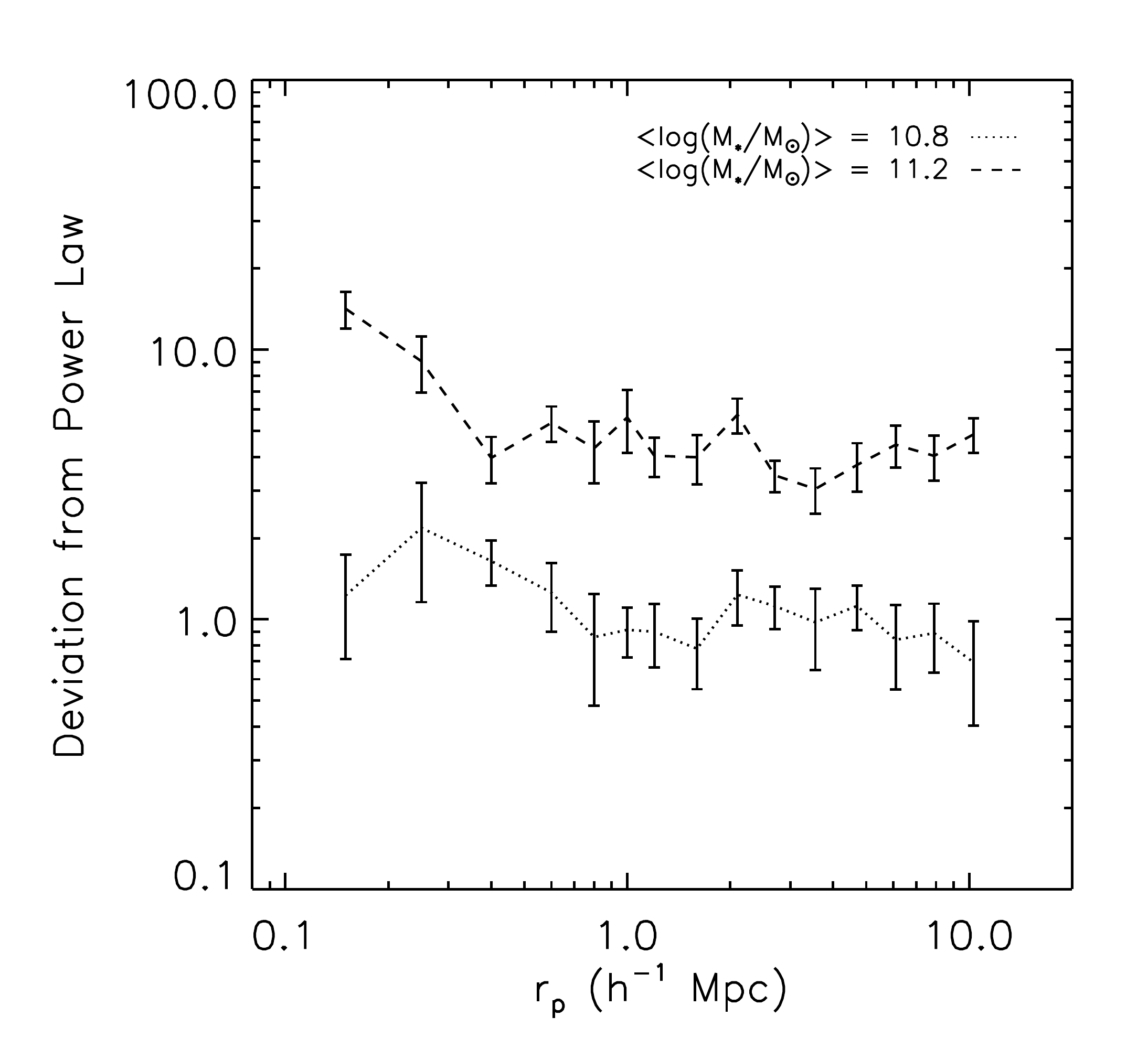}
\includegraphics[width=0.95\columnwidth]{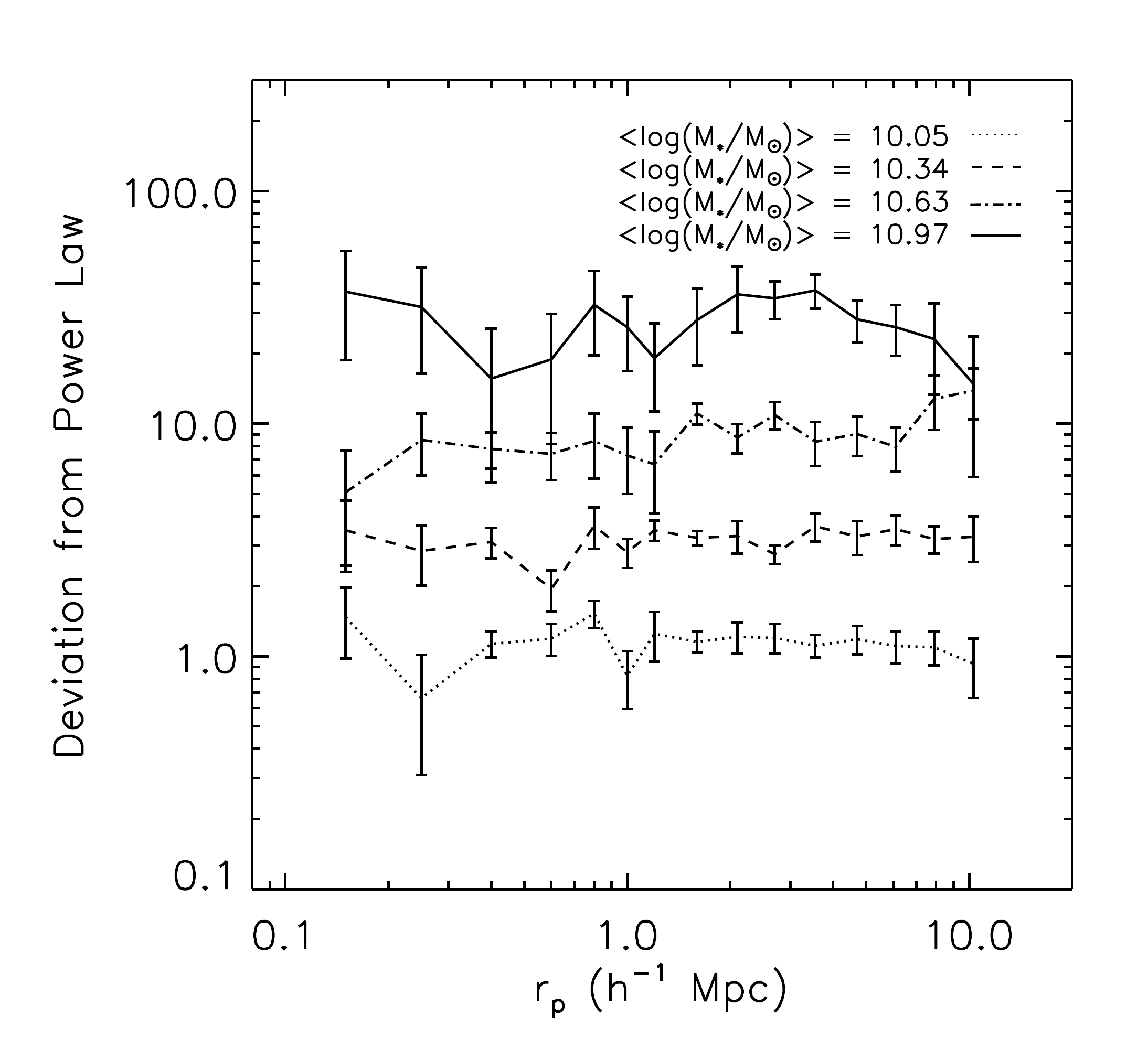}
\caption{The relative deviation in \wprp~from the best-fit power law for 
galaxy samples separated by red (left) and blue (right) restframe color 
and binned by stellar mass. The power law is fit to \wprp~on scales of 
1-10 $h^{-1}$ Mpc.  Each sample is offset by an arbitrary 0.3 dex
 for clarity in the figure. The mean deviation between the relative small-scale clustering 
 below $r_p$$\leq$0.3 $h^{-1}$ Mpc and large-scale
 clustering between 0.4$<$$r_p$$<$10.5 $h^{-1}$ Mpc
is significant in red galaxies at the $p<0.1$ level for 10.5$<$\smass$<$11.0 and $p<0.001$ level for \smass$>$11.0.
We find no significant increase in small-scale clustering for any of the blue galaxy mass samples beyond the best-fit large-scale power law. }
\label{fig:relmasscorrcolor}
\end{figure*}

\begin{figure*}[ht]
\centering
\includegraphics[width=1.0\textwidth]{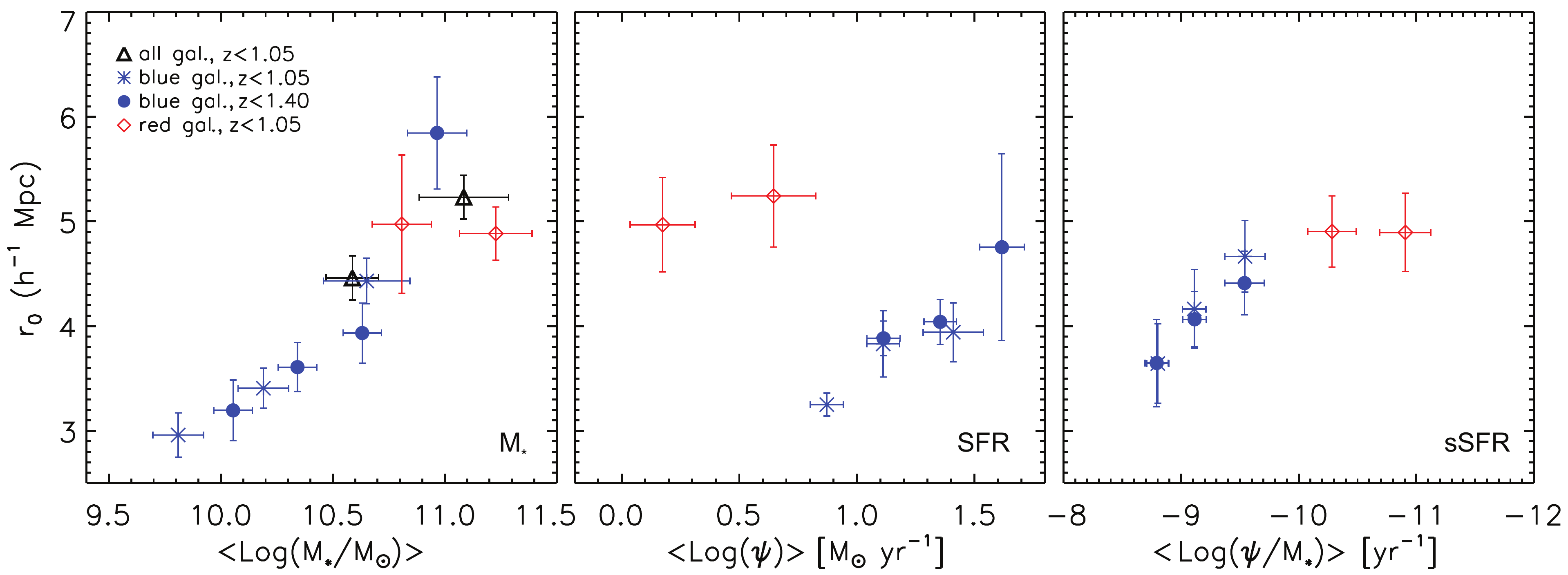}
\caption{The clustering scale length, $r_0$, measured by fitting a 
power law to \wprp~over 
linear scales of 1-10 $h^{-1}$ Mpc. For this figure, the power law slope is fixed to $\gamma$=1.6 
so as to remove degeneracies between $r_0$ and $\gamma$. The scale lengths are measured
for galaxy samples in stellar mass (left), SFR (center), and sSFR (right) as described in 
Sections~\ref{sec:smasssample}-\ref{sec:ssfrsample}. Vertical error bars are calculated from the rms
scatter of $r_0$ fit from 10 jackknifed sub-samples in each data bin, and horizontal error bars represent the 
rms scatter for each bin of selected galaxy sample values.  All measured values of absolute bias, $r_0$, and $\gamma$
are given in Tables~\ref{tab:smassresults}-\ref{tab:ssfrbinresults}.}
\label{fig:r0results}
\end{figure*}

Results for the two-dimensional correlation function, \corfp, as a function of 
stellar mass are shown in Figure~\ref{fig:massxisp} for the 0.74$<$\z$<$1.4 
blue galaxy samples.  The figure shows contours of constant correlation 
strength for three bins in stellar mass ranging from 
9.9$<$\smass$<$10.8 and a fourth sample with a 
lower stellar mass threshold of \smass$>$10.8 (lower right panel).  In the lowest stellar mass 
bin, \corfp~ is relatively symmetric both across and along the line of sight, 
while at higher stellar masses there is an increasing asymmetry in \corfp. 
At all masses there is evidence of a compression along the 
line of sight on large scales, due to coherent infall into gravitational 
potential wells known as the ``Kaiser effect" \citep{Kaiser87}. 
Redshift-space distortions on small scales (``Fingers of God'') also increase dramatically 
with stellar mass, indicating that high-stellar-mass blue galaxies 
have greater virial motions and therefore are more likely to be in systems 
with larger halo mass.  These results mirror those of C08, who found a similar 
increase in redshift-space distortions with increasing luminosity 
(see their Figure 7). 

As redshift-space distortions arise from peculiar velocities that are not directly correlated to the real-space clustering, 
we mitigate the distortions by integrating \corfp~
along the line of sight and projecting the clustering signal onto the 
$r_p$ axis. This results in the projected correlation function, \wprp, 
which we show for color-independent mass threshold samples in 
Figure~\ref{fig:masscorrall} and color-dependent mass threshold samples 
in Figure~\ref{fig:masscorrcolor}. We also plot the projected dark 
matter correlation function \citep{Smith03} generated for a 
$\Lambda$CDM universe ($\sigma_{8}$=0.8) at a mean redshift similar to that of 
the galaxy samples ($z=0.9$). 
In all cases, \wprp~is consistent with  
a power law on scales larger than $r_p$$>$1 $h^{-1}$ Mpc where 
the clustering of galaxies in independent halos (e.g the two-halo 
term) dominates the correlation function. 

On scales of $r_p$$\leq$0.3 $h^{-1}$ Mpc, we find a departure from a 
single power law for the highest stellar masses in the color-independent 
sample (Figure~\ref{fig:masscorrall}) and at similar masses 
in red galaxies (left side of Figure~\ref{fig:masscorrcolor}). The 
departure on small scales is seen more clearly in Figure~\ref{fig:relmasscorrcolor}, where we 
divide out the large-scale power 
law fit over 1-10 $h^{-1}$ Mpc  and plot the relative correlation function for red and blue 
galaxy mass samples.  For clarity, the 
figure shows these relative correlation functions with an offset of 0.3 
dex between different stellar mass samples. 

To determine the significance with which the observed clustering on small scales is a departure from the large-scale behavior, 
we perform a $t$-test between the relative large-scale and small-scale clustering amplitude (Figure~\ref{fig:relmasscorrcolor}). For each jackknife
subsample of a given galaxy sample, we fit a power law over large scales with  $r_p$=1-10 $h^{-1}$ Mpc and ratio the correlation function 
averaged over $r_p$$\leq$0.3 $h^{-1}$ Mpc to the best-fit power law averaged over the corresponding small scales.  
We then compute the mean and standard deviation 
among the 10 jackknife sample ratios and compute a $t$ score in which the null hypothesis ratio value is 1. Finally, we compute the corresponding
 $p$-value drawn from a $t$-distribution with 9 degrees of freedom. Assuming that there is minimal correlated error between $r_p$$\leq$0.3 $h^{-1}$ Mpc and $r_p>1$ $h^{-1}$ Mpc, this procedure encapsulates the error associated with both the large-scale power law fit and the covariance of \wprp~on small scales.
 
 For red galaxies with 10.5$<$\smass$<$11.0, the $t$-test
produces a $p$-value of $p$=0.005, meaning that a small-scale deviation as large as that observed or larger would occur only 0.5\% of the time 
if there were in fact no true deviation. 
Red galaxies with \smass$>$11.0 have a small-scale deviation from the large-scale power law fit with $p=0.042$, a significant deviation at the $p<0.05$ significance level. The lower significance in the higher mass galaxy sample is due in part to larger correlated variation in the large-scale power law slope $\gamma$ in the jackknifed samples. 
For blue galaxies (right side of Figure~\ref{fig:relmasscorrcolor}), we find no statistically 
significant increase in clustering on small scales relative to large scales at 
any stellar mass probed in this study.  The difference in the relative correlation function amplitude between 
red and blue galaxy samples on small scales could indicate a 
change in the central-to-satellite fraction or a change in the halo mass distribution 
in the halo mass function \citep{Zheng09}.
We will further discuss the physical interpretation of this result in Section~\ref{sec:sscale}.

The left panel of Figure~\ref{fig:r0results} shows the measured correlation length, $r_0$, 
when fitting \wprp~with a power law model of $\xi$=$(r/r_0)^{-\gamma}$. As the slope of the 
correlation function, $\gamma$, can be covariant with $r_0$, we 
fix $\gamma$=1.6 to study the trends in $r_0$ with stellar mass. 
We find a strong positive correlation between $r_0$ 
and stellar mass for \smass$<$11.0 for blue, red, and color-independent 
galaxy samples.  At 11.0$<$\smass$<$11.5, the clustering length appears to level off 
at $r_{0}\sim5$  $h^{-1}$ Mpc for all colors and does not depend on stellar mass, within the errors. 
The $r_{0}$ results could indicate that high-stellar-mass blue galaxies at \z$>$1 are precursors to similar mass 
red galaxies at \z$<$1, as they likely reside in similar mass halos. We note that 
the DEEP2 sample has poor constraints for \emph{all} stellar masses above \smass$>$11.5; the survey volume is 
not large enough to robustly sample the rarest, most massive galaxies. The measured values of $r_0$, 
$\gamma$, and the absolute bias for each stellar mass sample 
are given in Tables~\ref{tab:smassresults} and \ref{tab:smassbinresults}. By design, the absolute
bias has a similar trend to $r_0$ where the bias increases with stellar mass. We find no 
significant trend between the best fit correlation slope $\gamma$ and stellar mass.

For the threshold samples in Table~\ref{tab:smassresults}, we also estimate the minimum and mean halo mass 
using the measured large-scale galaxy bias and the halo mass function for central galaxies 
\citep{Sheth99, Jenkins01}, 
modified to accommodate the increased frequency of satellites 
at low halo mass (see the Appendix of \cite{Zheng07} for details). For each
galaxy sample, we use the halo mass function calculated for the mean redshift of the sample 
in a $\Lambda$CDM cosmology ($\sigma_8$=0.8),
and we assume a satellite fraction of $\sim17$\% as measured for DEEP2 $L^*$ galaxies \citep{Zheng07}. 
We caution that the minimum and mean halo masses are intended as estimates 
in lieu of a full HOD model fit to \wprp, which we reserve to a future study. 
Table~\ref{tab:smassresults} shows that the mean halo mass ranges from 
$<$log($M_{\rm{halo}}$)$>=12.3 \ h^{-1} M_{\sun}$ in the
lowest stellar mass blue galaxy sample to 
$<$log($M_{\rm{halo}}$)$>$$\sim13~h^{-1} M_{\sun}$ at the highest stellar masses. 
This mass range is particularly interesting as halo mass assembly models predict 
that galaxies at \z=1 transition from slow clustering growth 
below 10$^{12} M_{\sun}$ to near exponential clustering growth above 10$^{13}~M_{\sun}$ 
\citep{Moster10, Alexi12}. Although the 
halo masses presented here are estimates, we expect that they will help 
constrain the stellar-mass/halo-mass relationship in this important mass range
at $z\sim1$.

\subsection{Clustering Dependence on Star Formation Rate}
\label{sec:SFR}

We now investigate how clustering properties depend on SFR using the 
SFR-selected galaxy samples described in Section~\ref{sec:sfrsample}. Because 
complete samples of red and blue galaxies probe vastly different SFR ranges, 
we consider only galaxy samples separated by color 
(see Figure~\ref{fig:sfrbins}). Following the same 
procedure as with stellar mass, we investigate trends in the redshift 
space distortions for blue galaxies as a function of SFR by studying \corfp. 
We find that blue galaxies at all
SFRs display ``Kaiser infall'' on large scales, and all samples have 
relatively small ``Fingers of God'' on small scales.
This implies that large-scale infall dominates the \corfp~signal
in our SFR samples and contrasts with the significant redshift-space distortions seen as a function of 
stellar mass. The lack of  ``Fingers of God''  indicates that the SFR samples must span a broader
range of stellar masses than mass-selected samples, thus diluting the strong ``Fingers of God'' seen in the latter.

\begin{figure*}[ht]
\centering
\includegraphics[width=0.85\columnwidth]{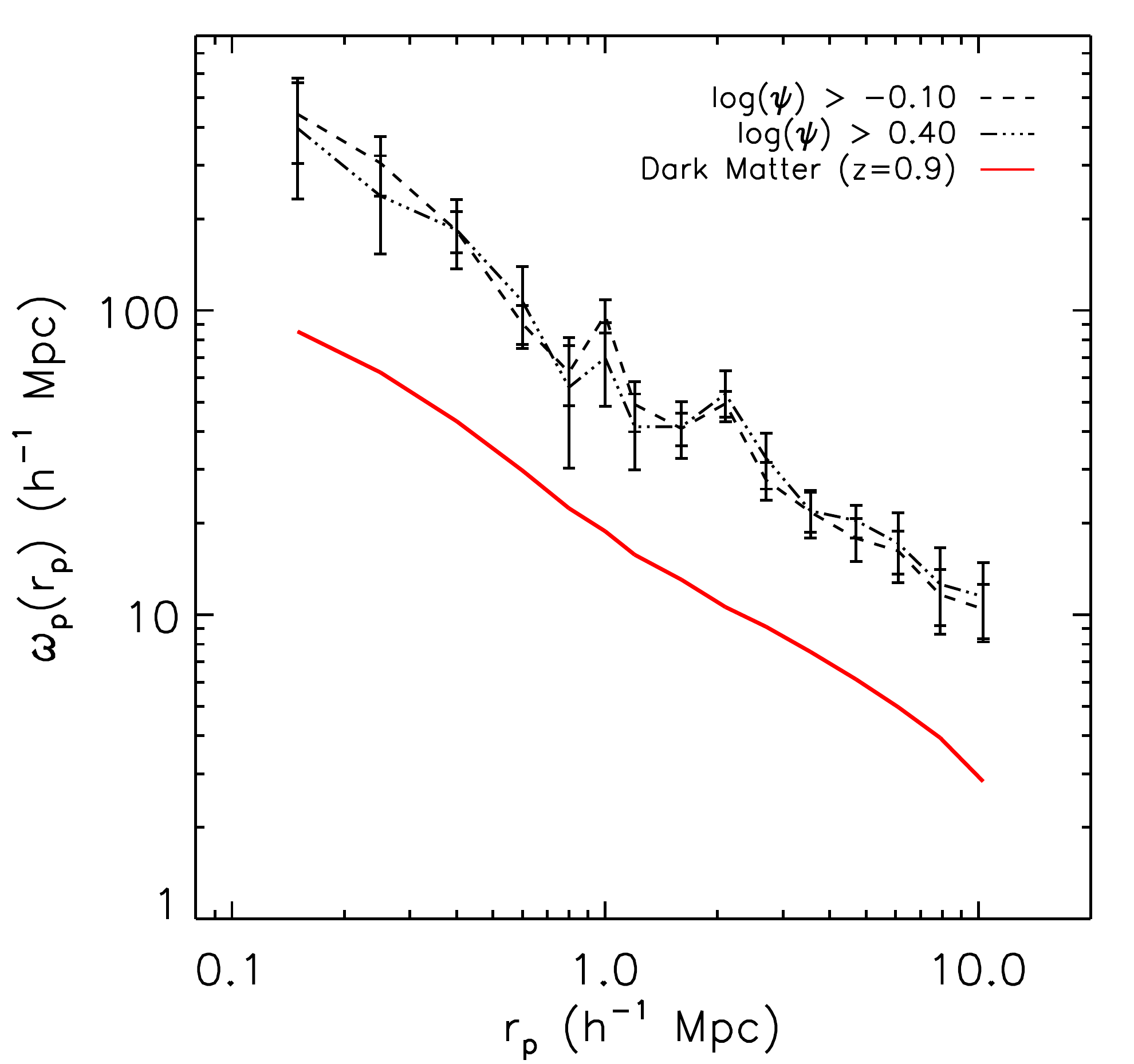}
\hspace{0.3in}
\includegraphics[width=0.85\columnwidth]{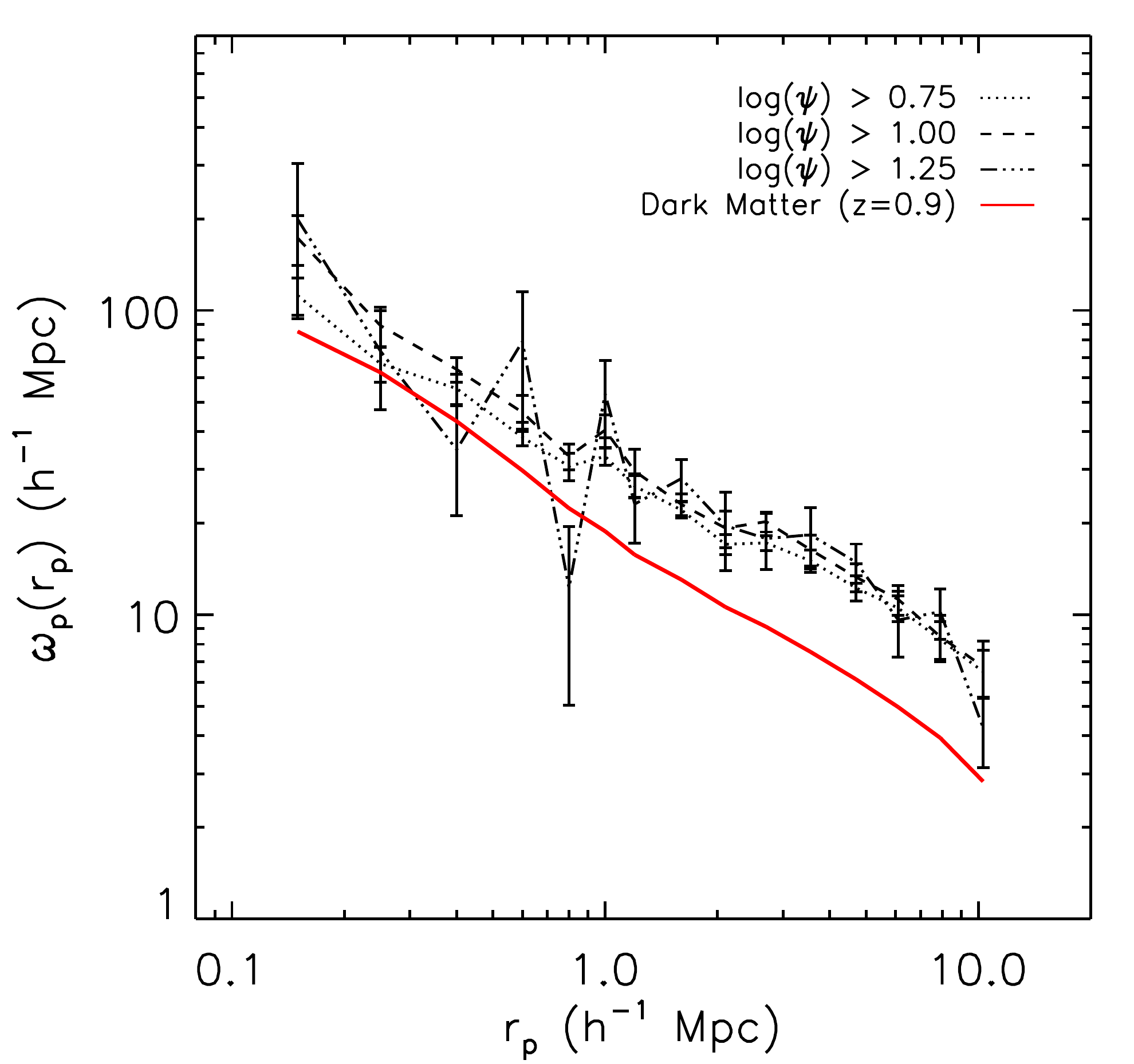}
\caption{
The projected correlation function, \wprp, for red (left) and blue (right) 
galaxy samples as a function of SFR threshold, limited to \z$<$1.05. The 
dark matter projected correlation function is shown for \z=0.9, 
near the mean redshift of all galaxy samples. }
\label{fig:sfrcorr}
\end{figure*}

\begin{figure*}[ht]
\centering
\includegraphics[width=0.95\columnwidth]{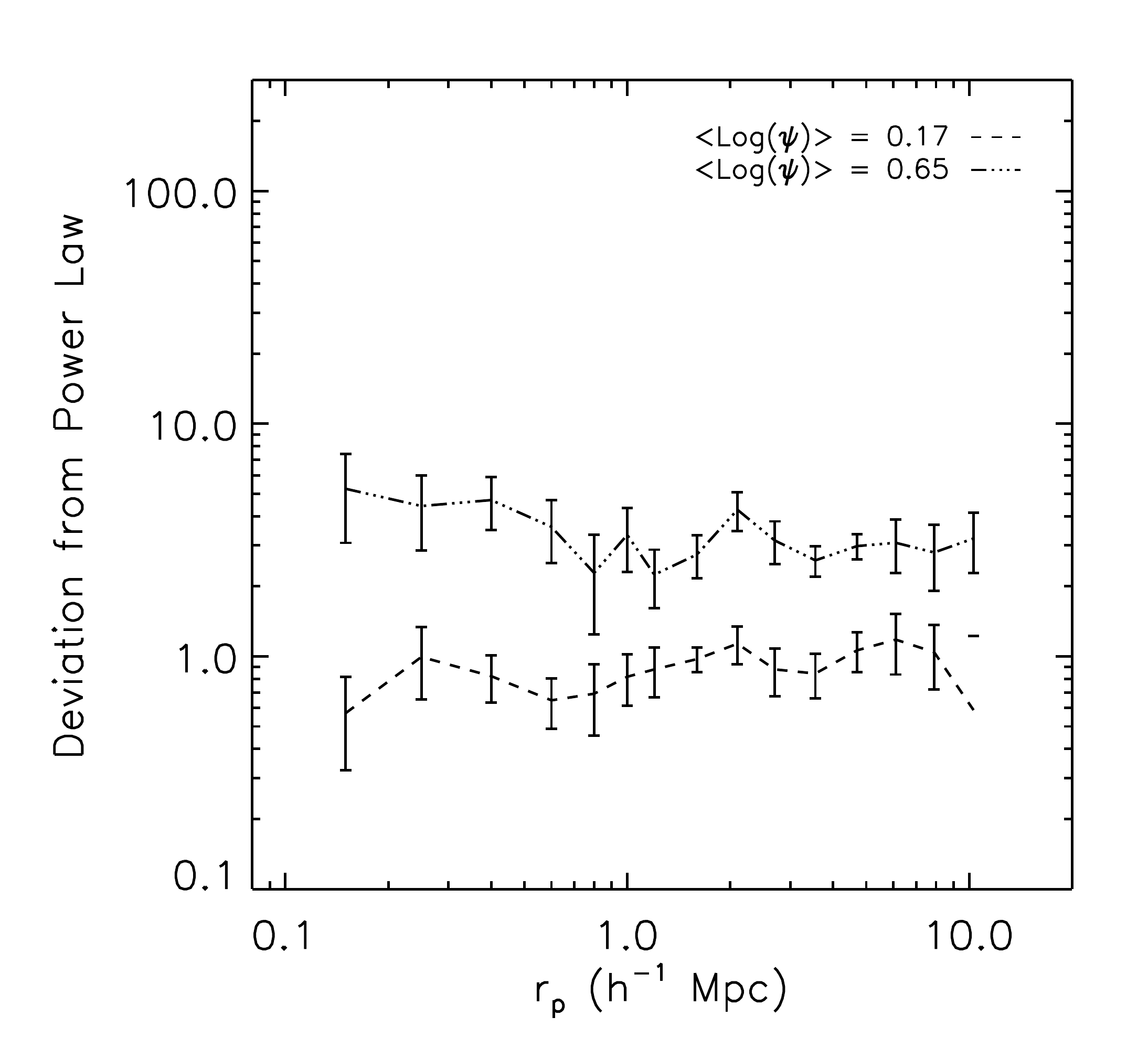}
\includegraphics[width=0.95\columnwidth]{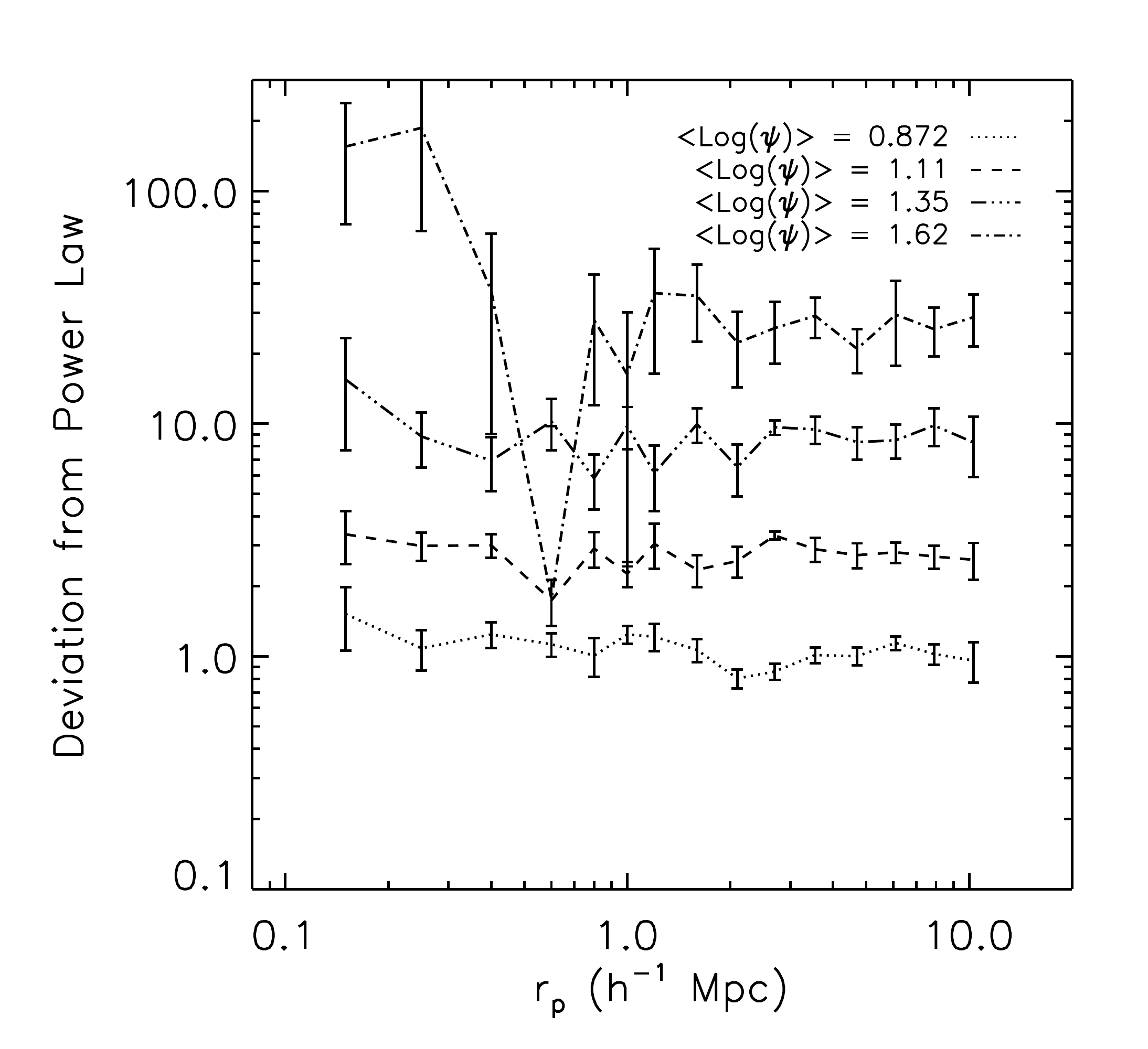}
\caption{The relative deviation in \wprp~from the best-fit power law for 
red (left) and blue (right) galaxy samples binned by SFR. Similar to stellar mass
in Figure~{ref:relmasscorrcolor}, the power law is 
fit to \wprp~on scales of 1-10 $h^{-1}$ Mpc, and each sample is offset 
by an arbitrary 0.3 dex in the figure for clarity. We find evidence of 
enhanced small-scale clustering in the highest-SFR blue galaxy sample 
but not for either red galaxy sample.}
\label{fig:relsfrcorrcolor}
\end{figure*}

Turning to the projected correlation function, Figure~\ref{fig:sfrcorr} 
shows \wprp~for two SFR threshold levels ranging from 
-0.1$<$\sfr$<$0.4 $M_\sun$ yr$^{-1}$ for red galaxies and three SFR threshold 
levels between 0.75$<$\sfr$<$1.25 $M_\sun$ yr$^{-1}$ for blue galaxies. 
The samples shown are restricted to \z$<$1.05 for a direct comparison 
between red and blue galaxy samples within the same volume.  
We find that the clustering amplitude changes much less as a function of SFR
within the red or blue sample compared to the difference in amplitude between
the red and blue samples. Blue galaxies with higher SFR have higher clustering 
amplitude, while there is no detected difference in clustering 
for the two red galaxy samples as a function of SFR, within the errors. 
Figure~\ref{fig:relsfrcorrcolor} shows the deviation of \wprp~ 
from a power law fit on scales of 1-10 $h^{-1}$ Mpc for red (left) 
and blue (right) galaxies binned by SFR. For all samples with 
\sfr$<$1.5 $M_\sun$ yr$^{-1}$, there is no significant deviation from 
a power law on small scales ($r_p$$\leq$0.3 $h^{-1}$ Mpc).  For the highest SFR 
blue galaxies with \sfr$>$1.5 $M_\sun$ yr$^{-1}$, however, we detect a factor 
of seven increase in the mean clustering amplitude 
below $r_p$$\le$0.3 $h^{-1}$ Mpc relative to the large-scale power law fit. Performing the same 
$t$-test as was done for stellar mass, the excess clustering signal has a $p$-value of
$p=0.038$ and therefore is significant at the $p<0.05$ level.

The measured $r_0$ values for color-separated galaxy samples binned by SFR are 
shown in the center panel of Figure~\ref{fig:r0results}.
The clustering scale length increases 
with increasing SFR for blue galaxies and is constant for red galaxies as a 
function of SFR, within the measurement error.   However, blue galaxies
 with the highest SFR have clustering amplitudes similar to red galaxies, 
suggesting that they occupy the same mass halos at these redshifts. 
 The overall trends in clustering amplitude agree with 
$z\sim1$ environment studies performed in CP08, who found that the galaxy 
overdensity increases at both extremes of galaxy SFR. 

Tables~\ref{tab:sfrthreshresults} and~\ref{tab:sfrbinresults} list the 
$r_0$, $\gamma$, and bias values for each SFR sample,
and Table~\ref{tab:sfrthreshresults} further lists the minimum and mean halo mass estimated for the 
SFR threshold samples. We note that while there is a significant rise in the clustering amplitude 
on small scales for the highest SFR galaxies, there is no trend between
SFR and $\gamma$, within the errors.  The
red galaxy SFR samples, which are just as clustered on large scales 
as the highest-SFR blue galaxy sample, have a consistent clustering slope 
as all of the blue SFR samples, though the errors on the slope are large.

Interestingly, when comparing the stellar mass and SFR samples in Figure~\ref{fig:r0results}, 
the \sfr$>$1.5 $M_\sun$ yr$^{-1}$ blue galaxy sample has 
similar correlation lengths as the high-stellar-mass 
red galaxy samples. Further, as shown in Figures~\ref{fig:relmasscorrcolor} and~\ref{fig:relsfrcorrcolor},
we see similar enhanced clustering on small 
scales relative to the best fit large-scale power law behavior in these galaxy samples.
The similarities  between high-SFR blue galaxies and high-stellar mass red galaxies 
on {\it both} large and small scales indicates that they are found in 
similar mass halos, perhaps linked to a shared star formation history, with 
high-SFR blue galaxies evolving to high-stellar mass, low-SFR red galaxies 
at later epochs after the star formation is quenched.

\subsection{Clustering Dependence on Specific Star Formation Rate}
\label{sec:sSFR}

We also measure galaxy correlation properties as a function of 
specific SFR (sSFR). The sSFR is the inverse of the star-formation timescale of a galaxy. 
In secular evolution scenarios, the measured sSFR timescale is proportional to the existing stellar mass
relative to the amount of cold interstellar gas available for production of stars. 
Shorter timescales indicate rapid star formation from large, cold interstellar 
gas reserves and vice versa for longer timescales. 
As shown in Figure~\ref{fig:colormasscontour}, sSFR is strongly correlated 
with restframe color as it reflects the star formation history of the 
aggregate stellar population within a galaxy.  

\begin{figure*}[ht]
\centering
\includegraphics[width=0.85\columnwidth]{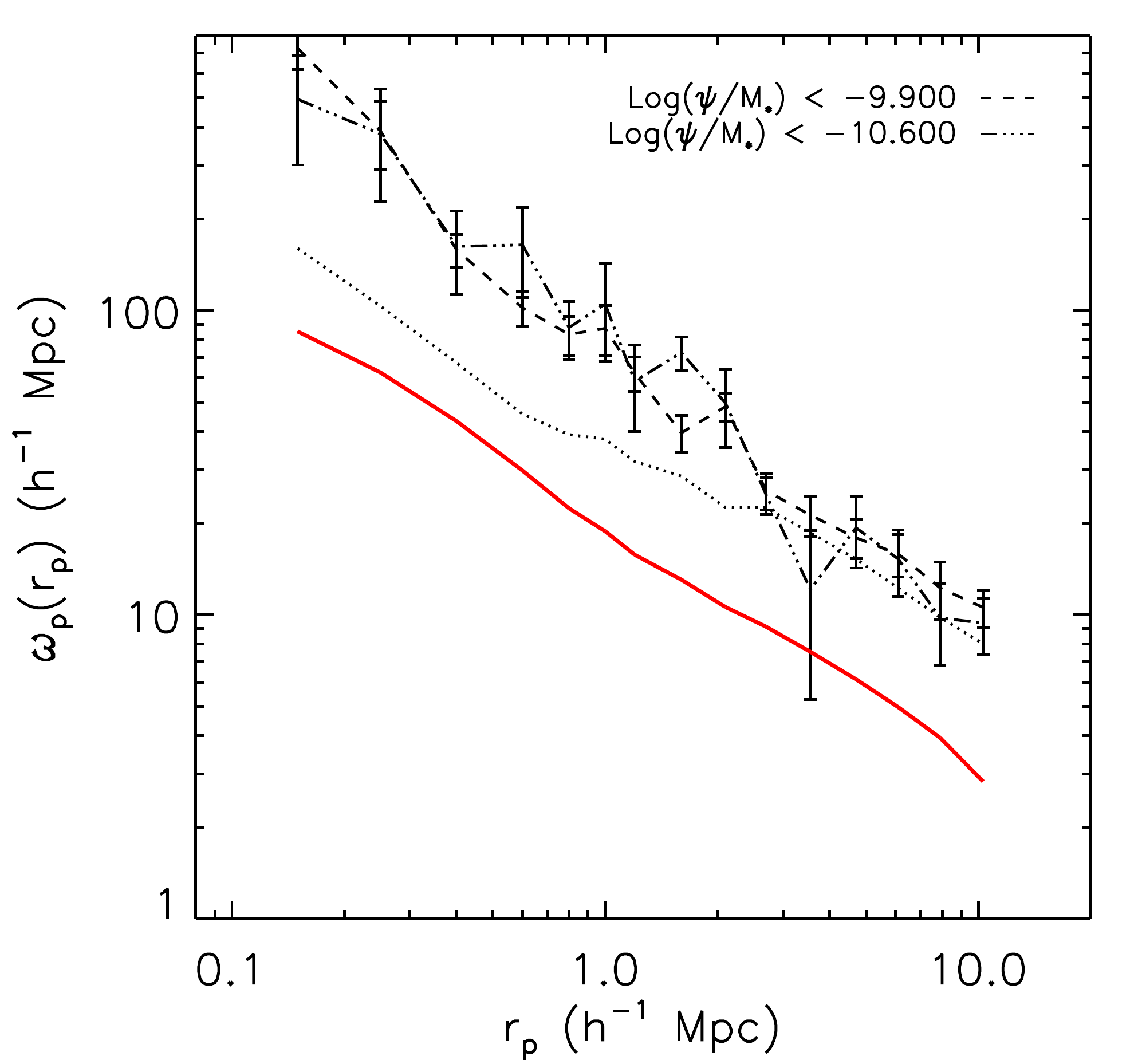}
\hspace{0.3in}
\includegraphics[width=0.85\columnwidth]{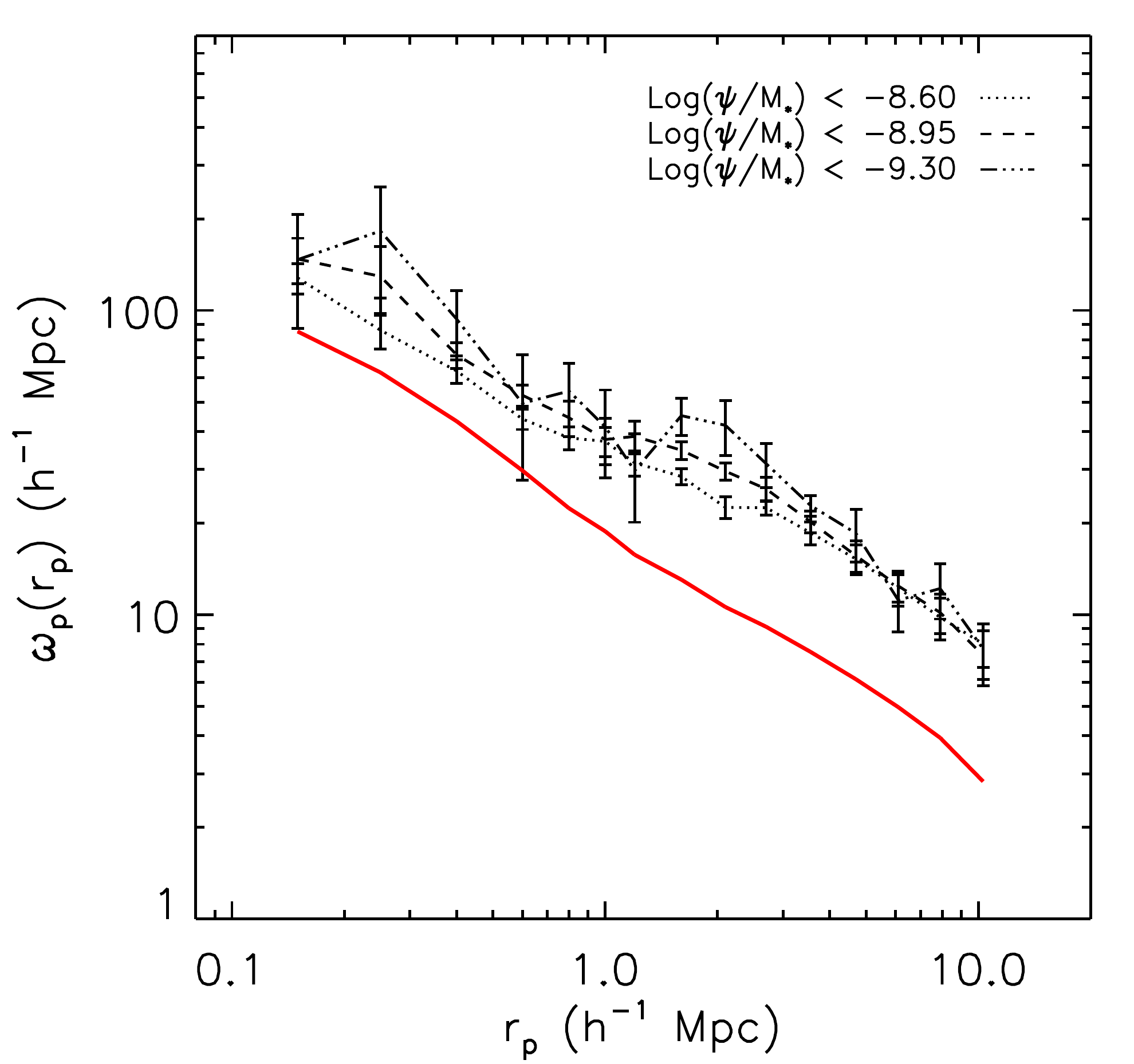}
\caption{The projected correlation function, \wprp, for red (left) and blue 
(right) galaxy samples selected using sSFR thresholds between 0.74$<$\z$<1.05$. The 
dark matter projected correlation function is shown for \z=0.9, 
near the mean redshift of all galaxy samples.
The dotted line in the left panel corresponds to \wprp~for 
blue galaxies with \ssfr$<$$-8.6$ yr$^{-1}$, shown in the right panel. }
\label{fig:ssfrcorr}
\end{figure*}
\begin{figure*}[ht]
\centering
\includegraphics[width=0.95\columnwidth]{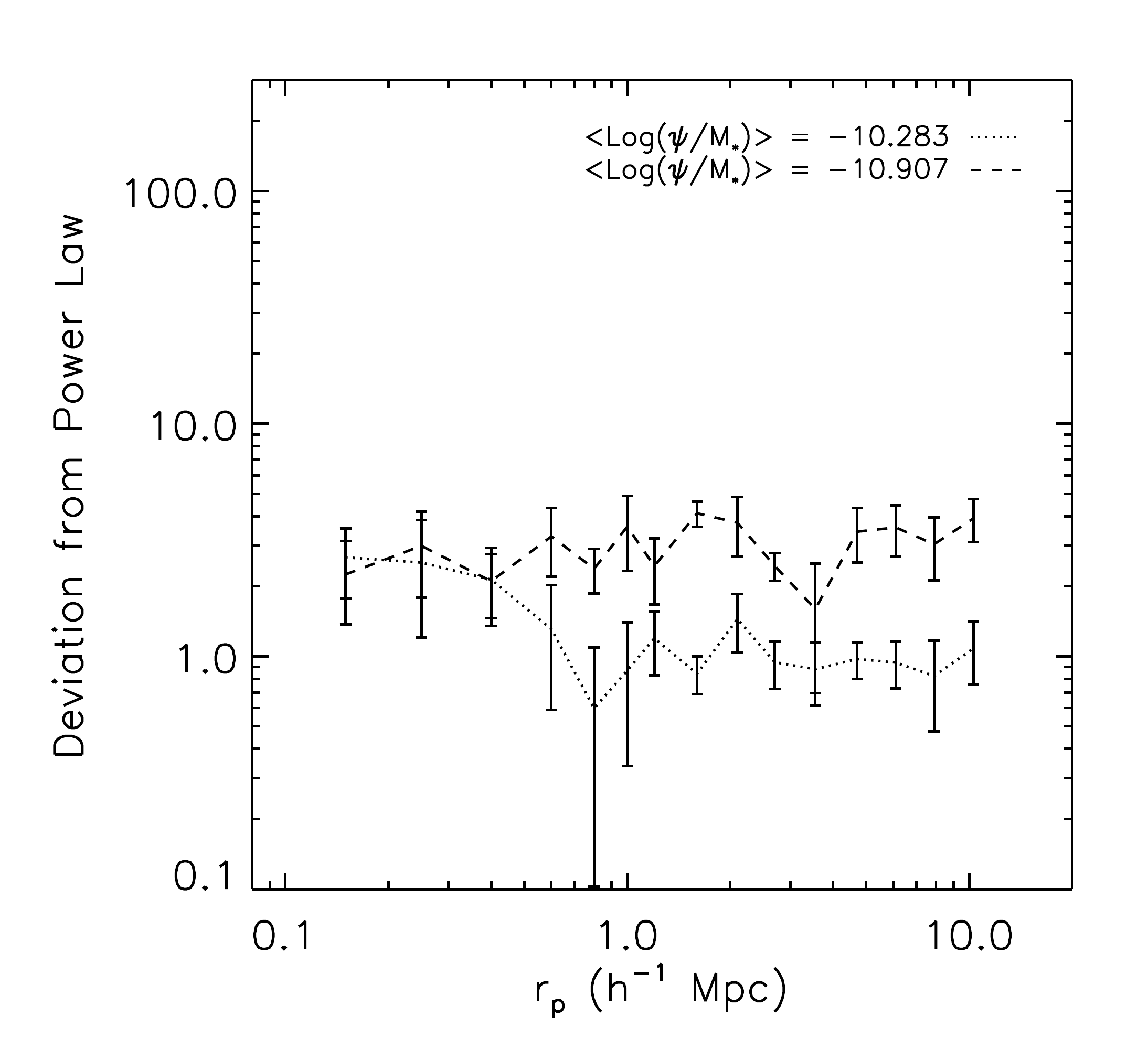}
\includegraphics[width=0.95\columnwidth]{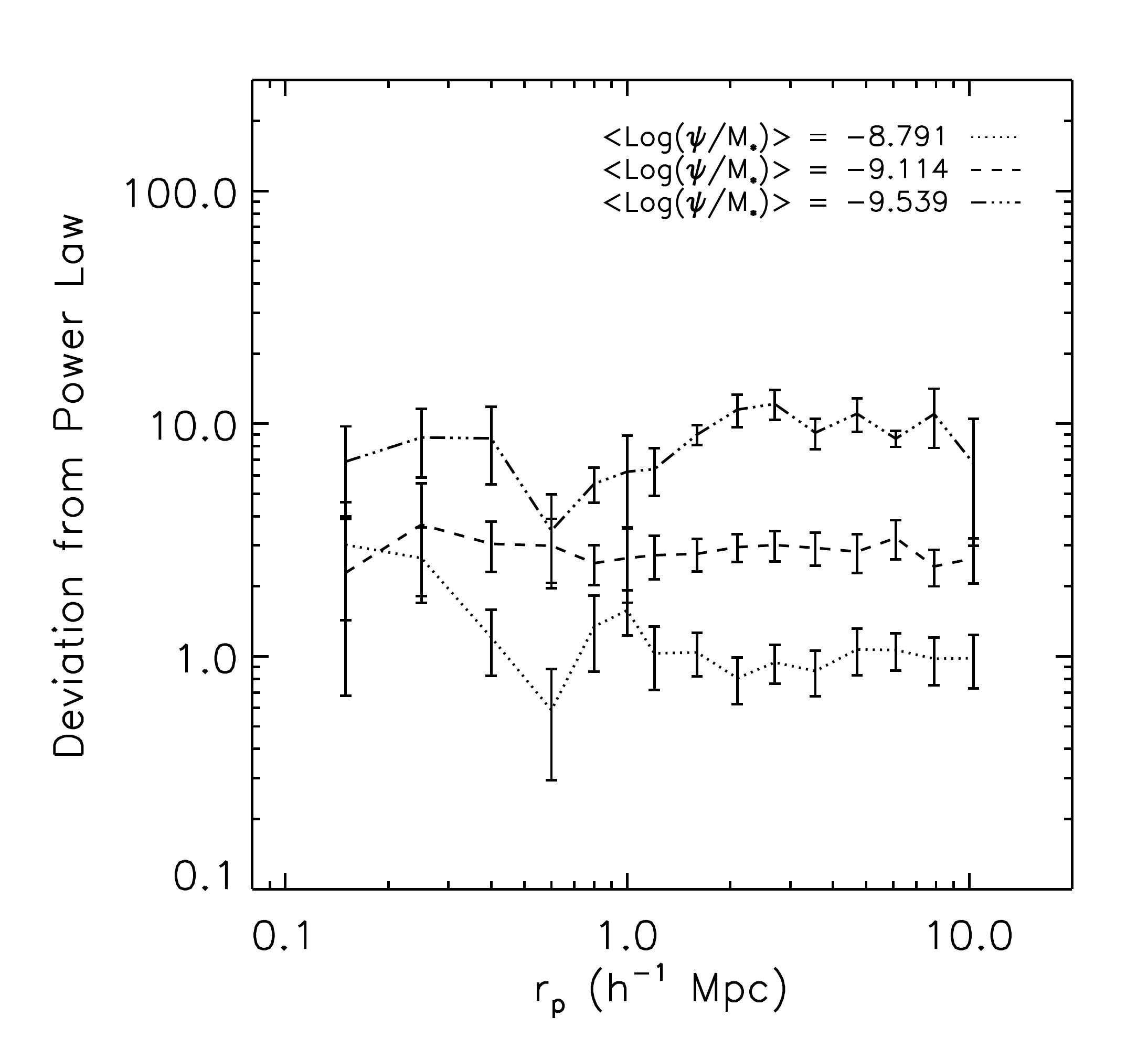}
\caption{
The relative deviation in \wprp~from the best-fit large-scale power law for 
red (left) and blue (right) galaxy samples binned by sSFR. 
A significant rise in the clustering amplitude is seen on small scales for the 
highest sSFR samples for both red and blue galaxies.}
\label{fig:relssfrcorrcolor}
\end{figure*}

Figure~\ref{fig:ssfrcorr} shows \wprp~for the color-dependent sSFR threshold 
samples described in Section~\ref{sec:ssfrsample}.  The sSFR samples are 
complete between 0.74$<$\z$<$1.05 for $M_B$$<$$-20.5$ blue galaxies and 
$M_B$$<$$-21.0$ red galaxies. In general, lower sSFR samples are more clustered
than higher sSFR samples, which is expected given the strong correlation
between clustering and galaxy color (C08).
Figure~\ref{fig:relssfrcorrcolor} shows the deviation from the best fit 
power law for samples binned by sSFR. We find that for the lowest sSFR galaxies in \emph{both} red 
and blue samples, \wprp~deviates significantly on small scales ($r_p$$\le$0.3 $h^{-1}$ Mpc) relative to the large-scale power law fit.
There is a factor of 4 increase in the small-scale clustering amplitude 
for $-10.6$$<$\ssfr$<$$-9.9$ red galaxies and a similar 
increase for $-8.95$$<$\ssfr$<$$-8.6$ blue galaxies.  The
$t$-test for both these samples produce $p$-values of $p=0.013$ and $p=0.018$, respectively.
Additionally, we find that the 0.74$<$\z$<$1.4 blue galaxy sample with $M_B$$<$$-21.0$ and $-8.95$$<$\ssfr$<$$-8.6$
also deviates from the large-scale power law fit by an average factor of 7 and has a highly significant increase 
in clustering ($p=2.8\times10^{-5}$) on $r_p$$\le$0.3 $h^{-1}$ Mpc scales.
The enhanced small-scale clustering indicates that the satellite fraction 
is related to how efficiently galaxies within a single halo, regardless of 
restframe color, process gas and produce stars.  We discuss the implications 
of this result further in Section~\ref{sec:discuss}.

The right panel of Figure~\ref{fig:r0results} shows the best fit $r_0$ values 
for galaxy samples binned by sSFR. 
The clustering length rises for blue galaxies with lower sSFR values 
and is constant for red galaxies with \ssfr$<$$-10$. 
Both red and blue galaxies with $-10.5$$<$\ssfr$<$$-9.5$ have similar 
clustering lengths and therefore likely occupy similar mass dark matter halos. 
The correlation slope, $\gamma$, also roughly trends towards higher 
values for lower sSFRs, although the trend is not highly significant given 
the errors. Tables~\ref{tab:ssfrthreshresults} and~\ref{tab:ssfrbinresults} list the $r_0$, $\gamma$, and bias values for each sSFR sample. 
Table ~\ref{tab:ssfrthreshresults} further records the minimum and mean halo mass estimated for the sSFR threshold samples.

\section{Discussion}
\label{sec:discuss}

The primary goal of this work is to measure the clustering properties of complete $z\sim1$ 
 galaxy samples selected by stellar mass, SFR, and sSFR .  We present 
a macroscopic view of our results in this section. To facilitate this 
discussion, we separate our findings into small- and large-scale behavior at 
$r_p\sim$1 $h^{-1}$ Mpc, which roughly corresponds to the transition between the 
one and two-halo terms in HOD models.  We first discuss the large-scale clustering 
results, including a comparison with previous work and relevant environment studies, 
and then discuss the small-scale clustering results.
\\
          
\subsection{Large-scale Clustering Behavior}
\label{sec:lscalemass} %6.1

\begin{figure*}[ht]
\centering
\includegraphics[width=0.33\textwidth]{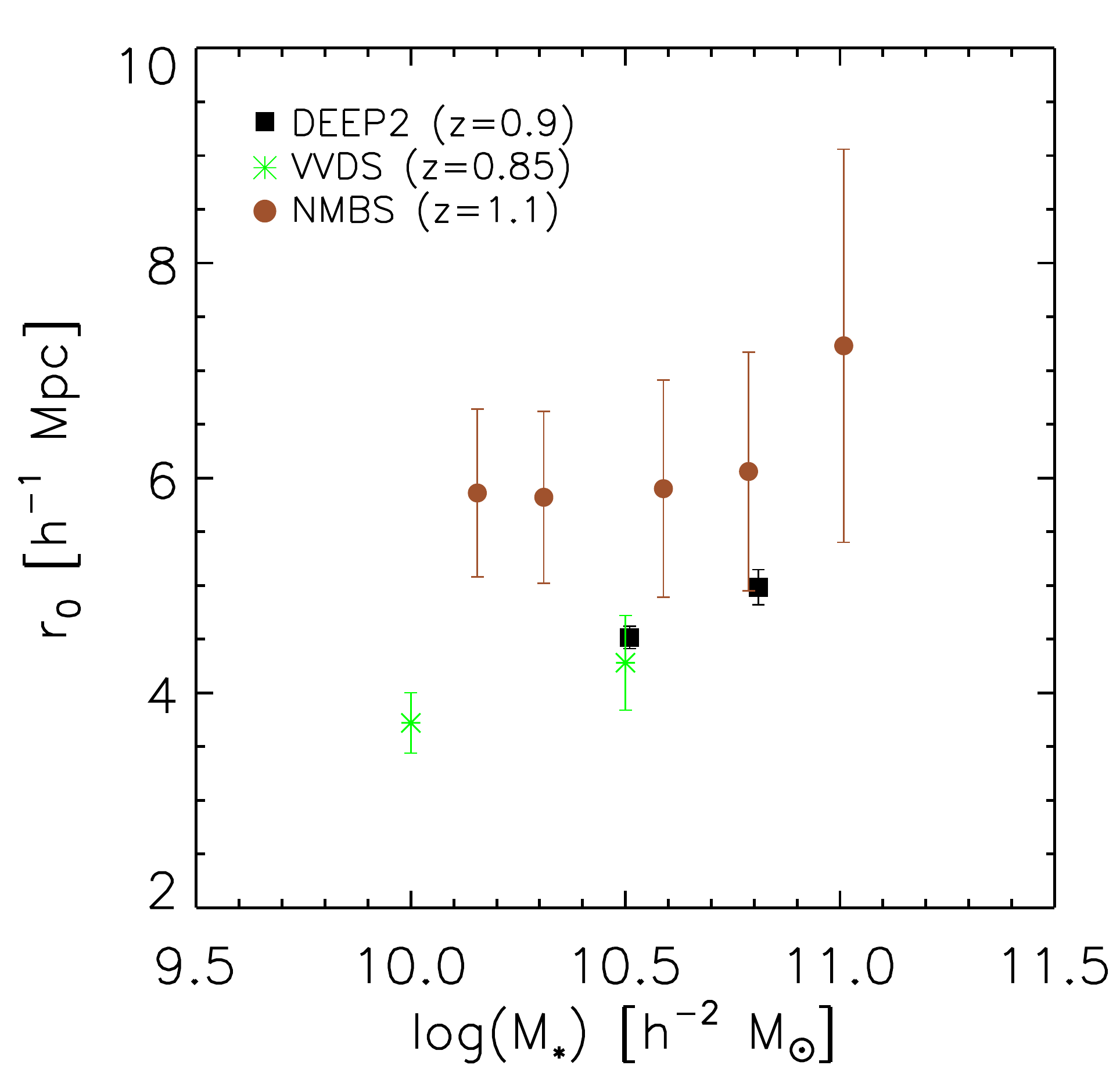}
\includegraphics[width=0.33\textwidth]{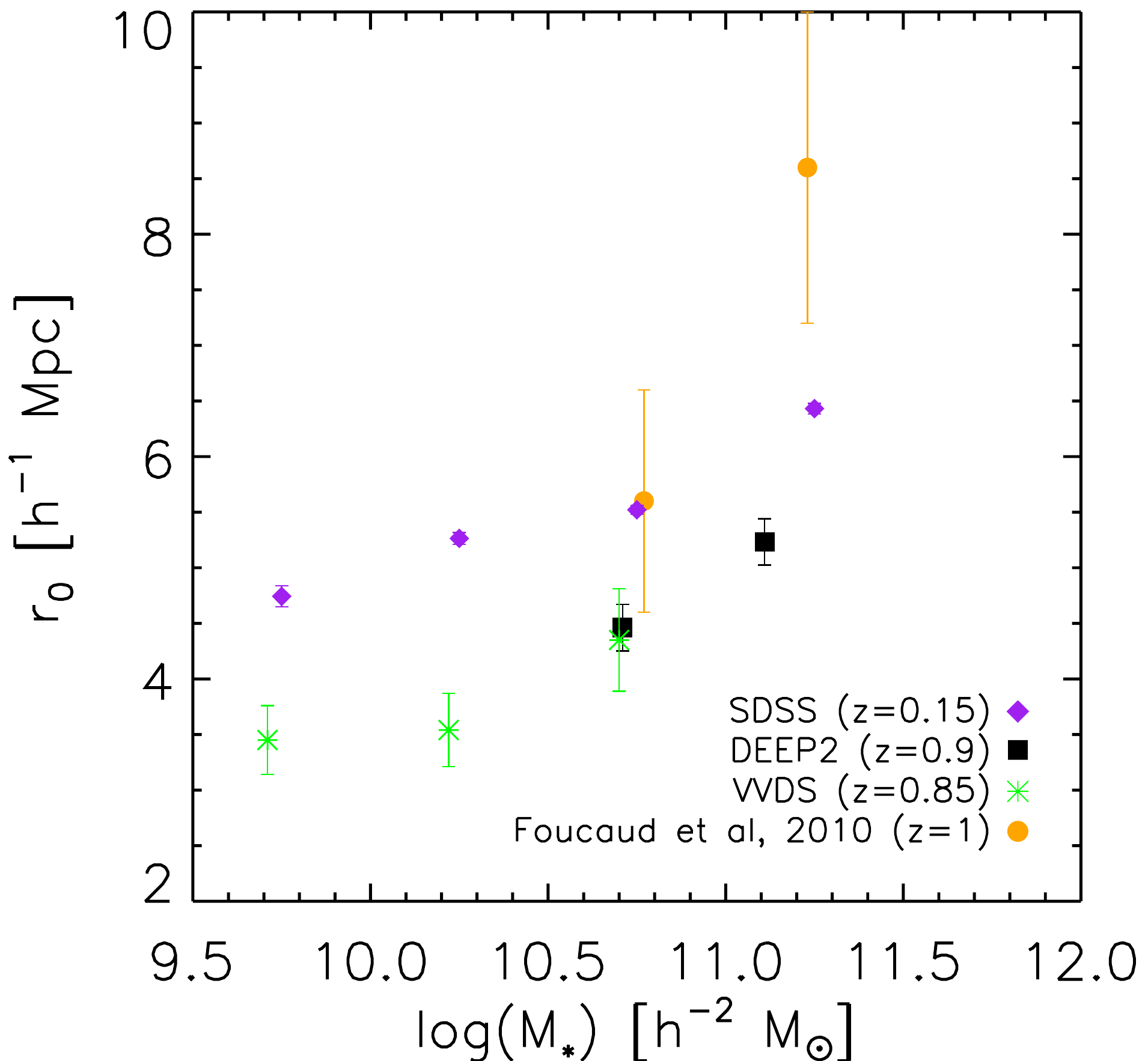}
\includegraphics[width=0.33\textwidth]{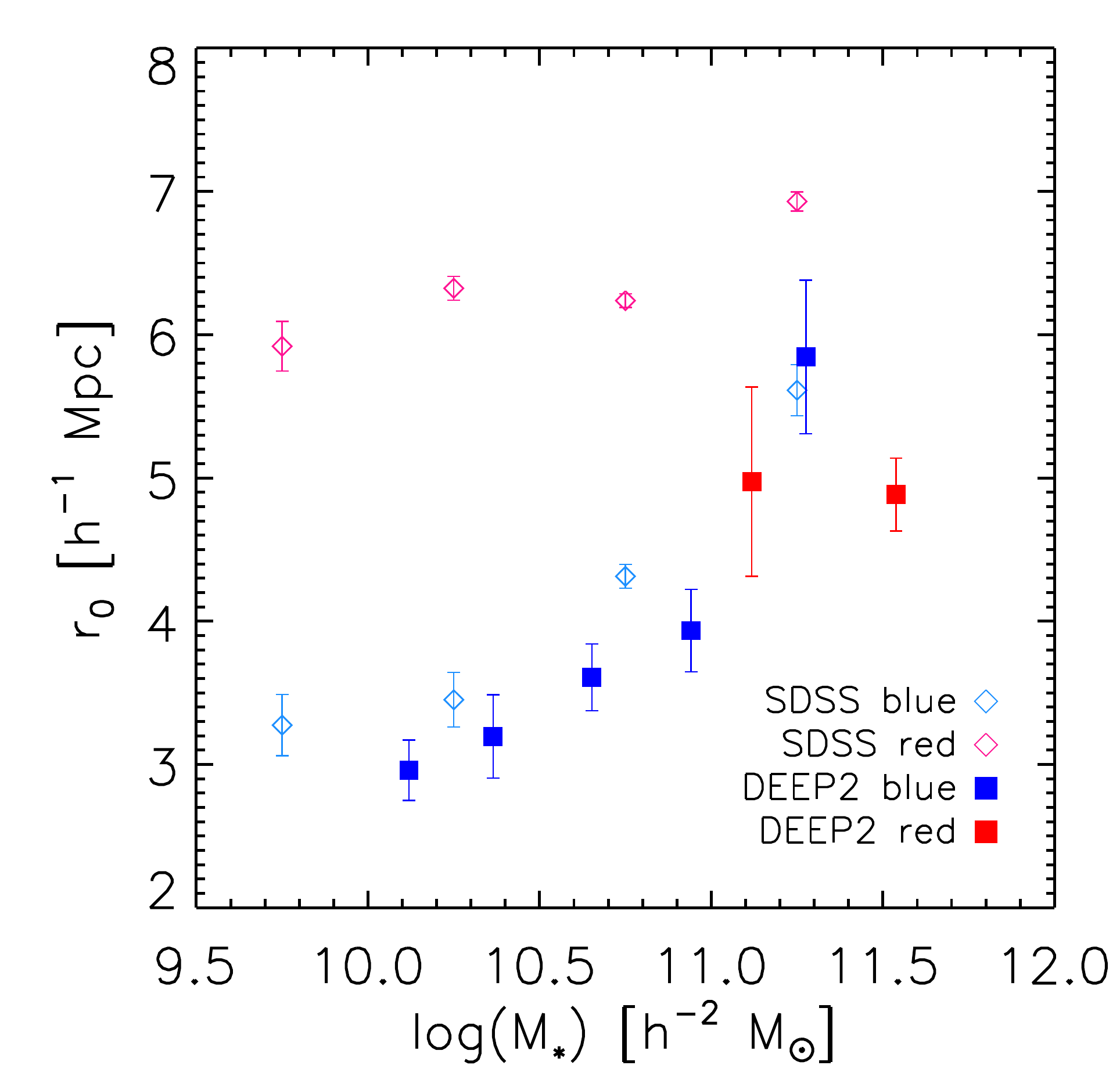}
\caption{(left) Clustering scale length, $r_0$, measured from color-independent stellar 
mass threshold samples in DEEP2 (black square), VVDS data (green star) from \cite{Meneux08}, 
and NMBS data (brown circle) from \cite{Wake11}. Data point locations indicate the lower mass 
threshold limit imposed on each sample. The Wake et al. measurements use NIR photometric 
redshifts to compute the angular correlation function at 0.9$<$\z$<$1.3. 
Both VVDS and DEEP2 have a lower clustering amplitude than the NMBS at log($M_*$)$<$11.0 $h^{-2} M_\sun$,
although the errors from the 0.4 deg$^2$ NMBS survey are relatively large in comparison.  
(center) $r_0$ resulting from power law fits to 
color-independent stellar mass bin samples in SDSS \citep{Li06}, VVDS, 
DEEP2, and \cite{Foucaud10}. Fits are performed over $r_p$=1-10 $h^{-1}$  Mpc in 
DEEP2, \citeauthor{Foucaud10} (orange circle), and SDSS (purple diamond) while VVDS is fit over 
scales of $r_p$=0.1-21 $h^{-1}$  Mpc. All points have similar mass bins sizes of 0.4-0.5 dex, and 
the power law slope $\gamma$ is not fixed in the \wprp~fits. The DEEP2 clustering scale lengths agree well with those from VVDS, both of 
which show lower clustering amplitudes at earlier times for all stellar 
masses probed. The angular clustering measured from  \citeauthor{Foucaud10} is in rough agreement with both early and late epochs, demonstrating
the improved statistical power of the 2PCF with spectroscopic redshifts.
(right) Best fit $r_0$ values for restframe color-separated red and 
blue galaxy stellar mass bin samples in SDSS and DEEP2 for a fixed $\gamma=1.6$ over 
$r_p$=1-10 $h^{-1}$  Mpc. DEEP2 mass bins range in size from 0.3 to 0.4 dex for blue and red galaxies, respectively. 
Blue galaxies in the local universe are somewhat more 
clustered at a given stellar mass than at $z\sim1$, while massive red galaxies 
are more clustered locally.  }
\label{fig:r0masscomp}
\end{figure*}

The projected 2PCFs measured in this study allow us to quantify the relationship 
between galaxy clustering and stellar mass and SFR at $z\sim1$. 
We first discuss how our power law fits to the projected correlation function on 
large scales compare with other studies.  
In the left hand panel of
 Figure~\ref{fig:r0masscomp}, we compare our $r_0$ results for 
color-independent, stellar mass threshold samples to results from NMBS \citep{Wake11} and VVDS \citep{Meneux08} 
at $z\sim1$. 
\citeauthor{Wake11} calculated the angular correlation function of galaxy samples as a function 
of stellar mass using NMBS photometric redshifts in portions of the COSMOS and AEGIS fields 
\citep{Scoville07, Davis07, NMBS}, covering 0.4 deg$^2$ in total. 
For NMBS, the angular correlation function is fit at 0.9$<$\z$<$1.3 over scales of 
$\sim0.05-10 h^{-1}$ Mpc with a fixed $\gamma=1.6$. The DEEP2 results shown also 
have a fixed $\gamma=1.6$ to facilitate comparison and remove degeneracies. \citeauthor{Meneux08}
used spectroscopic redshifts from the VVDS-Deep field \citep{VVDS05} to calculate \wprp~over 0.49 deg$^2$, 
fitting $r_0$ over $r_p$=0.1-21 $h^{-1}$ Mpc. While the VVDS results also allow $\gamma$ to vary in the fits,  
a slightly higher value of $\gamma=1.8$ is typical for their 2PCF data. We find good agreement with the VVDS clustering amplitudes; 
both VVDS and DEEP2 results are 
less clustered than the NMBS results, though the difference is not significant
given the NMBS errors.  
This may be due to cosmic variance within 
the NMBS $z=1.1$ sample, where independent correlation functions in the COSMOS and 
AEGIS samples are quite different on large scales (although they statistically agree 
given the size of the errors, see their Figure 2).  

We also note that the DEEP2 stellar mass threshold
sample with galaxies of all colors above \smass$>$10.5 is somewhat incomplete for red 
galaxies at the low mass end beyond \z$>$0.9. 
As stated in Section~\ref{sec:smasssample}, the red galaxy incompleteness between 
10.4$<$\smass$<$10.8 is estimated to be $\sim20\%$ between 0.74$<$\z$<$1.05.
As the clustering amplitude at a fixed stellar mass appears to be similar 
for red and blue galaxies above \smass$>$10.8, we expect the clustering strength 
for the  \smass$>$10.5 threshold sample to be relatively unaffected by 
the red galaxy incompleteness. 
If we assume that $r_0$ for the color-independent sample behaves as a weighted average 
between the red and blue galaxy samples (given in Table~\ref{tab:smassresults}), 
then increasing the number of red galaxies by 20\% would only increase the clustering 
strength by $\sim1\%$, much less than the quoted error.

We further compare our best fit clustering amplitudes to similar 
measurements in the local universe from SDSS \citep{Li06}. The center panel of Figure~\ref{fig:r0masscomp} 
shows $r_0$ for color-independent, binned stellar mass samples in SDSS\footnote{The SDSS and NMBS stellar masses have
assumed a Kroupa IMF, while DEEP2, \cite{Foucaud10}, and VVDS stellar masses use a Chabrier IMF; the differences in IMF produce
a negligible difference of 0.05 dex in stellar mass for our clustering comparisons.}. As \cite{Li06} did not
measure $r_0$ from their projected 2PCFs, we fit their \wprp~data over $r_p$=1-10 $h^{-1}$ Mpc scales 
in a similar manner to our DEEP2 data. The \cite{Li06} \wprp~data is drawn from a sample of 
$\sim200,000$ SDSS galaxies with an average statistical error of 5\% per measured scale over the fitted scale range and does 
not account for the full covariance between measured scales. Therefore, our power law fits to their 
published data have a small $1\%$ statistical error for each 
stellar mass sample and may not be fully representative of the true error in the SDSS \wprp~data. \cite{Zehavi11} estimate the 
$r_0$ error  to be $\sim$5\% from jackknifed samples of galaxy luminosity bins in SDSS, and therefore 
our conclusions comparing clustering amplitudes should be relatively robust. 
The same plot also shows the three most complete stellar mass samples in VVDS. Because the VVDS-Deep data becomes incomplete
for blue galaxies below \smass$<$10, the measured clustering amplitude is systematically biased to lower
values of $r_0$. At most, \cite{Meneux08} calculated that the (9.5$<$\smass$<$10.0) stellar mass bin underestimates the 
clustering amplitude by 10\% relative to mock catalogs. 
DEEP2 color-independent stellar mass samples are limited to \smass$>$10.5, 
while the VVDS data extends to somewhat lower stellar mass with an 
$I_{\rm{AB}}$=24 magnitude limit. Again, we find excellent agreement in clustering strength between DEEP2 and 
VVDS at $z\sim1$, both of which have significantly lower $r_0$ values than local SDSS 
galaxies for all stellar masses probed. At a fixed stellar mass, we estimate that the clustering amplitude in color-independent samples has
increased by roughly 35\% from $z=1$ to $z=0.1$.  The additional angular clustering data from \cite{Foucaud10} uses $K$-band stellar mass bins
and photo-$z$s at $z$=1 and is in rough statistical agreement with the NMBS mass threshold samples and \emph{both} the low and high redshift data. The measurements
from spectroscopic 2PCFs are clearly advantageous relative to angular clustering measures for equivalent surveys of a few square degrees.

The right panel of Figure~\ref{fig:r0masscomp} shows $r_0$ for stellar mass selected 
samples separated by red and blue restframe colors in both SDSS and DEEP2. On large scales,
we find that blue galaxies are somewhat more clustered ($\approx15\%$) locally when compared
 to $z\sim1$, 
although the $r_0$ values are completely consistent at the 
highest stellar mass, \smass=11.25.  Red galaxies at the same stellar mass, however, 
are much more clustered locally than at $z\sim1$.  We interpret this to mean that 
lower stellar mass blue galaxies have much of their current halo mass in place by 
$z\sim1$, and therefore there is little evolution in the clustering amplitude over the last
8 Gyrs. However, the halos that host massive red galaxies, which are more likely to 
be central galaxies, accumulate fractionally more halo mass relative to their existing stellar mass
over the same span of cosmic time. This effectively causes the clustering amplitude to increase 
at a fixed stellar mass and therefore become more clustered on large scales at recent epochs. 
These results support stellar-halo mass assembly history models where lower stellar mass blue galaxies 
have most of their final halo mass in place at $z\sim1$ but continue to add to their stellar 
mass, while higher stellar mass red galaxies form most of their final halo mass at \z$<$1 but have little evolution in their stellar
mass \citep{Zheng07, Conroy09, Coupon12}.

\begin{figure}[t]
\centering
\includegraphics[width=0.9\columnwidth]{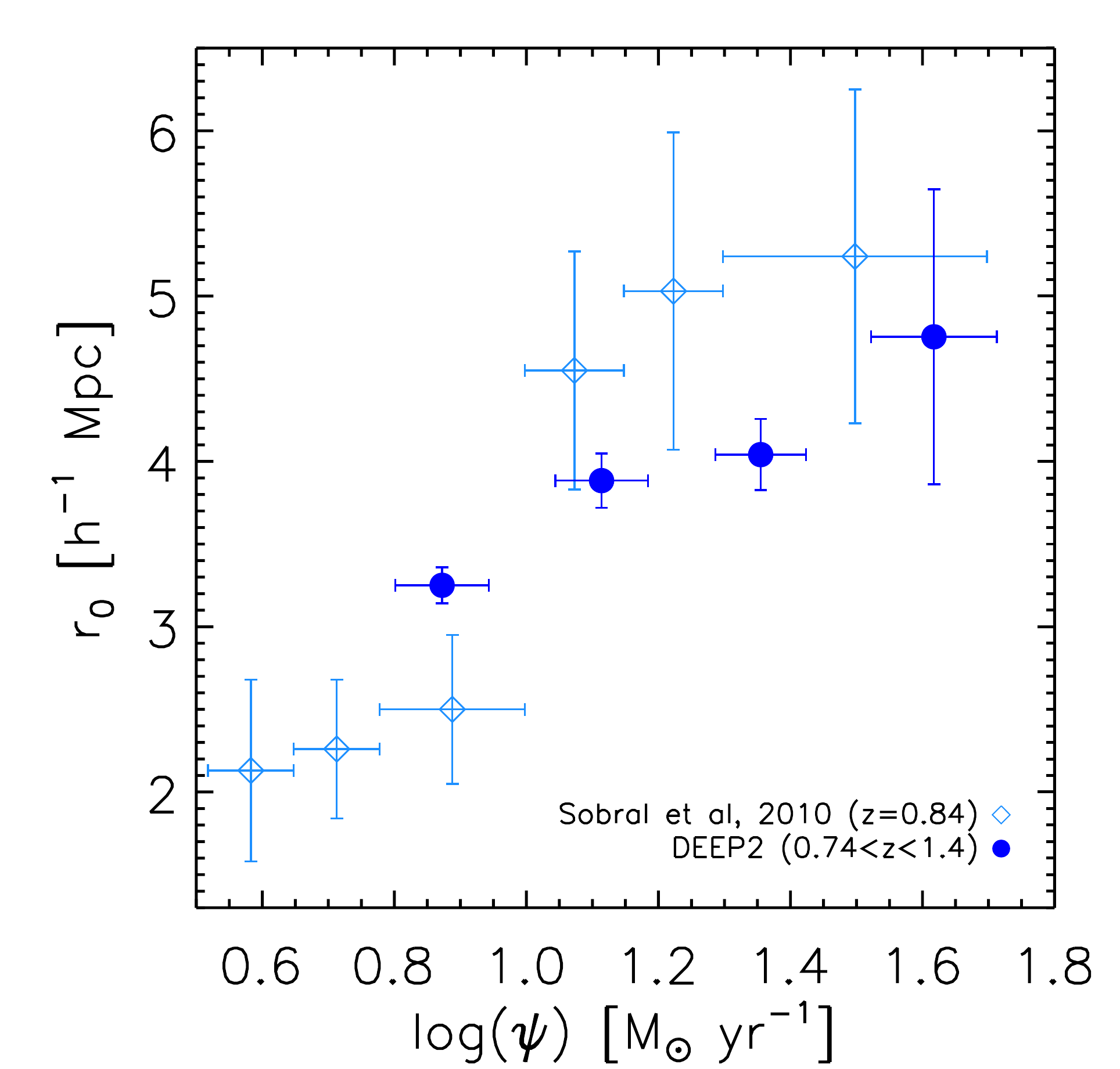}
\caption{The galaxy clustering with respect to SFR for $z$=0.84 H$\alpha$ emitters in the HiZELS survey (diamonds, \cite{Sobral10}) 
and the DEEP2 0.74$<$$z$$<$1.4 blue galaxy sample binned by SFR (filled circles). The results are in good agreement given the measurement errors
and different SFR calibration methods between the surveys.} %above \sfr$>$1.0 Mpc yr$^{-1}$, corresponding
%to the bright end of the H$\alpha$ luminosity function. We do not observe the large drop in clustering strength below $L^*_{\rm{H}\alpha}$ in the HiZELS data; the discrepancy is
 %likely due to the differences in SFR calibration methods. }
\label{fig:SobralSFR}
\end{figure}

Our clustering measurements also agree with the blue and red 
luminosity-dependent clustering results of C08, who 
found that red galaxies are more clustered than blue galaxies at $z=1$ at the 
same luminosity (which corresponds to a lower stellar mass for blue galaxies compared
to red). The angular clustering measurements for passive and star-forming galaxies presented in \cite{Hartley10} also broadly 
agree with C08 and our results for red and blue galaxies at z$\sim1$, albeit with larger errors due to the use of photometric redshifts. 
Both C08 and \citeauthor{Hartley10} find that
blue, star-forming galaxies are less clustered than their red, passive counterparts at the same luminosity at lower redshift
($r_{0}\approx5~h^{-1}$ Mpc at $z\sim1$ versus 
$r_{0}\approx6~h^{-1}$ Mpc locally, see Figure 11 in C08). 
Further, the results in the lower red galaxy stellar 
mass bin are consistent with the results found in C08, given the errors.
We find a slightly larger clustering difference between
$z\sim1$ red galaxies selected by stellar mass and those at $z\sim0.1$, using fits to 
the \citeauthor{Li06} \wprp~data. 

We note that our highest
stellar mass red galaxy sample is a \emph{threshold} sample limited by the total probed volume in DEEP2 volume. This upper stellar mass limit 
will not include the rarest, most massive galaxies
that would be present in the larger SDSS volume within an equivalent stellar mass range. 
%However, such rare galaxies should 
%contribute trivially to the clustering of the sample as a whole. 
Another concern is that less massive star-forming galaxies that have been
reddened by dust could be included in red galaxy samples, which may 
cause $r_0$ to be underestimated. 
Results from the PRIMUS survey \citep{Zhu11} show that such heavily reddened 
star-forming galaxies are rare at the bright end of the red sequence ($\leq10$\%) at intermediate redshift and therefore should have relatively 
little weight in the large-scale 2PCF amplitude.

\begin{figure}[t]
\centering
\includegraphics[width=0.95\columnwidth]{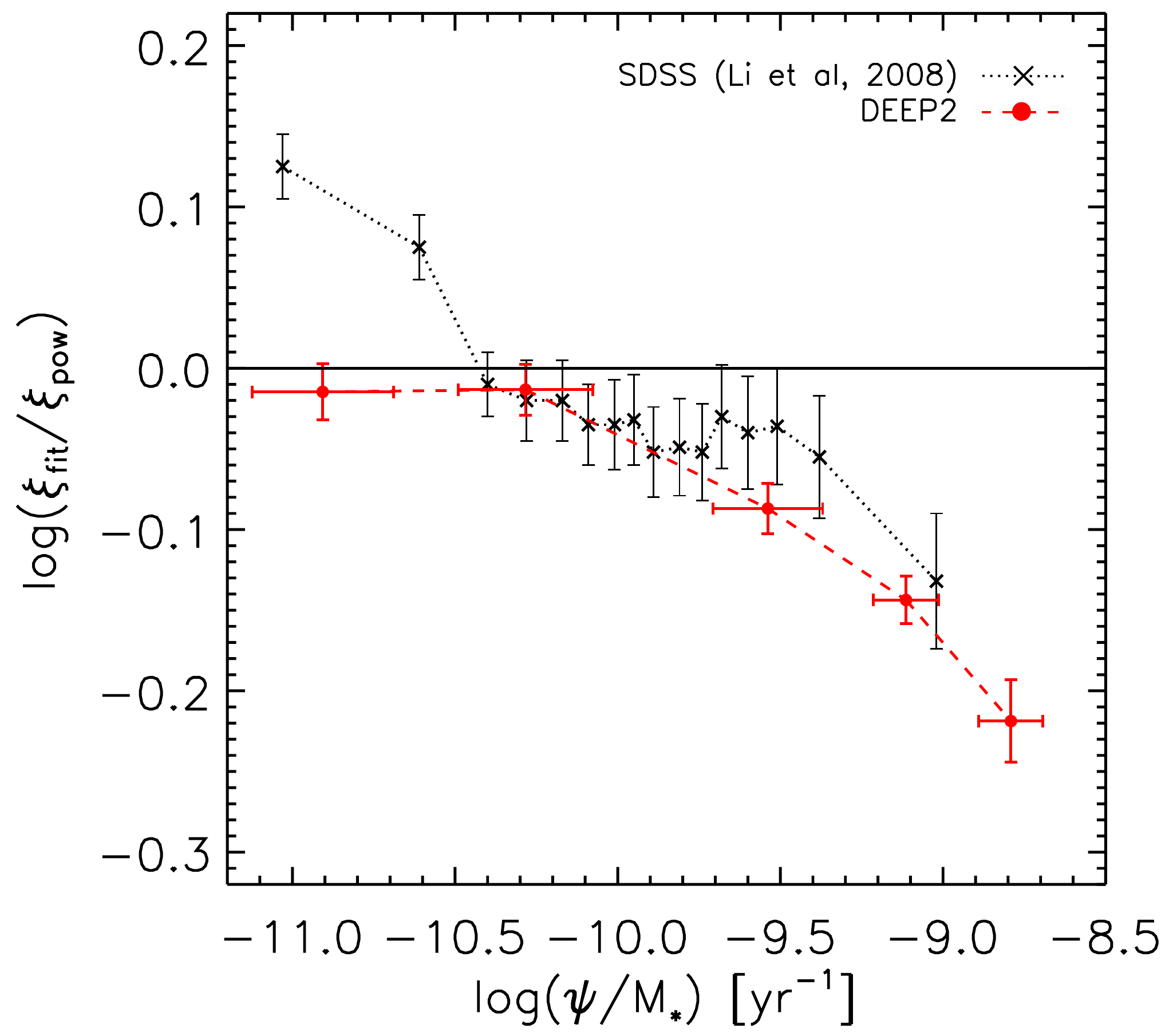}
\caption{The best fit power law for \corf~measured from the DEEP2 sSFR samples on 1-10 $h^{-1}$ 
Mpc scales (red circles). The real-space 2PCF is  normalized to a power law of 
$\xi(r)_{\rm{pow}}=(r / 5 h^{-1}$ Mpc)$^{-1.8}$ at a fixed scale of $r=5 h^{-1}$ Mpc to match the 
average scale for the DEEP2 data. \cite{Li08} produced similar measurements in SDSS at equivalent scales (black crosses). 
The relative trend in sSFR is similar between $z\sim1$ and local redshifts for \ssfr$>$$-10.5$; lower clustering amplitudes are observed at progressively 
higher sSFR values. For \ssfr$<$$-10.5$, the $z\sim1$ galaxies are less clustered than their local counterparts at fixed 
sSFR. }
\label{fig:ssfrli08}
\end{figure}

There are relatively few studies that have explored galaxy clustering at $z\sim1$ as a function of 
stellar mass, SFR, and sSFR to which we can compare directly. At a mean redshift of $z$=0.8, the HiZELS survey 
measured the angular clustering of a relatively small sample of H$\alpha$ emitters with respect to $K$-band magnitude and H$\alpha$ luminosity,
known tracers of stellar mass and SFR \citep{Sobral10}. Their results show similar trends as our $z\sim1$ blue galaxy samples, 
with a clustering strength ranging from $r_0$=2-5 $h^{-1}$ Mpc and a general trend for increased clustering with brighter $K$-band magnitude and H$\alpha$ luminosity.
Figure~\ref{fig:SobralSFR} shows the \cite{Sobral10} clustering with H$\alpha$ luminosity after converting to SFR \citep{Kennicutt98} and the results
from our 0.74$<$$z$$<$1.4 blue galaxy sample binned by SFR. We find good agreement with the HiZELS measurements given the errors, but we
do not observe a large drop to lower clustering at \sfr$<$1 corresponding to the break in H$\alpha$ luminosity function at $L^*_{\rm{H}\alpha}$. 
The disagreement may arise from the fact that our SFRs are calibrated by restframe magnitudes and colors of the bulk galaxy population at $z\sim1$, which may 
smooth such a rapid transition in SFR within the blue galaxy population. \cite{Sobral10} also found the clustering increased with increasing SFRs at a fixed $K$-band
magnitude, which is similar to our sSFR measurements in blue galaxies and will be further discussed in Section~\ref{sec:sfrmass}.

More recently, \cite{Lin12} performed an angular clustering analysis of $BzK$-selected 
star-forming galaxies at $z\sim2$. They also found that the measured clustering scale length increases strongly with increasing stellar mass
(\smass$>$9),  increasing SFR (\sfr$>$0.5), and decreasing sSFR (\ssfr$<$$-8.6$), in good agreement with
the trends found here. However, because we have restricted our blue galaxy sSFR samples to $M_B$$<$$-20.5$ 
and \ssfr$<$$-8.6$ for completeness to \z$<$1.4, we do not probe higher values of sSFR beyond \z$>$1 in this study.  
We therefore cannot confirm the existence of a turnover in the clustering behavior at
higher values of sSFR as presented in \cite{Lin12}.

We can also compare our results as a function
of sSFR to those of \cite{Li08}, who measured \wprp~for a large sample of SDSS 
galaxies with sSFRs ranging from  $-11$$<$\ssfr$<$$-9$. To facilitate 
this comparison, we follow \citeauthor{Li08} and compute the best fit \corf \ power law from 
10 jackknifed samples of our \wprp~data as a function of sSFR and normalize to a constant power law of 
$\xi(r)_{\rm{pow}}=(r / 5 \ h^{-1}$ Mpc)$^{-1.8}$. As we fit \wprp~over scales 1-10 $h^{-1}$ Mpc in each jackknife sample, 
we evaluate the power law $\xi(r)_{\rm{fit}}$ for the unweighted mean values of $r_0$ and $\gamma$ at a fixed scale of 5 $h^{-1}$ Mpc and plot 
the ratio of the 2PCFs as a function of sSFR (see Figure~\ref{fig:ssfrli08}). 
Compared to \citet{Li08} (also at $r_p=5 \ h^{-1}$ Mpc), we find good qualitative 
agreement with the trend seen for local clustering results. At \ssfr$>$-9.5, there 
is a significant trend to lower clustering amplitudes at higher sSFR, 
while the large-scale clustering amplitude is relatively independent of sSFR 
at \ssfr$<$$-9.5$.  At \ssfr$<$$-10.5$ we do not see a rise in the large-scale clustering 
amplitude, as is seen locally.  The trend of increased clustering at lower sSFRs should 
be expected given the C08 results, as 
sSFR is highly correlated with restframe color (see right 
panel of Figure~\ref{fig:colormasscontour}) and C08 found an increased clustering
amplitude with redder colors. 

Here again, as with the highest stellar mass red sample (\smass$>$11.0), 
the sample with the lowest sSFR (\ssfr$<$$-10.6$) is a threshold sample which is
limited by the DEEP2 survey volume. 
As the clustering signal from the rarest galaxies cannot be
measured from DEEP2 data, the clustering amplitude in this sample may be underestimated relative 
to larger survey volumes such as SDSS.  
The effect is expected to be minimal.
%The effect of missing these few galaxies is expected
%to minimally impact the pair-weighted clustering signal above the sSFR threshold.

\subsection{Clustering Amplitude and the SFR-$M_{*}$ Relationship}
\label{sec:sfrmass} %6.2

While galaxy samples selected by SFR probe a wide range of stellar mass, 
stellar mass and SFR are known to be correlated \citep[e.g.,][]{Noeske07a}
such that the mean stellar mass increases with increasing SFR. 
In Section~\ref{sec:smass}, we showed that the clustering amplitude $r_0$ increases 
monotonically with stellar mass, as
the amount of baryons processed into stars correlates with the halo mass. Therefore, 
it is possible that the change in mean stellar mass for our blue SFR samples is 
responsible for the observed increase in clustering amplitude with increasing SFR.   
Similar investigations in environmental studies \citep{Peng10, Sobral11} 
have found little to no difference between the 
SFR-density and stellar mass-density relationships, indicating that the SFR-density relation 
could be entirely due to differences in stellar mass.  

\begin{figure}[t]
\centering
\includegraphics[width=0.98\columnwidth]{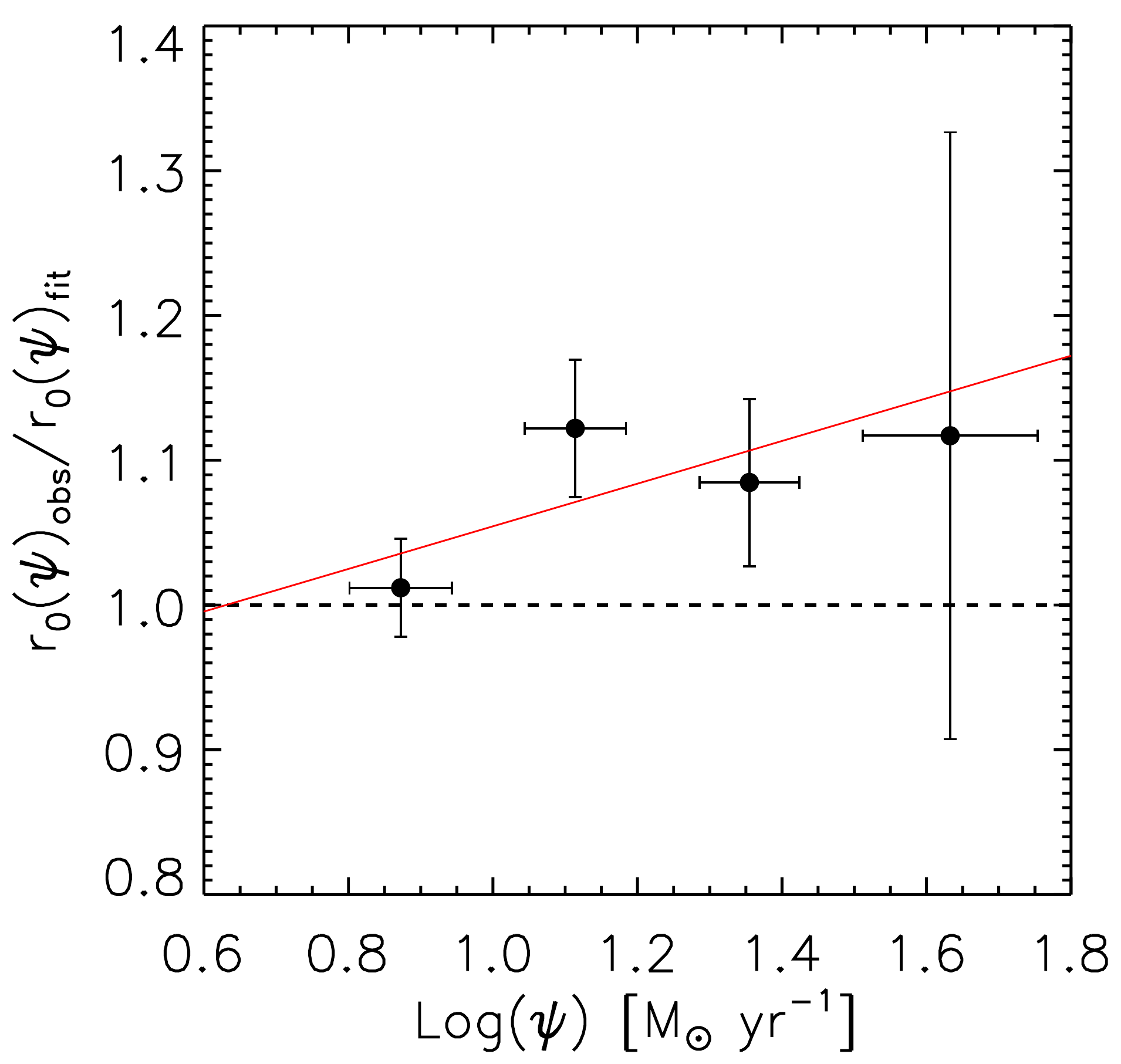}
\caption{The measured clustering scale length ($r_{0}(\psi)_{\rm{obs}}$) for 
blue galaxy SFR samples normalized to the scale length predicted by the correlation 
between $r_0$ and stellar mass. The predicted
scale length ($r_{0}(\psi)_{\rm{fit}}$) is generated from an exponential fit to the blue galaxy
stellar mass samples (Figure~\ref{fig:r0results}) and weighted by the stellar mass distribution of each SFR-selected sample.
The weighted mean over all SFRs is $r_{0}(\psi)_{\rm{obs}}$/$r_{0}(\psi)_{\rm{fit}}$=1.057$\pm$0.025, and 
a weighted linear fit (red, solid line) shows a trend for larger deviation from the stellar mass-predicted $r_0$ at higher SFR. 
}
\label{fig:sfrdetrend}
\end{figure}

To address this question, we generate a prediction for $r_0$ based on 
the measured clustering with stellar mass using a fixed slope 
of $\gamma$=1.6 (see left panel of Figure~\ref{fig:r0results}).  We fit an exponential relation between 
$r_0$ and \smass \ for binned blue galaxy stellar mass samples, finding a best fit relation
of 
\begin{equation}
r_{0}(M_*)=e^{2.57\rm{log}(M_*/M_\sun)-27.1}+2.88. 
\end{equation}
We then calculate a prediction for $r_0$ as a function of SFR by weighting the fit relation with the 
stellar mass distribution of each SFR-selected sample.  Because $\xi(r)$ is actually a pair-weighted statistic, this 
method is only an approximation to the $r_0$ that would be measured from a given stellar mass distribution.
The ratio of the measured $r_0$ to the predicted $r_0$ as a function of SFR is shown in Figure~\ref{fig:sfrdetrend}.  If the 
large-scale clustering amplitude observed can be predicted given the stellar mass distribution of each 
SFR sample and the observed relation between $r_0$ and stellar mass, 
$r_{0}(\psi)_{\rm{obs}}$/$r_{0}(\psi)_{\rm{fit}}$ would equal one.  
We find that the mean $r_{0}(\psi)_{\rm{obs}}$/$r_{0}(\psi)_{\rm{fit}}$ over all 
SFR is 1.057$\pm$0.025; the deviation
from unity has a $p$-value of $p=0.076$ and therefore is significant at the $p<0.1$ level.  We also find that 
there is a trend towards larger deviations from the mass-predicted $r_0$ at higher SFRs, suggesting that
most, but not all, of the SFR - $r_0$ relationship can be explained by the relationship between SFR and stellar mass.

%Because stellar mass is positively correlated with increasing SFR, 

\begin{figure}[t]
\centering
\includegraphics[width=1.0\columnwidth]{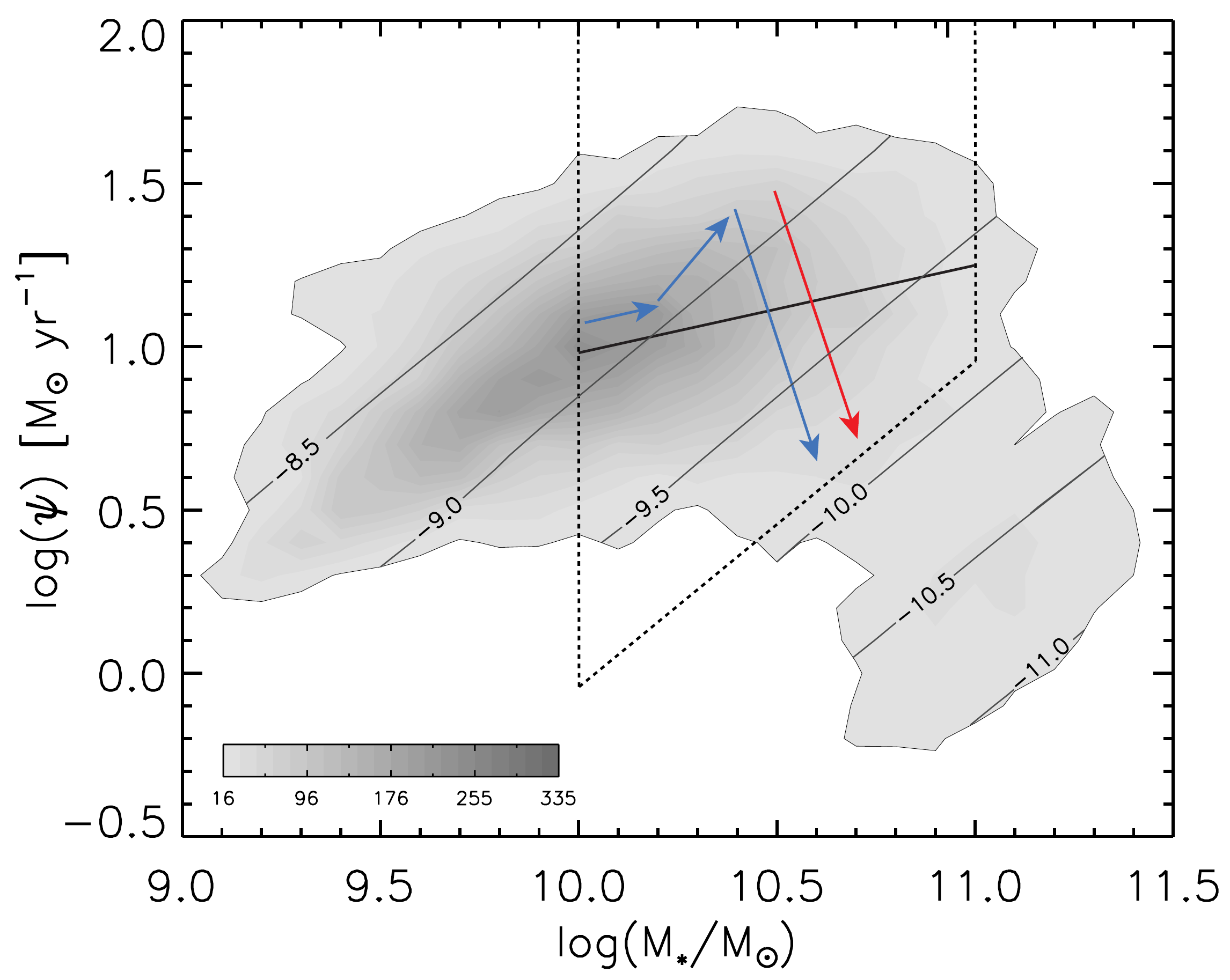}
\caption{The dependence of SFR on stellar mass, overlaid with contours of constant sSFR.  
Dashed lines show the selection of star-forming galaxies with \z$<$1.05
used to measure the clustering of galaxies above and below the star-forming
``main sequence''.  The black solid line shows a 
linear fit to the SFR-stellar mass relation in this sample. 
Colored lines show two possible evolution scenarios for star-forming galaxies. 
The major merger
scenario (blue arrows) predicts that galaxies evolve along the star-forming 
main sequence and undergo a merger event that initially boosts 
the SFR, followed quickly by a period of star formation quenching 
and progression onto the red sequence. The secular evolution scenario (red arrows) assumes 
that galaxies initially lie above the 
main sequence, during a period of intrinsically rapid star formation 
and are eventually quenched through internal or secular 
processes, moving through the main sequence in the process.
We find that galaxies above the main sequence have a lower clustering amplitude than
galaxies below the main sequence at a given stellar mass.  As the clustering of individual 
galaxies increases with time, our measurements favor the secular evolution scenario, as the galaxy 
population above the main sequence can not be dominated by galaxies 
that previously had a lower SFR at a given stellar mass, which are more clustered.}
\label{fig:sfrmasscont}
\end{figure}

While most of the clustering amplitude as a function of SFR may be explained 
by the $r_0$-$M_*$ relation, it is interesting to investigate
the SFR-dependent clustering behavior where the stellar mass distributions of 
SFR-selected samples are equal. In Figure~\ref{fig:sfrmasscont}, we plot SFR versus stellar mass for our parent DEEP2 sample with 
contours showing lines of constant \ssfr.  As we find that $r_0$ increases with decreasing
sSFR (see Figure~\ref{fig:r0results}), the clustering amplitude increases from the upper left 
(high sSFR) to the lower right (low sSFR) in this space.  To test whether high SFR galaxies are more 
clustered than low SFR galaxies with similar stellar masses, we construct a galaxy sample limited in stellar 
mass to 10$<$\smass$<$11 and sSFR to \ssfr$<$$-9.9$, which is complete in the volume between 0.74$<$\z$<$1.05. 
This sample selection is shown with a dotted line in Figure~\ref{fig:sfrmasscont}.  We then perform a linear fit to the 
SFR-$M_*$ dependence within this selected sample, weighting by the average errors of 0.3 dex 
in $M_*$ and 0.25 dex in SFR. The resulting fit (shown with a solid black line) roughly corresponds 
to the location of the star-forming ``main sequence'' 
\citep[(MS) e.g.,][]{Noeske07a}
for the selected sample.  
Separating the galaxy sample into populations above and below the MS, we find 
that the mean stellar mass is \smass=10.36 for both populations, while the average SFRs are 
\sfr=1.24 $M_{\sun}$ yr$^{-1}$ and \sfr=0.84 $M_{\sun}$ yr$^{-1}$, respectively. 
Comparing their clustering properties, we find galaxies above the 
MS have a clustering amplitude of $r_0$=3.73$\pm$0.18, while galaxies 
below the MS have $r_0$=4.36$\pm$0.21 and are therefore more clustered 
at a given stellar mass.  This confirms that
the $r_0$-sSFR results found above are likely not driven by stellar mass differences between the samples.

While the stellar mass and SFR values used in this study are not necessarily 
highly accurate on an individual galaxy basis, they do represent the global 
average well.  Any scatter in these values would wash out the observed 
SFR-$M_*$ relation and clustering correlations in this space. 
In particular, because
the highest SFR galaxies have the bluest restframe colors at a given $K$-band magnitude, it is possible that
our color-based stellar masses underestimate the true stellar mass as measured in the IR (Weiner et al., in preparation).
We have compared our color-M/L stellar masses to galaxies with measured $K$-band stellar masses both above and 
below the MS and find no statistically significant bias in the mass distributions.

The clustering dependence with SFR at a given stellar mass 
is particularly interesting in light of the differences in galaxy properties 
observed for galaxies above and below the MS.  In general, galaxies above the MS are 
found to have higher star-forming surface densities, smaller sizes, more dust attenuation, and 
somewhat higher Sersic indices than galaxies on or below 
 the MS \citep{Schiminovich07,Elbaz11, Wuyts11}.  In particular, both \citeauthor{Elbaz11} and \citeauthor{Wuyts11} interpret 
 their results to support major mergers as a dominant mechanism of star formation quenching, 
 where galaxies on the MS experience a merger event or some other instability, which leads to 
a central 
 starburst and bulge formation, and eventually results in a quiescent elliptical galaxy.  In this picture, 
 an individual galaxy would move from first being on the MS to being above the MS during the merger stage 
 and then eventually move below the MS as the galaxy becomes quiescent (blue arrows in Figure~\ref{fig:sfrmasscont}).  

However, the clustering results presented here do not support this general picture, as galaxies above 
the MS are \emph{less} clustered than those on or below the MS.  Therefore the majority of the galaxy population 
above the MS (i.e. with a higher sSFR) can not have previously been on the MS, as the clustering 
amplitude can not decrease with time.  Our clustering results instead support a picture 
in which most galaxies initially lie above the MS, where they have higher SFRs at a given 
stellar mass due to internal (non-interacting) processes, 
and undergo secular evolution down onto and eventually below the 
MS as their star formation is quenched.  While major merger events can certainly occur, their contribution to 
the overall clustering trends in field galaxies is likely a small effect. The lack of a strong clustering signal from major mergers above the MS
is also supported by recent evidence that mergers play a subdominant role in the build up of the stellar mass function at \z$<$1 \citep{Conselice13, Moustakas13}.
 The clustering properties as a function of 
restframe optical color in both SDSS \citep{Zehavi11} and DEEP2 (C08) also support this scenario, as restframe optical color is well correlated with 
sSFR at $z\sim0.1$ and $z\sim1$, and both studies find that within the star-forming population, bluer galaxies 
are less clustered than redder galaxies.  

The general evolutionary picture is one in which galaxies move from the upper left in the SFR-stellar mass plane down 
to the lower right to match the observed clustering measurements.  However, in addition to the clustering amplitude 
and therefore halo mass increasing below the MS, the satellite fraction may change as well, which could also 
influence the sSFR.  In particular, galaxies above the MS may have a higher satellite fraction, while galaxies below the 
MS may be more likely to be central galaxies in their parent dark matter halos. Therefore while secular evolution is a 
logical explanation for the clustering trend observed across the MS, a full HOD analysis is required to assess the 
importance of the changing satellite fraction above and below the MS and to more robustly determine the processes 
driving the observed change in sSFR.

\subsection{Comparison with Environment Studies}
\label{sec:environ} %6.3

An alternate way to study the dependence of galaxy properties on large-scale structure
is through overdensity measurements of a galaxy sample relative to the mean density of
galaxies at the same redshift.  Such studies measure how a particular galaxy property 
is correlated with environment and are inherently limited to probe physical scales larger than the 
density-defining population. At \z$\sim$1, a limiting scale of
 $\gtrsim1~h^{-1}$ Mpc is common. 
Several environment studies at \z$\sim$1 have found a strong correlation between stellar 
mass and density, in that galaxies in higher density environments tend to have 
higher stellar mass \citep{Kauffmann04, Cucciati06, Cooper06, 
Peng10,Sobral11}. Our results are in broad agreement with these studies, as we find higher clustering
amplitudes within increasing stellar mass between 9.6$<$\smass$<$11.0, above which 
the clustering amplitude remains constant. 
 
There has been some disagreement in the literature, however, as to whether the 
color-density relationship evolves as a function of redshift. Several previous 
studies of galaxy environments at $z\sim1$ have failed to detect any statistically significant 
relationship between color and density, in contrast to studies at lower redshifts 
%have shown that the mean 
%color-density relationship of $z\sim1$ samples does not seem to change significantly 
%from the local relation for a fixed stellar mass 
\citep{Cucciati06, Scodeggio09, Grutz11}. 
On the other hand, \cite{Cooper10} probed the most massive DEEP2 galaxies and found a
statistically-significant correlation between color and galaxy overdensity, with high mass red galaxies 
found preferentially in the most dense environments at $z\sim1$.  \citeauthor{Cooper10} emphasizes that
while no significant trend in the color-density relation may be \emph{detected} in several environment studies, 
it is possible that a color-density correlation may simply be missed due to larger statistical 
errors or observational systematics at 
higher redshifts (e.g due to low sample completeness in VVDS at \z$>$1).
In our analysis, we do not detect a difference between the clustering of red and blue galaxies at an 
equivalent stellar mass.  However, our samples are defined to probe the bulk color-density
relationship at $z\sim1$, and we do not probe the most over- or under-dense regions 
specifically, which was a primary focus of \cite{Cooper10}.  Our results are consistent with \cite{Cooper10} in that the 
color-density trend for high-mass galaxies at \z$\sim$1 is weaker than the stellar mass-density trend;
 the typical densities of red and blue galaxies at a given stellar mass are not significantly 
different.%, while the extreme ends of the population may differ, in that the most and least 
%clustered galaxies have a slightly different mean color.  The effect must be small, 
%however.

Other environment studies have measured the SFR-density relationship as a function of 
redshift and stellar mass.  Locally, this relation is monotonic, with lower-SFR galaxies 
found in higher density environments \cite{Gomez03}. At earlier times, 
the SFR-density relationship was quite different; dense environments are associated 
with both low and high SFRs (\citealt{Elbaz07}, CP08, \citealt{Sobral11}). Our SFR-clustering results at 
$z\sim1$ are in broad agreement with these studies, with the highest-SFR blue galaxies
having roughly the same clustering strength as the lowest-SFR red galaxies. 
The similar environments found for galaxy populations with vastly different SFRs 
(\sfr$<$ 0.5 $M_\sun$ yr$^{-1}$ and \sfr$>$1.5 $M_\sun$ yr$^{-1}$) indicate that the 
processes leading to the large-scale clustering and quenching of star formation in massive galaxies are 
primarily independent. One possible scenario is that rapidly star-forming blue galaxies at these redshifts are 
progenitors of the quiescent red galaxies found at later epochs \citep[][C08]{Cooper06}. 
The preponderance of quiescent galaxies found in high density regions at 
$z\sim1$ and 
the relative paucity of high SFR blue galaxies at low redshifts 
gives the appearance that the large-scale environment and quenching are related.

Mechanisms proposed to explain the evolution of the observed 
SFR-density relation involve the quenching of star formation in a variety of 
isolated and merging systems \citep{Sobral11, Peng10} and include 
effects such as virial heating of cold gas needed for star formation and the
rate of gas exhaustion. However, as pointed out in the previous section,
it seems likely that quenching specifically from major mergers plays a sub-dominant
role at $z\sim1$; the clustering of star-forming galaxies at a fixed stellar mass increases monotonically
with decreasing SFR, not vice-versa. Our results support the theory that a simple 
 sSFR-$M_*$ relation, such as proposed ``mass-quenching" models where the quenching
 rate for galaxies at or above $M^*$ is proportional to their stellar mass \citep{Noeske07b, Peng10}, 
 should reproduce the gross relationship between star-formation, stellar mass, and
clustering on large scales. However, we caution that not 
\emph{all} quenching in star-forming galaxies must follow this form, 
as our samples probe only the average clustering properties of the $z\sim1$ galaxy distribution.

\subsection{Small-scale Clustering Behavior}
\label{sec:sscale} %6.4

We now turn to our clustering results on small scales, below $r_p\sim$1 $h^{-1}$ Mpc. 
In Section~\ref{sec:smass}, we found that \corfp~ for 
blue galaxies selected by stellar mass exhibits stronger ``Fingers of God" on small
scales at higher stellar mass (see Figure~\ref{fig:massxisp}), indicating that 
higher-stellar mass blue galaxies reside in halos with higher velocity dispersions 
and therefore likely reside in higher mass halos than lower-stellar mass blue galaxies. 
Such an obvious increase in the ``Fingers of God" within \corfp~is not seen for 
blue galaxy samples selected by SFR or sSFR, suggesting that these properties do not correlate with the dark matter halo mass 
as directly as stellar mass does.  Although we do not show them here, we find that all red galaxy samples 
exhibit strong ``Fingers of God", independent of the stellar mass or SFR and similar to the 
red luminosity samples in C08 (see their Figure 8).  

In order to reveal any relative change in the small-scale clustering amplitude, we divide \wprp~on all scales 
by the best fit power law on scales 1-10 $h^{-1}$ Mpc and look for excess clustering  below $r_{p}$$\leq$0.3 $h^{-1}$ Mpc relative to this best fit power law.
In HOD models, the slope of \wprp~on these small scales, where the one-halo term 
dominates the correlation function, is governed by the relative number 
of central and satellite galaxy pairs within an individual halo, as 
well as the location of the probed halo mass scale
on the halo mass function \citep{Berlind02, Zehavi05, Zheng05, Skibba09}. 
For massive galaxy samples with a minimum halo mass corresponding to $\gtrsim10^{12.5}~
h^{-1} M_{\sun}$ at $z=1$, 
the \wprp~amplitude rapidly increases on small scales due to a greater contribution 
from central-satellite and satellite-satellite pairs (\cite{Sheth02, Tinker08};  see also Appendix A of \cite{Zheng09} for further details). 
Such massive halos are near the exponential tail of the halo mass function, and higher-mass halos 
are increasingly rare in the volume surveyed by DEEP2.  

For our galaxy samples defined by stellar mass, we find no 
significant increase in clustering on small scales for 
any of the blue galaxy samples. However, both mass-selected red galaxy samples exhibit 
enhanced clustering on scales below $r_{p}$$\leq$0.3 $h^{-1}$ Mpc ($p=0.005$ for 10.5$<$\smass$<$11.0 
and $p=0.04$ for \smass$>$11.0).
The enhanced small-scale clustering observed in these massive red galaxies 
may reflect a higher prevalence of central galaxies relative to satellite galaxies in the population 
and the higher minimum halo mass ($\sim10^{12.5} h^{-1} M_{\sun}$), 
which contributes more correlated power in the one-halo term as discussed above.
The lack of a detectable small-scale rise for the blue galaxy sample at an 
equivalent stellar mass may be due in part to larger measurement errors in the highest mass sample
 (see the right panel of Figure~\ref{fig:relmasscorrcolor}) or 
a larger contribution from satellite-satellite pairs than central-satellite pairs, which smoothes
the transition in \wprp~between the one-halo to two-halo terms.
The small-scale clustering in color-independent stellar mass samples 
is driven 
primarily by the fraction of red to blue galaxies at a fixed stellar mass.
Accordingly, we observe that the color-independent 
samples exhibit a similar small-scale rise at high stellar mass due to the 
high fraction of red galaxies present in the sample. 

When binned by SFR, we find that red galaxies with low SFRs do not 
show a rise on small scales, while blue galaxies do exhibit a 7-fold increase in 
small-scale clustering amplitude over the best-fit large-scale power law 
 below $r_{p}$$\leq$0.3 $h^{-1}$ Mpc. While the \wprp \ error in the highest-SFR blue galaxy sample
 (\sfr$>$1.5 $M_{\sun}$ yr$^{-1}$) is the largest of all measured blue galaxy samples, 
 the deviation from the large-scale power law has a $p$ value of $p=0.038$ and is 
 statistically significant at the $p<0.05$ level.
One possible explanation for the correlation between high SFR and increased small-scale clustering amplitude is that
star formation is being triggered by galaxy interactions, either through merger events or tidal
 interactions \citep[e.g.][]{Barton07}. Late-type galaxies in the 
blue cloud contain large gas 
reservoirs which can support enhanced levels of star formation triggered though 
close interactions of galaxy pairs. Enhanced IR luminosity, 
which is tightly correlated with star formation, has been previously observed in extremely blue 
DEEP2 galaxies in close kinematic galaxy pairs ($\delta r_p=50$ $h^{-1}$ kpc) within the EGS \citep{Lin07}. 
Further, \cite{Robaina09} show enhanced clustering with SFR on slightly larger scales (40$<$$r_p$$<$180 $h^{-1}$ kpc)
between 0.4$<$\z$<$0.8 in a sample of COMBO-17 galaxies \citep{Wolf03} matched to \emph{Spitzer} 24$\mu$m measurements.

Our results show a similar increase in small-scale clustering power at high SFR,  on scales $r_{p}$$\leq$0.3 $h^{-1}$ Mpc.  However, we have demonstrated
that an increased SFR is also correlated with higher stellar mass, and therefore it is also possible that galaxies with higher stellar mass 
have a stronger one-halo term due to their higher halo mass. 
It is possible that both stellar mass and SFR enhancement are correlated with the increase clustering seen on small-scales in SFR-selected blue galaxies.
 We note, however, that red galaxies with low SFR have similar halo masses to the high-SFR blue galaxies and do not show a 
similar rise, indicating that SFR enhancement is likely more strongly correlated with the increased clustering signal. %The deviation has a $<$5\% chance of being 
 %due to a random fluctuation in the relative correlation function. 
% Given the volume density 
%of this sample, $2\times10^{-4}~h^{3}$ Mpc$^{-3}$, it is likely that the observed
%triggered star formation extends beyond disruptive merger events
%and includes more mild tidal interactions.

These small-scale clustering trends allow us to interpret the behavior seen as a function 
of luminosity in C08 (their Figure 10), who detected a relative small-scale
rise for the brightest blue galaxies and a marginal increase for bright 
red galaxies.  
Figure~\ref{fig:colormasscontour} shows that 
blue galaxies with the highest SFRs also have the highest
luminosities, and therefore the observed small-scale rise for bright blue
galaxies in C08 may have been due in part to both higher stellar masses and SFR enhancement.
The small-scale rise observed here for massive red galaxies is likely 
reflected in the marginally-significant increase seen for bright red galaxies in C08. As 
the mass-to-light ratio varies across the red sequence, a given luminosity range corresponds to a wider range in stellar mass;
if stellar mass is more strongly linked to halo mass, luminosity-selected samples will have a broader host mass range
and hence exhibit a weaker small-scale clustering signal.

Turning to sSFR, 
we find that galaxies that are forming stars most rapidly relative to their existing
stellar mass have an increased clustering amplitude on small scales.
Interestingly, \emph{both} the high-sSFR sub-samples of the blue and red galaxy population
show a small-scale rise at $r_{p}$$\leq$0.3 $h^{-1}$ Mpc. In each sample, we measure a 4-fold 
increase in \wprp~relative to the large-scale power law fit over $r_{p}$=1-10 $h^{-1}$
Mpc scales. The increase in clustering is significant at the $p<0.05$ level in both galaxy samples.
Unlike the previous small-scale clustering results with SFR, the sSFR samples have normalized the SFR
relative to the stellar mass, and therefore the measured small-scale rise will be more directly connected to 
enhanced star formation at a fixed stellar mass. Because sSFR is highly correlated with 
star formation history and restframe color, the galaxies demonstrating this increased small-scale clustering 
are the bluest members of their respective populations (e.g. the bluest blue galaxies and the bluest red
galaxies). 

\cite{Li08} measured the clustering of SDSS galaxies
selected by sSFR and found enhanced small-scale galaxy
clustering in the local Universe, with $>$40\% of these galaxies
having close ($r_p<$100 kpc) companions. They conclude that the
observed clustering behavior is a signature of tidal interactions
between galaxy pairs inside the same dark matter halo, which leads to
an inflow of cold gas and enhanced star formation.  Evidence
for enhanced sSFR on small scales was also measured from close pairs of
blue galaxies ($r_{p}$$<$50 $h^{-1}$ kpc) at higher redshift in PRIMUS
between 0.25$<$\z$<$0.5 \citep{Wong11} and in DEEP2 between
0.1$<$\z$<$1.1 \citep{Lin07}.  
\cite{Robaina09} also found enhanced clustering on scales $r_{p}$$<$40 kpc for star-forming galaxies with 
$M_{*}>10^{10}~M_{\sun}$ between 0.4$<$\z$<$0.8. 
While we define ``small-scales'' at a larger scale threshold of $r_{p}$$\leq$0.3 $h^{-1}$ Mpc, 
we also find statistically significant evidence for enhanced small-scale clustering with sSFR. 
These results support our interpretation that enhanced star formation, possibly triggered from galaxy tidal interactions, 
is correlated with the observed small-scale rise in our blue galaxy sample with high sSFR.

However, enhanced small-scale clustering for high-sSFR 
red galaxies may be somewhat surprising. The
range of sSFR for which we find this enhancement 
($-10$$<$\ssfr$<$$-11$) is often referred to as the ``green
valley", a transition region from blue star-forming galaxies to red
quiescent galaxies \citep{Martin07, Salim09, Mendez11}.  
The enhanced clustering could be due to 
recent merger events or tidal disruptions between red and blue 
galaxy pairs where the blue galaxy has provided
additional gas for star formation in the quiescent galaxy. 
\cite{Robaina09} found that major interactions can contribute significantly
to the small-scale clustering amplitude for 
the most dust-obscured starbursts, which may be
selected in our DEEP2 sample as red with the restframe color cut of Equation~\ref{eq:color}.
Alternatively, close interactions between
blue galaxy pairs could temporarily boost the SFR and eventually
lead to quenching, which would move
galaxies from the blue cloud to the red sequence.  As we argued
in the last subsection, however, such events must be sub-dominant to the
general quenching trend in individual halos. 

\section{Conclusions}
\label{sec:conclusions}

In this study, we measure the two-point correlation function of complete DEEP2 
galaxy samples selected by stellar mass, SFR and sSFR and separated by restframe color.  We fit a power law to the 
projected correlation function and measure the bias and clustering scale length and slope 
on scales of 1-10 $h^{-1}$ Mpc ($\bar{r}_p$=4.1 $h^{-1}$ Mpc), where 
the two-halo term dominates the 2PCF. 
We also study the relative shape of the correlation 
function on small and large scales, which depends on the satellite fraction and
 dark matter halo mass. We summarize our findings as follows:

\begin{enumerate}
\item The large-scale galaxy clustering amplitude increases monotonically as a function 
of stellar mass for star-forming galaxies with 9.6$<$\smass$<$11, indicating that stellar 
mass closely tracks halo mass at $z\sim1$. Within the limited mass range probed for red 
galaxies (10.5$<$\smass$<$11.5), we find no significant trend between stellar mass and 
clustering amplitude.  Within the measurement errors, our results agree with a $z\sim1$ clustering analysis from VVDS 
\citep{Meneux08} in the stellar mass range where both DEEP2 and VVDS samples are complete. 

\item Within the blue, star-forming galaxy population, the large-scale clustering amplitude
increases as a function of increasing SFR.  Red galaxies with low SFRs are strongly clustered 
and show no dependence on SFR within the range probed.  We find the highest-SFR blue galaxies 
have the same clustering amplitude as red galaxies, as seen in previous DEEP2 environment 
studies.

\item The observed trend between large-scale clustering amplitude and SFR can
primarily be accounted for by the observed correlation between clustering 
amplitude and stellar mass. However, the $r_0$-stellar mass relation alone 
does not fully predict the clustering amplitude at all SFRs;
there is a small clustering excess at high SFR above what is predicted by 
the stellar mass of each sample alone.
This suggests that most, but not all, of the correlation between large-scale 
clustering amplitude and SFR can be attributed to the SFR-stellar mass relation.

\item Galaxy samples selected by sSFR show similar 
large-scale clustering trends as samples selected by stellar mass. 
The clustering amplitude decreases with increasing sSFR (corresponding to bluer restframe color) 
while red galaxies with low sSFRs have a constant clustering strength in the range probed.
While the large-scale clustering amplitude only mildly depends on sSFR for 
$-11.0$$<$\ssfr$<$$-9.5$,
$r_0$ drops significantly between $-9.5$$<$\ssfr$<$$-8.5$, similar to the trend observed locally in SDSS.

\item By constructing stellar mass-limited star-forming galaxy 
samples above and below the SFR-stellar mass ``main sequence'',
 we find that the clustering amplitude increases with decreasing SFR at a fixed stellar mass 
(i.e. below the main sequence), which confirms that the clustering trends seen with sSFR are not driven entirely by
stellar mass.  Given that galaxies below the main sequence are more clustered 
than those above, the bulk of the population 
%undergo secular evolution from above the main sequence
%to below the main sequence, 
above the main sequence can not be dominated by galaxies that used to be 
on the main sequence and are currently undergoing major merger events.  Instead,
galaxies must smoothly evolve from above the main sequence to below it, 
as the clustering of galaxies can only increase with time.
%Therefore the bulk of the population above the star-forming main sequence can not have started on the main 
%sequence and later evolved above it, due to, for example, a merger event. 

\item We detect enhanced clustering on scales less than $r_{p}$$\leq$0.3 $h^{-1}$ Mpc, relative to 
larger scales, for high stellar mass red galaxies, high SFR blue galaxies, and the highest sSFR 
sub-samples of both red and blue galaxy populations. The increased small-scale clustering may reflect 
a combination of effects, including a changing satellite fraction, higher halo mass, and/or enhanced SFR. 
We conclude that triggered star formation due to close galaxy interactions is a likely explanation for the enhanced clustering seen at high sSFR.
 \end{enumerate}

As our analysis uses the two-point correlation function formalism and measures  
clustering properties for several important galaxy properties, we expect these results
to be highly useful for galaxy evolution models.  In particular, HOD and abundance 
matching models should greatly benefit from measurements of the clustering amplitude at 
$z\sim1$ with respect to stellar mass, SFR, and sSFR.  We also anticipate that the bias values 
calculated here as a function of stellar mass and SFR will be used in projections of 
future BAO surveys that will probe the \z$>$1 universe. \linebreak

We thank Guangtun Zhu and Jeremy Tinker for providing their code to calculate $V_{\rm{max}}$ values and dark matter correlation functions, and we thank Kevin Bundy for providing $K$-band stellar masses to crosscheck DEEP2 stellar masses. We also appreciate the useful comments from Ramin Skibba, Lihwai Lin, and the referee of this article. This work is supported in part by the Director, Office of Science, High Energy Physics, of the U.S. Department of Energy, under contract number DE-AC03-76SF00098. ALC gratefully acknowledges support
from NSF CAREER award AST-1055081. JAN acknowledges support from DOE Early Career grant DE-SC0003960. DEEP2 survey funding has been provided by NSF grants AST95-09298, AST-0071048, AST-0071198, AST-0507428, AST-0507483, and AST-0806732 as well as NASA LTSA grant NNG04GC89G. 
Some of the data presented herein were obtained at the W. M. Keck Observatory, which is operated as a scientific partnership among the California Institute of Technology, the University of California and the National Aeronautics and Space Administration. The Observatory was made possible by financial support of the W. M. Keck Foundation. The DEEP2 team and Keck Observatory acknowledge the significant cultural role that the summit of Mauna Kea has within the indigenous Hawaiian community and appreciate the opportunity to conduct observations from this mountain.

\clearpage
\begin{table*}[h]
\caption{Results for Stellar Mass Threshold Samples }
\begin{center}
%\begin{tabularx}{0.98\textwidth}{llllccccccccccccccc}
\begin{tabularx}{0.98\textwidth}{lccrccccccccc}
\hline
\hline
Color \T & $M_*$ & $<$$M_*$$>$ & $N_{gal}$ & $n$ & $\sigma_{\rm{cv}}$ &  $z$ Range & $<$\z$>$ &  $r_0$ & $\gamma$ & Bias & $M_{\rm{min}}$ & $<$$M_{\rm{halo}}$$>$ \\ \cmidrule(r){1-1} \cmidrule(lr){2-3} \cmidrule(lr){4-4} \cmidrule(lr){5-6} \cmidrule(lr){7-8} \cmidrule(lr){9-9} \cmidrule(lr){10-10}  \cmidrule(lr){11-11} \cmidrule(lr){12-13}
& \multicolumn{2}{c}{\footnotesize{[log($M_\sun$)]}} &  & \multicolumn{2}{c}{\footnotesize{[$10^{-4}~h^3$ Mpc$^{-3}$]}} &  &  & \footnotesize{[$h^{-1}$ Mpc]} &  &   & \multicolumn{2}{c}{\footnotesize{[log($h^{-1} M_\sun$)]}} \B\\
\hline
%All\T & $~>$9.6 & 10.34 & 12195 & --%{$10.1\pm1.9$
%							 & -- & 0.74--1.05 & 0.89 & $3.86\pm0.10$ & $1.64\pm0.07$  & $1.44\pm0.05$ & 12.13 & 12.70~~\\%[3pt]
%All\T & $~>$9.9 & 10.50 & 9578 & --%$11.3\pm1.3$ 
%							& --& 0.74--1.05 & 0.89 & $4.12\pm0.14$ & $1.61\pm0.07$ & $1.53\pm0.06$ & 12.26 & 12.80~~\\%[3pt]
%All\T & $>$10.2 & 10.69 & 6637 & $8.1\pm1.0$ & 3.4 & 0.74--1.05 & 0.90 & $4.52\pm0.10$ & $1.67\pm0.09$ & $1.64\pm0.04$ & 12.42 & 12.92~~\\%[3pt]
All\T & $>$10.5 & 10.90 & 4201 & $16.9\pm0.7$ & 8.2 & 0.74--1.05 & 0.89 & $4.99\pm0.14$ & $1.78\pm0.12$ & $1.77\pm0.06$ & 12.48 & 12.95~~\\%[3pt]
All\T & $>$10.8 & 11.08 & 2400 & $11.4\pm0.5$ & 4.7 & 0.74--1.05 & 0.89 & $5.33\pm0.16$ & $1.88\pm0.14$ & $1.87\pm0.06$ & 12.61 & 13.05\B ~~\\%[3pt]
\hline
Blue\T & $~>$9.6 & 10.18 & 9885 & $23.0\pm1.7$ & 7.2 & 0.74--1.05 & 0.89 & $3.46\pm0.12$ & $1.59\pm0.06$ & $1.35\pm0.05$ & 11.67 & 12.32~~\\%[3pt]
Blue\T & $~>$9.9 & 10.33 & 7268 & $29.4\pm1.0$ & 8.5 & 0.74--1.05 & 0.90 & $3.66\pm0.14$& $1.53\pm0.08$ & $1.42\pm0.07$ & 11.85 & 12.47~~\\%[3pt]
Blue\T & $>$10.2 & 10.51 & 4331 & $21.0\pm0.6$ & 8.2 & 0.74--1.05 & 0.91 & $4.13\pm0.14$ & $1.61\pm0.08$ & $1.50\pm0.06$ & 12.03 & 12.60~~\\%[3pt]
Blue\T & $>$10.5 & 10.73 & 1969 & $9.9\pm0.4$ & 3.8 & 0.74--1.05 & 0.91 & $4.29\pm 0.34$ & $1.60\pm0.10$ & $1.60\pm0.10$ & 12.22 & 12.75\B~~\\%[3pt]
\hline
Blue\T & $~>$9.9 & 10.34 & 12613 & $14.0\pm0.9$ & 3.2 & 0.74--1.40 & 1.03 & $3.78\pm0.10$ & $1.60\pm0.06$  & $1.50\pm0.05$ & 12.03 & 12.60~~\\%[3pt]
Blue\T & $>$10.2 & 10.51 & 7844 & $12.5\pm0.5$ & 4.1 & 0.74--1.40 & 1.04 & $4.24\pm0.09$ & $1.77\pm0.07$ & $1.59\pm0.04$ & 12.20 & 12.74~~\\%[3pt]
Blue\T & $>$10.5 & 10.73 & 3567 & $6.7\pm0.3$ & 2.0 & 0.74--1.40 & 1.04 & $4.40\pm0.18$ &  $1.73\pm0.15$ & $1.71\pm0.08$  & 12.40 & 12.89~~\\%[3pt]
Blue\T & $>$10.8 & 10.97 & 1134 & $2.3\pm0.1$ & 0.9 & 0.74--1.40 & 1.04 & $5.89\pm0.50$ & $1.63\pm0.38$ & $2.11\pm0.13$ & 12.87 & 13.26\B~~\\%[3pt]
\hline
Red\T & $>$10.5 & 11.04 & 2232 & $8.5\pm0.6$ & 4.4 & 0.74--1.05 & 0.88 & $5.28\pm0.26$ & $1.98\pm0.19$ & $1.84\pm0.05$ & 12.57 & 13.02~~\\%[3pt]
Red\T & $>$11.0 & 11.23 & 1200 & $5.8\pm0.3$ & 3.2 &0.74--1.05 & 0.89 & $4.96\pm0.24$ & $1.72\pm0.16$ & $1.81\pm0.08$ & 12.53 & 12.99\B~~\\%[3pt]
\hline
\hline
\end{tabularx}
\end{center}
%\hspace{0.3in}$^{a}$  Units of log($M_\sun$).

%\hspace{0.3in}$^{b}$  Units of $10^{-4}~h^3$ Mpc$^{-3}$.

%\hspace{0.3in}$^{c}$  Units of log($h^{-1} M_\sun$).
%\hspace{0.8in}$^a$ Fixed $\gamma=1.6$ in fit.
\label{tab:smassresults}
\end{table*}

\begin{table*}[h]
\caption{Results for Stellar Mass Bin Samples }
\begin{center}
%\begin{tabularx}{0.98\textwidth}{llllccccccccccccccc}
\begin{tabularx}{0.86\textwidth}{lccrccccccc}
\hline
\hline
%Color\T & $M_*$$^a$ & $<$$M_*$$>^a$ & $N_{gal}$ & $n^b$&  $\sigma_{\rm{cv}}^{b}$ &$z$ Range & $<$$z$$>$ &  $r_0$ & $\gamma$ & Bias \B\\%& $M_{\rm{min}}^a$ & $<$$M_{\rm{halo}}$$>^a$ \B\\
Color \T & $M_*$ & $<$$M_*$$>$ & $N_{gal}$ & $n$ & $\sigma_{\rm{cv}}$ &  $z$ Range & $<$\z$>$ &  $r_0$ & $\gamma$ & Bias \\ \cmidrule(r){1-1} \cmidrule(lr){2-3} \cmidrule(lr){4-4} \cmidrule(lr){5-6} \cmidrule(lr){7-8} \cmidrule(lr){9-9} \cmidrule(lr){10-10}  \cmidrule(lr){11-11}
& \multicolumn{2}{c}{\footnotesize{[log($M_\sun$)]}} &  & \multicolumn{2}{c}{\footnotesize{[$10^{-4}~h^3$ Mpc$^{-3}$]}} &  &  & \footnotesize{[$h^{-1}$ Mpc]} &  &  \B\\
\hline
%All\T & ~9.6--10.0 & 9.81 & 3573 & --%{$10.1\pm1.9$
%							 & -- & 0.74--1.05 & 0.86 & $3.01\pm0.18$ & $1.64\pm0.14$  & $1.18\pm0.06$ & 11.63 & 12.32~~\\%[3pt]
%All\T & 10.0--10.4 & 10.19 & 3679 & --%$11.3\pm1.3$ 
%							& -- & 0.74--1.05 & 0.90 & $3.34\pm0.14$ & $1.52\pm0.12$ & $1.35\pm0.07$ & 11.95 & 12.56~~\\%[3pt]
All\T &10.4--10.8 & 10.59 & 2554 & -- %$3.0\pm0.7$ & 1.4 
							& -- & 0.74--1.05 & 0.90 & $4.52\pm0.21$ & $1.70\pm0.15$ & $1.60\pm0.05$~~\\%& 12.35 & 12.87~~\\%[3pt]
All\T & $>$10.8 & 11.08 & 2400 & $11.4\pm0.5$ & 4.7 & 0.74--1.05 & 0.89 & $5.33\pm0.16$ & $1.88\pm0.14$ & $1.87\pm0.06$\B~~\\%& 12.68 & 13.13~~\\%[3pt]
\hline
Blue\T & ~9.6--10.0 & 9.81 & 3573 & $7.0\pm1.5$ & 2.9 & 0.74--1.05 & 0.86 & $3.02\pm0.18$ & $1.65\pm0.14$ & $1.18\pm0.06$~~\\%& 11.63 & 12.32~~\\%[3pt]
Blue\T & 10.0--10.4 & 10.19 & 3643 & $14.9\pm0.7$ & 3.8 & 0.74--1.05 & 0.90 & $3.37\pm0.14$& $1.55\pm0.12$ & $1.34\pm0.07$~~\\%& 11.94 & 12.56~~\\%[3pt]
Blue\T & $>$10.4 & 10.65 & 2676 & $13.4\pm0.5$ & 5.2 & 0.74--1.05 & 0.91 & $4.50\pm0.17$ & $1.68\pm0.11$ & $1.62\pm0.08$\B~~\\%& 12.36 & 12.87~~\\%[3pt]
\hline
Blue\T & ~9.9--10.2 & 10.05 & 4774 & $4.7\pm0.7$ & 1.7 & 0.74--1.40 & 1.02 & $2.80\pm0.41$ & $1.30\pm0.16$  & $1.35\pm0.10$~~\\%& 11.83 & 12.44~~\\%[3pt]
Blue\T & ~10.2--10.5 & 10.34 & 4279 & $6.4\pm0.4$ & 2.0 & 0.74--1.40 & 1.04 & $3.71\pm0.21$ & $1.76\pm0.09$ & $1.48\pm0.07$~~\\%& 12.03 & 12.59~~\\%[3pt]
Blue\T & ~10.5--10.8 & 10.63 & 2435 & $4.4\pm0.2$ & 1.5 & 0.74--1.40 & 1.04 & $4.14\pm0.21$ &  $1.90\pm0.19$ & $1.57\pm0.09$~~\\%& 12.15 & 12.69~~\\%[3pt]
Blue\T & $>$10.8 & 10.97 & 1134 & $2.6\pm0.1$ & 0.9 & 0.74--1.40 & 1.04 & $5.89\pm0.50$ & $1.63\pm0.38$ & $2.11\pm0.13$\B~~\\%& 12.74 & 13.15\B~~\\%[3pt]
\hline
Red\T & 10.5-11.0 & 10.81 & 1034 & $3.8\pm0.6$ & 2.0 & 0.74--1.05 & 0.87 & $5.00\pm0.63$ & $1.65\pm0.17$ & $1.76\pm0.14$~~\\%& 12.57 & 13.05~~\\%[3pt]
Red\T & $>$11.0 & 11.23 & 1200 & $5.8\pm0.3$ & 3.2 & 0.74--1.05 & 0.89 & $4.96\pm0.24$ & $1.72\pm0.16$ & $1.81\pm0.08$\B~~\\%& 12.62 %& 13.08\B~~\\%[3pt]
\hline
\hline
\end{tabularx}
\end{center}
%\hspace{0.3in}$^{a}$  Units of log($M_\sun$).

%\hspace{0.3in}$^{b}$  Units of $10^{-4}~h^3$ Mpc$^{-3}$.
%\hspace{0.8in}$^a$ Fixed $\gamma=1.6$ in fit.
\label{tab:smassbinresults}
\end{table*}

\begin{table*}[h]
\caption{Results for Star Formation Rate Threshold Samples }
\begin{center}
%\begin{tabularx}{0.98\textwidth}{llllccccccccccccccc}
\begin{tabularx}{0.98\textwidth}{lccrccccccccc}
\hline
\hline
Color \T & SFR & $<$SFR$>$ & $N_{gal}$ & $n$ & $\sigma_{\rm{cv}}$ &  $z$ Range & $<$\z$>$ &  $r_0$ & $\gamma$ & Bias & $M_{\rm{min}}$ & $<$$M_{\rm{halo}}$$>$ \\ \cmidrule(r){1-1} \cmidrule(lr){2-3} \cmidrule(lr){4-4} \cmidrule(lr){5-6} \cmidrule(lr){7-8} \cmidrule(lr){9-9} \cmidrule(lr){10-10}  \cmidrule(lr){11-11} \cmidrule(lr){12-13}
& \multicolumn{2}{c}{\footnotesize{[log($M_\sun$ yr$^{-1}$)]}} &  & \multicolumn{2}{c}{\footnotesize{[$10^{-4}~h^3$ Mpc$^{-3}$]}} &  &  & \footnotesize{[$h^{-1}$ Mpc]} &  &   & \multicolumn{2}{c}{\footnotesize{[log($h^{-1} M_\sun$)]}} \B\\
\hline
Blue\T & $>$0.75 & 1.05 & 8542 & $39.4\pm0.9$ & 8.7 & 0.74--1.05 & 0.90 & $3.46\pm0.14$ & $1.53\pm0.04$ & $1.38\pm0.05$ & 11.75 & 12.39~~\\%[3pt]
Blue\T & $>$1.00 & 1.22 & 4456 & $22.2\pm0.6$ & 5.2 & 0.74--1.05 & 0.91 & $3.70\pm0.27$& $1.53\pm0.17$ & $1.44\pm0.08$ & 11.90 & 12.50~~\\%[3pt]
Blue\T & $>$1.25 & 1.41 & 1693 & $8.2\pm0.4$ & 1.7 & 0.74--1.05 & 0.91 & $3.80\pm0.26$ & $1.60\pm0.22$ & $1.48\pm0.09$ & 11.99 & 12.57\B~~\\%[3pt]
\hline
Blue\T & $>$1.00 & 1.22 & 7507 & $17.8\pm0.5$ & 5.2 & 0.74--1.25 & 1.01 & $3.90\pm0.20$ & $1.62\pm0.08$ & $1.52\pm0.07$ & 12.24 & 12.64~~\\%[3pt]
Blue\T & $>$1.25 & 1.41 & 4355 & $7.3\pm0.3$ & 3.5 & 0.74--1.40 & 1.12 & $3.99\pm0.25$ & $1.37\pm0.13$ & $1.72\pm0.08$ & 12.41 & 12.90~~\\%[3pt]
Blue\T & $>$1.50 & 1.62 & 1114 & $2.3\pm0.1$ & 0.3 & 0.74--1.40 & 1.13 & $4.45\pm1.19$ & $1.38\pm0.27$ & $1.97\pm0.32$ & 12.64 & 13.07\B~~\\%[3pt]
\hline
%Red\T & $<$0.4$^{c}$ & 1162 & $1.9\pm0.4$ & 0.74--1.05 & 0.87 & $5.18\pm0.14$ & $2.24\pm0.24$ & $1.83\pm0.33$~~\\%[3pt]
Red\T & $>$-0.1 & 0.39 & 2041 & $6.4\pm0.7$ & 2.9 & 0.74--1.05 & 0.88 & $5.28\pm0.28$ & $1.84\pm0.24$ & $1.84\pm0.06$ & 12.58 & 13.02~~\\%[3pt]
Red\T & $>$0.4 & 0.65 & 954 & $4.7\pm 0.3$ & 1.7 & 0.74--1.05 & 0.90 & $5.26\pm0.63$ & $1.65\pm0.31$ & $1.90\pm0.13$ & 12.64 & 13.08\B~~\\%[3pt]
\hline
\hline
\end{tabularx}
\end{center}
%\hspace{0.3in}$^{a}$  Units of log($M_{\sun}$ yr$^{-1}$).

%\hspace{0.3in}$^{b}$  Units of $10^{-4}~h^3$ Mpc$^{-3}$.

%\hspace{0.3in}$^{c}$  Units of log($h^{-1} M_\sun$).
\label{tab:sfrthreshresults}
\end{table*}

\begin{table*}[h]
\caption{Results for Star Formation Rate Binned Samples }
\begin{center}
%\begin{tabularx}{0.98\textwidth}{llllccccccccccccccc}
\begin{tabularx}{0.86\textwidth}{lccrccccccc}
\hline
\hline
Color \T & SFR & $<$SFR$>$ & $N_{gal}$ & $n$ & $\sigma_{\rm{cv}}$ &  $z$ Range & $<$\z$>$ &  $r_0$ & $\gamma$ & Bias \\ \cmidrule(r){1-1} \cmidrule(lr){2-3} \cmidrule(lr){4-4} \cmidrule(lr){5-6} \cmidrule(lr){7-8} \cmidrule(lr){9-9} \cmidrule(lr){10-10}  \cmidrule(lr){11-11}
& \multicolumn{2}{c}{\footnotesize{[log($M_\sun$ yr$^{-1}$)]}} &  & \multicolumn{2}{c}{\footnotesize{[$10^{-4}~h^3$ Mpc$^{-3}$]}} &  &  & \footnotesize{[$h^{-1}$ Mpc]} &  &  \B\\
\hline
%Color\T & SFR$^a$ & $<$SFR$>^a$ & $N_{gal}$ & $n^b$& $\sigma_{\rm{cv}}^{b}$ & $z$ Range & $<$$z$$>$ &  $r_0$ & $\gamma$ & Bias \B\\%& $M_{\rm{min}}^c$ & $<$$M_{\rm{halo}}$$>^c$ \B\\
\hline
Blue\T & 0.75--1.00 & 1.05 & 4091 & $17.8\pm0.7$ & 5.5 & 0.74--1.05 & 0.90 & $3.05\pm0.13$ & $1.44\pm0.11$ & $1.31\pm0.04$~~\\% & 11.88 & 12.51~~\\%[3pt]
Blue\T & 1.00--1.25 & 1.22 & 2765 & $14.0\pm0.5$ & 3.8 & 0.74--1.05 & 0.91 & $3.61\pm0.47$& $1.37\pm0.18$ & $1.47\pm0.08$~~\\% & 12.15 & 12.71~~\\%[3pt]
Blue\T & $>$1.25 & 1.41 & 1693 & $8.2\pm0.4$ & 1.7 & 0.74--1.05 & 0.91 & $3.88\pm0.27$ & $1.53\pm0.16$ & $1.53\pm0.10$\B~~\\% & 12.25 & 12.79\B~~\\%[3pt]
\hline
Blue\T & 1.00--1.25 & 1.11 & 4634 & $12.8\pm0.4$ & 3.2 & 0.74--1.25 & 1.01 & $3.91\pm0.27$ & $1.62\pm0.15$ & $1.48\pm0.07$~~\\% & 12.06 & 12.62~~\\%[3pt]
Blue\T & 1.25--1.50 & 1.35 & 3243 & $5.2\pm0.2$ & 3.2 & 0.74--1.40 & 1.11 & $3.93\pm0.18$ & $1.48\pm0.22$ & $1.63\pm0.07$~~\\% & 12.17 & 12.69~~\\%[3pt]
Blue\T & $>$1.50 & 1.62 & 1114 & $2.3\pm0.1$ & 0.4 & 0.74--1.40 & 1.13 & $4.45\pm1.19$ & $1.38\pm0.27$ & $1.97\pm0.32$\B~~\\% & 12.53 & 12.97\B~~\\%[3pt]
\hline
%Red\T & $<$0.4$^{c}$ & 1162 & $1.9\pm0.4$ & 0.74--1.05 & 0.87 & $5.18\pm0.14$ & $2.24\pm0.24$ & $1.83\pm0.33$~~\\%[3pt]
Red\T & -0.1--0.4 & 0.14 & 1081 & $2.9\pm0.6$ & 2.6 & 0.74--1.05 & 0.87 & $5.00\pm0.29$ & $2.10\pm0.40$ & $1.72\pm0.08$~~\\% & 12.53 & 13.01~~\\%[3pt]
Red\T & $>$0.4 & 0.65 & 954 & $4.7\pm 0.3$ & 1.7 & 0.74--1.05 & 0.90 & $5.26\pm0.63$ & $1.65\pm0.31$ & $1.90\pm0.13$\B~~\\% & 12.69 & 13.13\B~~\\%[3pt]
\hline
\hline
\end{tabularx}
\end{center}
%\hspace{0.3in}$^{a}$  Units of log($M_{\sun}$ yr$^{-1}$).

%\hspace{0.3in}$^{b}$  Units of $10^{-4}~h^3$ Mpc$^{-3}$.

%\hspace{0.3in}$^{c}$  Units of log($M_{\sun}$).
\label{tab:sfrbinresults}
\end{table*}

\begin{table*}[h]
\caption{Results for Specific Star Formation Rate Threshold Samples }
\begin{center}
%\begin{tabularx}{0.98\textwidth}{llllccccccccccccccc}
\begin{tabularx}{0.99\textwidth}{lccrccccccccc}
\hline
\hline
Color \T & sSFR & $<$sSFR$>$ & $N_{gal}$ & $n$ & $\sigma_{\rm{cv}}$ &  $z$ Range & $<$\z$>$ &  $r_0$ & $\gamma$ & Bias & $M_{\rm{min}}$ & $<$$M_{\rm{halo}}$$>$ \\ \cmidrule(r){1-1} \cmidrule(lr){2-3} \cmidrule(lr){4-4} \cmidrule(lr){5-6} \cmidrule(lr){7-8} \cmidrule(lr){9-9} \cmidrule(lr){10-10}  \cmidrule(lr){11-11} \cmidrule(lr){12-13}
& \multicolumn{2}{c}{\footnotesize{[log(yr$^{-1}$)]}} &  & \multicolumn{2}{c}{\footnotesize{[$10^{-4}~h^3$ Mpc$^{-3}$]}} &  &  & \footnotesize{[$h^{-1}$ Mpc]} &  &   & \multicolumn{2}{c}{\footnotesize{[log($h^{-1} M_\sun$)]}} \B\\
\hline
Blue$^a$\T & $<$-8.60 & -9.13 & 5777 & $28.9\pm0.7$ & 8.5 & 0.74--1.05 & 0.91 & $3.97\pm0.16$ & $1.47\pm0.08$ & $1.52\pm0.06$ & 12.07 & 12.64~~\\%[3pt]
Blue\T & $<$-8.95 & -9.30 & 3819 & $19.2\pm0.5$ & 8.2 & 0.74--1.05 & 0.91 & $4.43\pm0.20$& $1.63\pm0.07$ & $1.59\pm0.07$ & 12.20 & 12.74~~\\%[3pt]
Blue\T & $<$-9.30 &-9.54 & 1651 & $8.2\pm0.4$ & 4.7 & 0.74--1.05 & 0.91 & $4.66\pm0.35$ & $1.66\pm0.14$ & $1.67\pm0.09$ & 12.34 &  12.84\B~~\\%[3pt]
\hline
Blue$^b$\T & $<$-8.60 & -8.79 & 7144 & $13.1\pm0.4$ & 3.2 & 0.74--1.40 & 1.11 & $4.21\pm0.11$ & $1.70\pm0.07$ & $1.64\pm0.05$ & 12.29 & 12.80~~\\%[3pt]
Blue\T & $<$-8.95 & -9.11 & 4693 & $9.3\pm0.3$ & 2.9 & 0.74--1.40 & 1.07 & $4.54\pm0.10$ & $1.82\pm0.07$ & $1.70\pm0.04$ & 12.38 & 12.88~~\\%[3pt]
Blue\T & $<$-9.30 & -9.54 & 2013 & $3.8\pm0.2$ & 1.5 & 0.74--1.40 & 1.03 & $4.62\pm0.24$ & $1.84\pm0.18$ & $1.76\pm0.08$ & 12.47 & 12.94\B~~\\%[3pt]
\hline
Red$^a$\T & $<$-9.90 & -10.59 & 1641 & $8.2\pm0.4$ & 3.8 & 0.74--1.05 & 0.89 & $5.17\pm0.23$ & $1.80\pm0.13$ & $1.84\pm0.06$ & 12.58 & 13.02~~\\%[3pt]
Red\T & $<$-10.6 & -10.91 & 795 & $4.1\pm0.3$ & 2.6 & 0.74--1.05 & 0.89 & $5.04\pm0.32$ & $2.04\pm0.25$ & $1.81\pm0.11$ & 12.54 & 12.99\B~~\\%[3pt]
\hline
\hline
\end{tabularx}
\end{center}
%\hspace{0.3in}$^{a}$  Units of log(yr$^{-1}$).

%\hspace{0.3in}$^{b}$  Units of $10^{-4}~h^3$ Mpc$^{-3}$.

%\hspace{0.3in}$^{c}$  Units of log($h^{-1} M_\sun$).

\hspace{0.3in}$^{a}$  $z$$<$1.05 galaxy samples are limited to $M_{B}$$<$-20.5.

\hspace{0.3in}$^{b}$  $z$$<$1.4 blue galaxy samples are limited to $M_{B}$$<$-21.0.

\label{tab:ssfrthreshresults}
\end{table*}

\begin{table*}[h]
\caption{Results for Specific Star Formation Rate Binned Samples }
\begin{center}
%\begin{tabularx}{0.98\textwidth}{llllccccccccccccccc}
\begin{tabularx}{0.89\textwidth}{lccrccccccc}
\hline
\hline
Color \T & sSFR & $<$sSFR$>$ & $N_{gal}$ & $n$ & $\sigma_{\rm{cv}}$ &  $z$ Range & $<$\z$>$ &  $r_0$ & $\gamma$ & Bias \\ \cmidrule(r){1-1} \cmidrule(lr){2-3} \cmidrule(lr){4-4} \cmidrule(lr){5-6} \cmidrule(lr){7-8} \cmidrule(lr){9-9} \cmidrule(lr){10-10}  \cmidrule(lr){11-11}
& \multicolumn{2}{c}{\footnotesize{[log(yr$^{-1}$)]}} &  & \multicolumn{2}{c}{\footnotesize{[$10^{-4}~h^3$ Mpc$^{-3}$]}} &  &  & \footnotesize{[$h^{-1}$ Mpc]} &  &  \B\\
%Color\T & sSFR$^a$ & $<$sSFR$>^a$ & $N_{gal}$ & $n^b$& $\sigma_{\rm{cv}}^{b}$ & $z$ Range & $<$$z$$>$ &  $r_0$ & $\gamma$ & Bias \B\\%& $M_{\rm{min}}^c$ & $<$$M_{\rm{halo}}$$>^c$ \B\\
\hline
Blue$^a$\T &  -8.60 -- -8.95 & -8.80 & 1959 & $9.6\pm0.4$ & 3.2 & 0.74--1.05 & 0.91 & $3.64\pm0.36$ & $1.59\pm0.09$ & $1.42\pm0.11$~~\\% & 12.08 & 12.66~~\\%[3pt]
Blue\T & -8.95 -- -9.30 & -9.11 & 2170 & $11.1\pm0.4$ & 3.8 & 0.74--1.05 & 0.91 & $4.12\pm0.43$& $1.54\pm0.16$ & $1.54\pm0.10$~~\\% & 12.26 & 12.79~~\\%[3pt]
Blue\T & $<$-9.30 & -9.54 & 1651 &  $8.2\pm0.4$ & 5.0 & 0.74--1.05 & 0.91 & $4.70\pm0.35$ & $1.62\pm0.13$ & $1.67\pm0.09$\B~~\\% & 12.44 & 12.94\B~~\\%[3pt]
\hline
Blue$^b$\T & -8.60 -- -8.95 & -8.79 & 2451 & $3.9\pm0.2$ & 2.9 & 0.74--1.40 & 1.12 & $3.29\pm0.35$ & $1.30\pm0.28$ & $1.59\pm0.15$~~\\% & 12.10 & 12.64~~\\%[3pt]
Blue\T & -8.95 -- -9.30 & -9.11 & 2681 & $5.2\pm0.2$ & 1.4 & 0.74--1.40 & 1.07 & $4.13\pm0.21$ & $1.66\pm0.16$ & $1.64\pm0.07$~~\\% & 12.23 & 12.74~~\\%[3pt]
Blue\T & $<$-9.30 & -9.54 & 2013 & $3.9\pm0.2$ & 1.6 & 0.74--1.40 & 1.03 & $4.64\pm0.26$ & $1.83\pm0.17$ & $1.76\pm0.08$\B~~\\% & 12.42 & 12.89\B~~\\%[3pt]
\hline
Red$^a$\T & -9.90 -- -10.6 & -10.28 & 847 & $4.2\pm0.3$ & 2.0 & 0.74--1.05 & 0.90 & $4.82\pm0.35$ & $1.43\pm0.32$ & $1.86\pm0.13$~~\\% & 12.66 & 13.10~~\\%[3pt]
Red\T & $<$-10.6 & -10.91 & 795 & $4.0\pm0.3$ & 2.7 & 0.74--1.05 & 0.89 & $5.04\pm0.32$ & $2.04\pm0.25$ & $1.81\pm0.11$\B~~\\% & 12.62 & 13.08\B~~\\%[3pt]
\hline
\hline
\end{tabularx}
\end{center}
%\hspace{0.3in}$^{a}$  Units of log(yr$^{-1}$).

%\hspace{0.3in}$^{b}$  Units of $10^{-4}~h^3$ Mpc$^{-3}$.

%\hspace{0.3in}$^{c}$  Units of log($M_{\sun}$).

\hspace{0.3in}$^{a}$  $z$$<$1.05 galaxy samples are limited to $M_{B}$$<$-20.5.

\hspace{0.3in}$^{b}$  $z$$<$1.4 galaxy samples are limited to $M_{B}$$<$-21.0 .

\label{tab:ssfrbinresults}
\end{table*}

\end{document}